\documentclass[sigconf]{acmart}

\AtBeginDocument{%
  \providecommand\BibTeX{{%
    \normalfont B\kern-0.5em{\scshape i\kern-0.25em b}\kern-0.8em\TeX}}}

\copyrightyear{2025}
\acmYear{2025}
\setcopyright{cc}
\setcctype{by-nc-sa}
\acmConference[IUI '25]{30th International Conference on Intelligent User Interfaces}{March 24--27, 2025}{Cagliari, Italy}
\acmBooktitle{30th International Conference on Intelligent User Interfaces (IUI '25), March 24--27, 2025, Cagliari, Italy}\acmDOI{10.1145/3708359.3712151}
\acmISBN{979-8-4007-1306-4/25/03}

\usepackage[final,commandnameprefix=always]{changes}
\usepackage{xcolor}
\definechangesauthor[name={Your Name}, color=orange]{YN}

\begin{document}

\title{MemPal: Leveraging Multimodal AI and LLMs for Voice-Activated Object Retrieval in Homes of Older Adults} 


\author{Natasha Maniar}
\orcid{0000-0002-7490-3718}
\affiliation{%
  \institution{MIT Media Lab}
  \city{Cambridge}
    \country{USA}}    
%  \streetaddress{P.O. Box 1212}
%%\city{Cambridge}
%%\state{Massachusetts}
%%\country{United States}
\email{nmaniar@media.mit.edu}

\author{Samantha Chan}
\orcid{0000-0003-1159-0467}
\affiliation{%
  \institution{MIT Media Lab}
  \city{Cambridge}
    \country{USA}}    
%  \streetaddress{P.O. Box 1212}
%%\city{Cambridge}
%%\state{Massachusetts}
%%\country{United States}
\email{swtchan@media.mit.edu}

\author{Wazeer Zulfikar}
\orcid{0009-0008-7753-8817}
\affiliation{%
  \institution{MIT Media Lab}
  \city{Cambridge}
    \country{USA}}    
%  \streetaddress{P.O. Box 1212}
%%\city{Cambridge}
%%\state{Massachusetts}
%%\country{United States}
\email{wazeer@media.mit.edu}

\author{Scott Ren}
\orcid{0009-0002-8725-1648}
\affiliation{%
  \institution{MIT Media Lab}
  \city{Cambridge}
    \country{USA}}    
%  \streetaddress{P.O. Box 1212}
%%\city{Cambridge}
%%\state{Massachusetts}
%%\country{United States}
\email{scottren@media.mit.edu}

\author{Christine Xu}
\orcid{}
\affiliation{%
  \institution{MIT Media Lab}
  \city{Cambridge}
    \country{USA}}    
%  \streetaddress{P.O. Box 1212}
%%\city{Cambridge}
%%\state{Massachusetts}
%%\country{United States}
\email{cjx@mit.edu}

\author{Pattie Maes}
\orcid{0000-0002-7722-6038}
\affiliation{%
  \institution{MIT Media Lab}
  \city{Cambridge}
    \country{USA}}    
%  \streetaddress{P.O. Box 1212}
%%\city{Cambridge}
%%\state{Massachusetts}
%%\country{United States}
\email{pattie@media.mit.edu}

\renewcommand{\shortauthors}{Maniar, et al.}

\begin{abstract}
% ONLY FOCUSING ON RETROSPECTIVE MEMORY LOSS!
 Older adults have increasing difficulty with retrospective memory, hindering their abilities to perform daily activities and posing stress on caregivers to ensure their wellbeing. Recent developments in Artificial Intelligence (AI) and large context-aware multimodal models offer an opportunity to create memory support systems that assist older adults with common issues like object finding. This paper discusses the development of an AI-based, wearable memory assistant, MemPal, that helps older adults with a common problem, finding lost objects at home, and presents results from tests of the system in older adults' own homes. Using visual context from a wearable camera, the multimodal LLM system creates a real-time automated text diary of the person's activities for memory support purposes, offering object retrieval assistance using a voice-based interface. The system is designed to support additional use cases like context-based proactive safety reminders and recall of past actions. We report on a quantitative and qualitative study with N=15 older adults within their own homes that showed improved performance of object finding with audio-based assistance compared to no aid and positive overall user perceptions on the designed system. We discuss further applications of MemPal’s design as a multi-purpose memory aid and future design guidelines to adapt memory assistants to older adults’ unique needs. 

\end{abstract}

\begin{CCSXML}
<ccs2012>
   <concept>
       <concept_id>10003120.10003121.10003128</concept_id>
       <concept_desc>Human-centered computing~Interaction techniques</concept_desc>
       <concept_significance>500</concept_significance>
       </concept>
    <concept>
            <concept_id>10003120.10003121.10003124.10010870</concept_id>
           <concept_desc>Human-centered computing~Natural language interfaces</concept_desc>
           <concept_significance>500</concept_significance>
           </concept>
   <concept>
       <concept_id>10003120.10003121.10011748</concept_id>
       <concept_desc>Human-centered computing~Empirical studies in HCI</concept_desc>
       <concept_significance>300</concept_significance>
       </concept>
 </ccs2012>
\end{CCSXML}

\ccsdesc[500]{Human-centered computing~Interaction techniques}
\ccsdesc[500]{Human-centered computing~Natural language interfaces}
\ccsdesc[300]{Human-centered computing~Empirical studies in HCI}
\keywords{memory assistant, large language models, large visual language models, voice interfaces, context-aware agent, multimodal systems, wearables, older adults}

\begin{teaserfigure}
    \centering
    \includegraphics[scale=0.25]{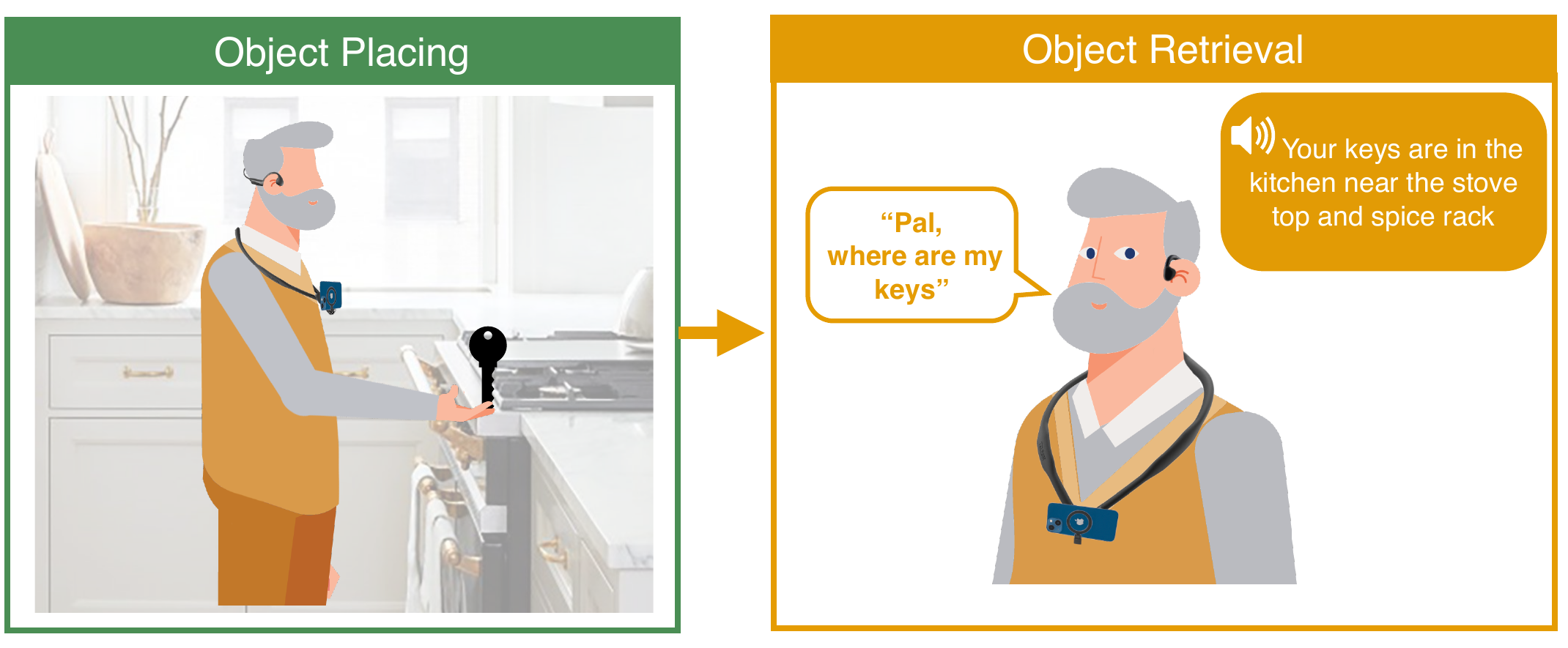}
    \caption{MemPal object location tracking and retrieval feature}
    \label{fig:mempalsystem}
\end{teaserfigure}


\maketitle

\section{Introduction}

The global population of older adults is projected to reach 2.1 billion by 2050, with their ratio of the population in the USA expected to increase from 16\% to 22\% \cite{WHO2024, statista2022agepopulation}. Memory impairments, including Subjective Cognitive Decline (SCD), Mild Cognitive Impairment (MCI), and diagnosed dementia, affect about one-third of this demographic \cite{AlzheimersFactsFigures2023}. Despite these challenges, most older adults prefer to age in place rather than in a nursing home \cite{kasper2015disproportionate} with many living alone \cite{harris2006experience}. Memory function is critical for their independence to comfortably live at home with minimal assistance\cite{marchant2011memory}.

Older adults experience declines in retrospective memory, significantly impacting daily functions like object retrieval \cite{mogle2023individual, ramos2016designing}, significantly more than younger adults \cite{balota2000memory}. Such memory issues not only strain individuals but also their caregivers—mostly unpaid family members—who face burnout from constant monitoring \cite{kasper2015disproportionate}. The economic impact of unpaid caregiving was valued at 196 billion in 1997 \cite{arno1999economic}.

%An opportunity exists to leverage technology to address some of the memory issues of older adults. A crucial challenge, however, is tailoring the technology to match their preferences, such as natural voice interaction rather than text-based communication on non-wearable external devices like smartphones or tablets \cite{stigall2019older}. Advances in AI and natural language processing, specifically multi-modal foundation models (LLMs), can be used to understand context and human intent, and facilitate easier interaction through voice-based interfaces \cite{brown2020language,wang2023enabling}.

There is an opportunity to use advances in AI and conversational interfaces to alleviate some of these memory challenges. Specifically VLMs (Vision language models) can better transform visual and text data for contextual assistance like creating accurate activity descriptions and LLMs can understand complex queries, necessary for memory impaired individuals. Successful development of such technology requires focusing on key needs and usability preferences of older adults, such as integrating natural voice interaction instead of text-based interfaces to support independent living \cite{stigall2019older, brown2020language, wang2023enabling} or simplifying the onboarding process to reduce complexity \cite{collerton2014exploratory}.  

%Tailoring technology to older adults' preferences, such as using natural voice interaction instead of traditional text-based interfaces, could enhance usability and easier interaction to support independent living \cite{stigall2019older, brown2020language, wang2023enabling}.

%We began by interviewing a diverse group of participants, including older adults, individuals with and without dementia, caregivers, and physicians, to identify key user needs and pain points. These interviews helped shape the design of a new technological intervention aimed at addressing these  challenges. Based on the insights gathered, we developed targeted features that were incorporated into the development of MemPal. 

This paper presents a multimodal wearable system for older adults that leverages LLM technology to assist their retrospective memory within their homes. Our development and testing of this system begins with a common pain point—assisting with finding lost objects. MemPal utilizes visual context and a voice-based, open-ended natural language interface to help users locate misplaced objects (Figure \ref{fig:mempalsystem}). It automatically logs visual information about the user’s environment (object held, location) and activities into a text-based diary which then integrates with a voice-queried LLM for retrospective memory assistance, for example to assist with finding misplaced objects. The system can possibly be further adapted to address other memory issues like recalling past actions and assisting with proactive memory, like providing proactive context-based safety reminders.

The questions we aim to address in this research are as follows: 

\begin{itemize}
    \item \textbf{RQ1}: What are the effects of using a voice-enabled multimodal LLM system for object retrieval on retrieval accuracy, path length, cognitive task load, retrieval confidence and recall difficulty for older adults, compared to using visual cues or no system?
    
    \item \textbf{RQ2}: What are older adults' perceptions and experiences of using a voice-enabled wearable system like MemPal for object retrieval and generally memory assistance?  
\end{itemize}

To evaluate MemPal, we conducted a study with 15 older adults (ages 62-96) in their own homes, simulating a near real-life setting. The study focused on assessing overall user perceptions and the technical accuracy of the system in varying home environments. Findings indicated that participants' objective performance in object finding tasks significantly improved with MemPal's audio assistance in terms of decreased search time and number of objects retrieved as compared to no assistance, but similar to using visual aid. Perceived difficulty of recall also significantly reduced with MemPal's assistance. 

The contributions of this paper include: 
\begin{itemize}
    \item (1) Design of an LLM-based multimodal, wearable system, MemPal, that assists older adults with memory issues such as finding lost objects within their own home through a natural conversational voice interface and contextual visual information, requiring limited system onboarding through a user-centered approach.  
    \item (2) A within-subject user study with older adult subjects in their own homes that validates helpfulness of the object finding system, increased objective performance and decreased recall difficulty for object finding, and qualitative analysis of user preferences informing future design guidelines.
\end{itemize}

\section{Related Work}
Our work is inspired by previous work on wearable assistants for older adults as well as object-finding and lifelogging tools. Since memory aids are typically underutilized in older adults than in younger adults \cite{schryer2013use}, we explore the interaction of a multimodal user interface within the older adult population to assist with memory. 
% reframe to (1) overview of memory aids for older adults- all types. Then talk about how younger vs older people approach memory aids - https://onlinelibrary.wiley.com/doi/abs/10.1002/acp.2946?casa_token=a5HTMdaIx80AAAAA%3AR1iSMU16PA3BgxD26399vRlZiTH6_q5Z-dHSYt4QWvQjP5fKEzBO9JRASPsfxHo3Mvajj0k-Loc77do 

% clearly state the novelty in the work- automated diary, voice based, AI context based and home tour (limited onboarding) 

\subsection{Voice-based, Camera-based and Wearable Assistants for Older Adults}
\label{relatedworks_assistants}
%voice-based 

Prior works showed positive perceptions towards voice user interfaces (VUIs) by older adults and benefits of VUIs (e.g., Google Home, Amazon Alexa, social robots) in their daily lives more so than younger adults \cite{stigall2019older}. These benefits include being assistants for various functions such as seeking information and controlling home devices~\cite{mehrotra2016embodied,opfermann2017communicative,tsiourti2014virtual,o2022optimizing}, companions to alleviate loneliness~\cite{sidner2018creating,jones2021reducing,pradhan2019phantom,corbett2021voice,gasteiger2021friends}, helping users develop skills~\cite{brewer2017xpress,ali2018aging}, encouraging physical activity~\cite{bickmore2005s}, \textcolor{black}{and providing personalized reminders through AR~\cite{10.1145/3463914.3463918}}. Wearable conversational agents have been used to provide memory training for older adults~\cite{chan2020prompto,chan2019prospero,10.1145/3334480.3375031} and monitor their daily physical activity~\cite{romanvoice} but these agents do not use visual data for \textcolor{black}{general} memory assistance. 

%It finds that optimal trigger times for these voice based prompts/reminders are during lower cognitive loads and idle times, suggesting that assisting with prospective memory loss should take voice-notification timing into consideration. 
%As MemPal aims to provide just-in-time safety reminders based on location context, we investigate the potential timing for these prompts during tasks. 
%The language of voice queries is closer to natural language than typed queries~\cite{guy2016searching}, thus, voice-based assistants can be highly usable compared to screen-based/typing-based interfaces. 
With the integration of language models in voice assistants and advancements in language understanding, users can converse more naturally with these agents. LLMs now have an increased ability to process larger unstructured text and provide a contextualized response. In this work, we implement a multi-modal real-time LLM voice-based assistant that includes image input as context and uses vision language models for intelligent understanding. \textcolor{black}{Although the use of LLMs on older adults has recently been explored for fostering conversations such as A-CONECT \cite{hong2024aconect} and Mindtalker \cite{xygkou2024mindtalker}, the use of multimodal LLMs that can process greater context has not.}

%Camera-based [research lifelogging literature] Remote tracking of the older adults using c
Previous Human-Computer Interaction (HCI) studies have mainly explored camera-based systems to support remote tracking in older adults' homes and could reduce caregiver burden via a smart-connected home with camera sensors. These systems could provide health monitoring~\cite{abowd2002aware}, detecting anomalies in activities of daily living~\cite{buzzelli2020vision} and fall detection~\cite{de2017home}.
%Prior Human-Computer Interaction (HCI) studies and commercial products aim to solve this issue via a smart connected home with camera sensors or wearable camera systems. 
%Since 2002, the proposed idea of the aware-home has been an ongoing study to transform the home of an older adults individual into one that is integrated with sensors for better patient monitoring \cite{abowd2002aware}. 
%Applications for in home camera sensors primarily include fall detection systems \cite{de2017home} or activity recognition to detect anomalies in activities of daily living \cite{buzzelli2020vision}. 
Despite this, the technology readiness level of smart home and health monitoring technologies for older adults has been considered low and there is limited evidence that they help improve quality of life~\cite{liu2016smart}. Other limitations include the need to deploy multiple cameras to broaden monitoring coverage~\cite{hasan2019real}. %The extent of tracking is directly proportional to the number of installed cameras, with limited cameras potentially leading to gaps in activity identification.  
%Recent wearable camera systems have become more prevalent in the market, suggesting wider spread adoption of these technologies among the public. Examples include the Humane AI clip, Narrative clip camera, and smart glasses like RayBan Meta smart glasses. These devices are more lightweight and useful in tracking egocentric activity but not currently designed for older adults and especially people with memory struggles.  
Wearable camera-based systems have primarily been used as lifelogging devices for supporting retrospective memory in older adults (i.e. remembering past events or people's names), such as the SenseCam~\cite{hodges2006sensecam,dubourg2016sensecam} and Autographer plus Flo~\cite{molesworth2016evaluation} (both of which are neck-worn devices), but store images as memory, and approaches to text-based memory diaries require integration with smart homes for activity monitoring \cite{dahmen2018design}.
%Smart glasses, like Google Glass, could support older adults in checking if they performed tasks by reviewing previous captured image on the heads-up display and for providing time-based reminders~\cite{kunze2014wearable}.% Devices like lifelogging clips (e.g., Narrative Clip~\cite{farajsport}) are more likely to be worn by older adults compared to head-worn form factors (e.g., smart glasses, VR headsets, LED glasses)~\cite{schwind2020anticipated}.

Building upon literature, we present a combined wearable camera-based and voice-based system that helps older adults with object tracking and activity monitoring. To allow for both modalities, our proposed system would store text-only data as memory unlike other systems and prioritize older adult preferences for simplistic interfaces \cite{farivar2020wearable} such as voice-based and limited onboarding. %Building upon literature, we use a neck-worn instead of a head-worn form factor.

\subsection{Object Finding Systems}
 
There are several wearable camera systems that help users find misplaced objects but all provide visual aid either on a tablet or AR display and do not store long term memory to allow for chat-based interfaces. Audio based assistance has yet to be explored. Fiducial Marker Tracker (FMT) had a neck-worn camera and was tested with older adults; it required manual registration of objects (placing marker tags on objects) and captured videos anytime the objects are interacted with~\cite{fmt}. To recall the object's last state (e.g., light: on/off), users sorted through video footage grouped by object, which might not provide a seamless experience for object tracking. FMT also required installation and markers around the house, and marker detection was heavily dependent on user’s height (limited personalization). GoFinder also had a neck-worn camera but was registration-free and allowed users to review automatically-grouped images to find objects~\cite{gofinder} but not categorizing the name of the object. It was, however, only tested with young adults (18-28 years old) and has not yet been studied with older adults. Our proposed solution would not store any image data and provide a more seamless user experience for older adults by only voice based querying.  %It allows users to explore	video clips that capture objects of interest and can help user determine if they have completed certain actions with these objects. 
%Yagi, Takuma, et al. "GO-finder: a registration-free wearable system for assisting users in finding lost objects via hand-held object discovery." 26th International Conference on Intelligent User Interfaces. 2021.[Go-Finder]
Systems like LocatAR \cite{oshimi2023locatar} or Overthere \cite{seo2021overthere} either simplified or removed the object registration process entirely based on gesturing or user motions but required an AR headset system for continuous use posing questions about social acceptability which our solution would avoid. %was an augmented reality headset system that automatically registered objects and object movement based on user's grasping and placing motions, and then visually displayed directions to find the objects. 
%Oshimi, H., Perusquía-Hernández, M., Isoyama, N., Uchiyama, H., & Kiyokawa, K. (2023, March). LocatAR: An AR Object Search Assistance System for a Shared Space. In Proceedings of the Augmented Humans International Conference 2023 (pp. 66-76).
%The automatic registration is performed by image-based item movement recognition from the user's grasping and placing motions. Tracks item movement
%Another notable wearable system~\cite{funk2014representing} showed that displaying the last seen image of the object on head-mounted displays was more effective at helping users find objects compared to map view of the objects' location.

There are other camera-based systems, like CamFi, an AI-driven system to help find lost objects in multi-user scenarios that uses stationary camera and displays objects and last-seen users of the objects on a smartphone~\cite{yan2022camfi}. It has also only been tested with younger adults who are generally more technologically savvy \cite{olson2011diffusion}. 
Other commercially-available sensor systems, like RFID tags~\cite{lin2005object,abowd2002aware}, Ultra-Wideband (UWB) Systems ~\cite{fontana2002ultra} or Wi-Fi Fingerprint-Based Indoor Positioning ~\cite{he2015wi, nabati2023real}, can track objects but require manually tagging of these devices to the objects.

%Yan, Ge, et al. "CamFi: An AI-driven and Camera-based System for Assisting Users in Finding Lost Objects in Multi-Person Scenarios." CHI Conference on Human Factors in Computing Systems Extended Abstracts. 2022. [Lost objects found with multi-person scenario] 
%[Other commercially-available systems: Rfid tags + airtags] 
%Gofinder 
%Li, Franklin Mingzhe, et al. "FMT: A wearable camera-based object tracking memory aid for older adults." Proceedings of the ACM on Interactive, Mobile, Wearable and Ubiquitous Technologies 3.3 (2019): 1-25. [explore whether	video clips captures from a body-worn camera every time objects	 of interest are	 found within its field of view can help older adults determine if they have completed certain actions with these objects and	 what their states are]

While LLMs have been used to identify the attributes of objects being found by a mobile robot in the FindThis system~\cite{majumdar2023findthis}, none of the works, to date, have used LLMs for helping users themselves to find the objects and enabled a voice-based chat interface for querying and response. This work presents a system which uses multi-modal LLM AI capabilities to support one of its key features of object finding designed specifically for older adults allowing them to also ask follow-up questions on the whereabouts and action of the user during misplacement. It uses visual inputs as context to the LLM so that no explicit object markers or tags are required. It stores object location information as text instead of images to be mindful of users' privacy and to enable audio descriptions of where any hand-held object is located. We also build upon Go-Finder's study design~\cite{gofinder} to test our proposed object-finding feature.

\section{Design Considerations}
% Shorter version: 

We conducted a user needs analysis through live interviews, caregiver and memory support groups (facilitated by a hospital network) and asynchronous online forums with a diverse pool of older adults, caregivers, and physicians that would provide initial design specifications for a prototype system that supports the most common memory issues among the older adult population. Transcripts and forum posts were coded independently by two researchers following thematic analysis method~\cite{braun2006using} to generate initial themes. The researchers then reviewed the coded data and themes to come up with our final themes and analysis.

Full compiled themes and quotes are found in Appendix \ref{appendixinterviews}.

\subsection{Interviews}

\subsubsection{Issues} \label{designconsiderations}

\begin{itemize}
    \item \textbf{\underline{Issue 1}}: Finding misplaced objects is a time consuming issue. Caregivers listed the items that they often have trouble remembering and the emotional toll it takes to find them. These insights suggest that location detection must be accurate and objects may be left in enclosed areas (OC8-11). For example: "My [mother-in-law] MIL was always missing her purse and was sure someone had stolen it. She would ransack her room all night long. It would be in the closet. This happened often." (OC8).
    \item \textbf{\underline{Issue 2}}: \textcolor{black} {Safety hazards and concerns necessitate continuous caregiver monitoring, which is often managed through inefficient manual written reminders (D1, SP1, E6-8, C10, C12, OC1-8). For example, "We have notes everywhere, but he doesn't pay attention to any of them. Whenever he is performing any action, I just watch over him and make sure he has what he needs" (OC6).}
    \item \textbf{\underline{Issue 3}}: Confabulation and over-reliance on subjective questionnaires lead to inaccurate diagnosis of memory conditions and misconstrued memories (D1, D2, E6, E7, E8, SP1). For example, "My biggest issue with his memory is that he'll tell us he has something when he doesn't. He lies daily." (E6).
    
\end{itemize}

\subsubsection{Proposed Solutions}
\begin{itemize}
    \item \textbf{Solution 1: Object Finder} Inclusion of an object finding system that is voice activated and identifies within-home location with low latency. Overall, older adults want to feel independent and "they don’t want to go to a nursing home" (C11) so designing features that enables that is essential. 
    \item \textbf{Solution 2: Activity Diary}: We suggest the implementation of an automated version of a memory diary commonly used among older individuals with poor memory (SP1), that passively logs daily activities. Objective data from at-home activity monitoring can provide crucial insights to physicians, who are currently reliant on subjective often biased memory recollections to diagnose potential memory conditions.

    % Subjective
    % https://onlinelibrary.wiley.com/doi/full/10.1111/j.1479-8301.2011.00354.x
    % https://karger.com/dem/article-abstract/27/4/310/99175/Development-of-the-Subjective-Memory-Complaints
    % https://www.ncbi.nlm.nih.gov/pmc/articles/PMC8579566/
    
    %\item \textbf{Solution 3: Context-Based Safety Reminders}: We propose a context-based safety reminder system with a voice interface and a caregiver app for remote tracking based on passive task completion. Additionally, proactive reminders and the ability to inquire about past tasks via a voice interface can enhance older adults' independence by reducing their reliance on constant caregiver monitoring.   
    
\end{itemize}

\subsection{Form factor}

Based on previous works, we prioritized key design considerations to enhance user acceptance. 
Neck-worn devices are more likely worn by older adults compared to other form factors (e.g., smart glasses) \cite{schwind2020anticipated} due to higher social acceptability. \textcolor{black}{Research by BMC geriatrics proves that specifically lifelogging tools for the older adults are considered acceptable unless individuals are using the camera in public outside of the home \cite{gelonch2019acceptability}. MT and Go-finder also performed acceptability studies and found that individuals were more concerned with the functionality and how it could assist them rather than social acceptability \cite{fmt} \cite{gofinder}.} However ideal form factor was not the focus of this paper as we just used existing hardware to test our methods and concept. 
    % \textbf{Device Form:} Neck-worn devices have a higher social acceptability compared to other form factors (e.g. smart glasses).
    %Using the stereotype content model (SCM) that predicts a device's stereotypical perception and social acceptability, a study shows that devices near the neck like the lifelogging clip (ex. Narrative clip) are more likely to be worn by the older adults compared to other form factors that could sense camera input (smart glasses, VR headsets, LED glasses) \cite{schwind2020anticipated} since “they usually take their glasses off” (D4). %Research by BMC geriatrics proves that specifically lifelogging tools for the older adults are considered acceptable unless individuals are using the camera in public outside of the home \cite{gelonch2019acceptability}. FMT and Go-finder also performed acceptability studies and found that individuals were more concerned with the functionality and how it could assist them rather than social acceptability \cite{fmt} \cite{gofinder}. 

\section{System Design and Implementation}

MemPal is a wearable, multimodal memory assistant designed to help older adults with self-reported memory decline locate misplaced objects and remember past actions within their homes through visual context and voice-based interface. The prototype system features an egocentric camera, that captures images regularly. These images are analyzed in real-time to automatically generate a diary of all activities. An audio bone-conduction headset then enables users to voice-query this activity diary. The wearable assistant is accompanied by the MemPal smartphone app for initial onboarding. System architecture illustrated in Figure \ref{fig:overall}. \textbf{ More details found in Appendix \ref{systemimp}.}

\subsection{System Onboarding} 

\begin{figure*}
    \centering
    \includegraphics[scale=0.25]{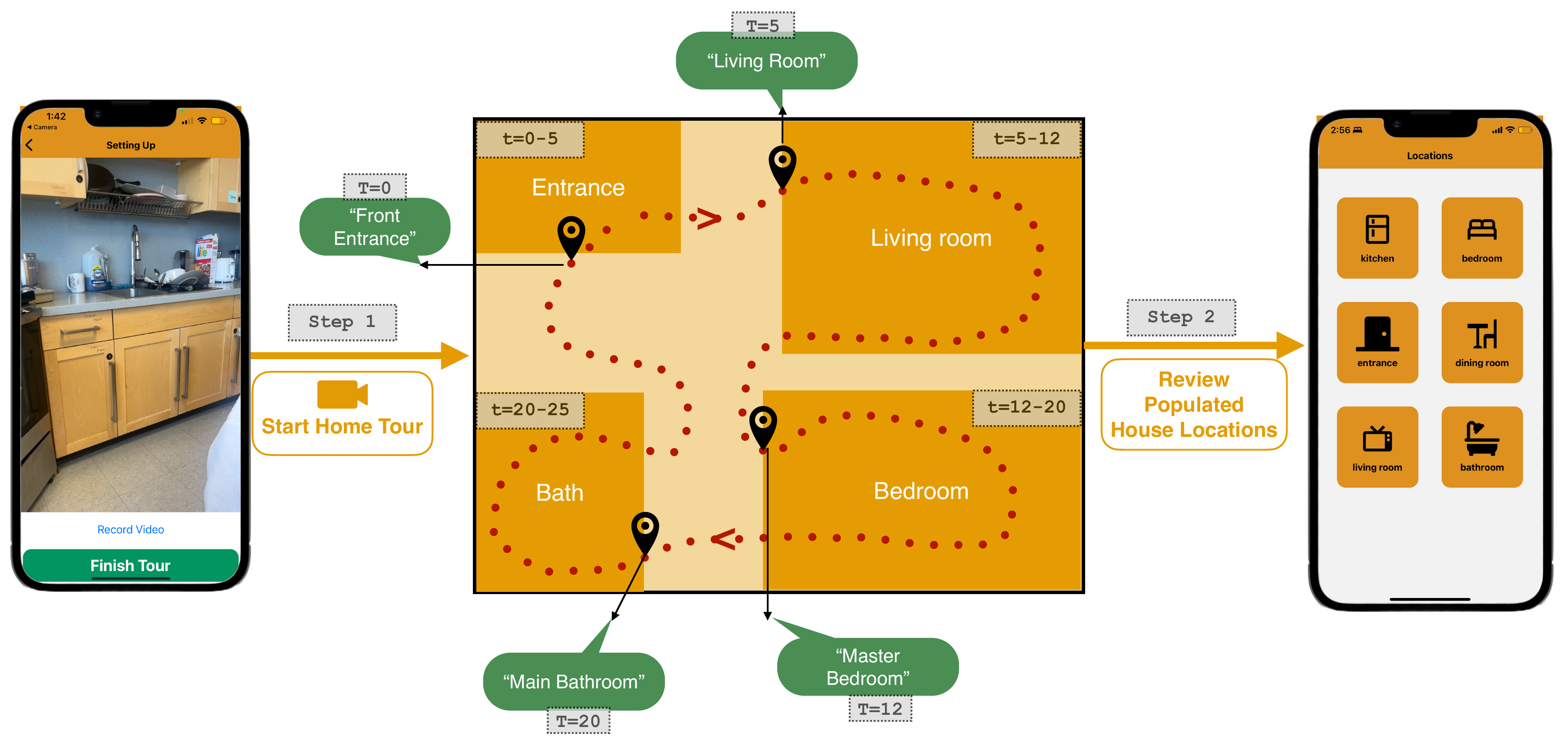}
    \caption{System onboarding phase: Using the MemPal app, the user first watches an instructional demo video before beginning the home tour video walk-through. Once the video is processed, the verbally labeled locations are populated in the app.} 
    \label{fig:location-system}
   
\end{figure*}

To create a seamless and personalized setup experience of the MemPal system for older adults, the user would take a 1-5 minute house-tour video while wearing the camera, which creates a spatial representation of their home later used for room localization. This initial information provides location context critical for object retrieval. Upon entering a new area, the user verbally labels and scans each room before moving to the next. After about 2 minutes of processing, the labeled locations appear for review in the MemPal app. To accommodate different house configurations and to enable a voice interface, MemPal allows users to voice-label rooms in a personalized manner (for instance labeling the entrance as parlor) to provide a more familiar experience as opposed to auto-labeling rooms in a standard format for all users. Figure \ref{fig:location-system} displays the user flow of the system onboarding process. 

\subsection{Activity Log}

\begin{figure*}
    \centering
    \includegraphics[scale=0.27]{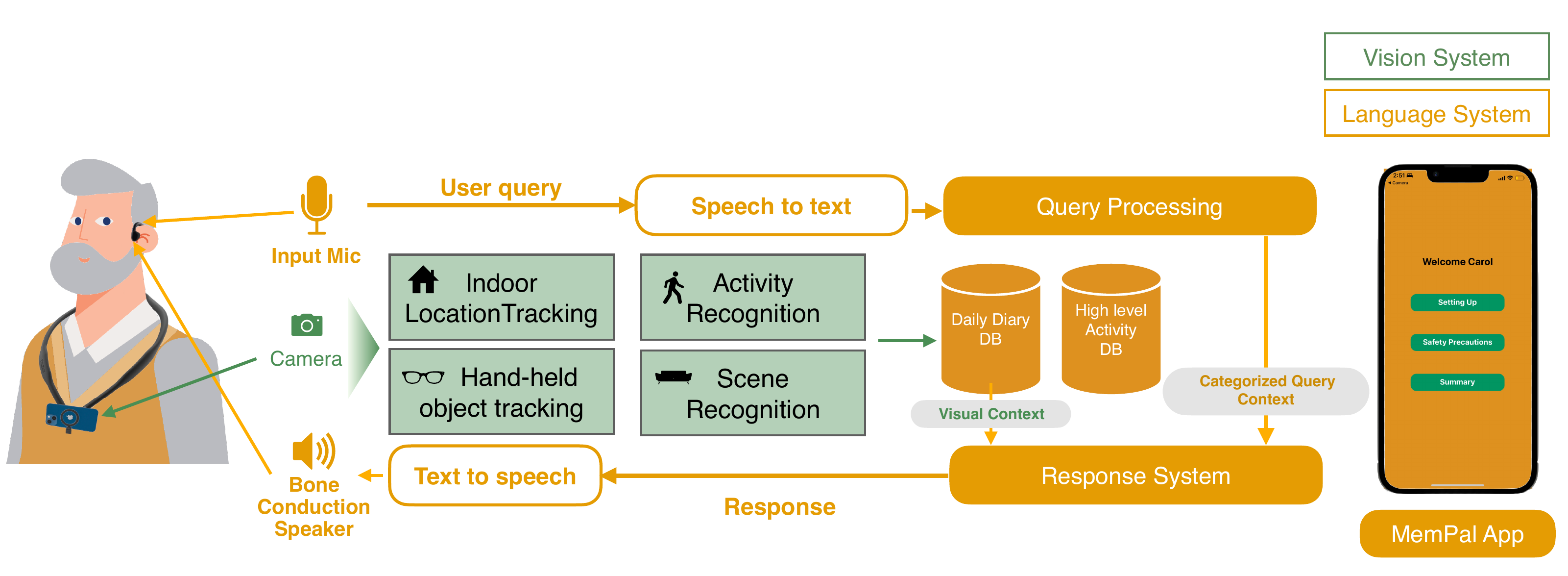}
    \caption{System implementation overview highlights the use of real-time visual context through a wearable camera for a question-answering language system using wearable audio I/O. The visual context from the camera consists of location tracking, object tracking, and activity and scene recognition to create a virtual diary (Daily Diary DB) which is used later for querying.} 
    \label{fig:overall}
\end{figure*}

To enable a multi-purpose memory augmentation system, MemPal consists of a vision system and language system (Figure \ref{fig:overall}) to create an activity log. The vision system comprises of parallel real-time detection AI models responsible for identifying the real-time location, generating background scene descriptions, and describing both hand-held objects and the current activity of the user. \textcolor{black}{Since MemPal uses a monocular egocentric camera for real-time indoor location tracking, it avoids the need for sensors like GPS or high-compute methods like SLAM. It generates an embedding map using CLIP embeddings and a room adjacency list from a calibration video, creating a personalized spatial map of the user's home. This map provides context for real-time localization. Activity is only detected once the user's hand(s) are observed as opposed to continuous "video-on" to prevent unnecessary video processing since activity detection models are expensive (ex. user is walking through a doorway). A Vision Language Model (GPT-4V)takes an input of tiled frames, text context of the previous activity, and a prompt, and outputs activity descriptions, objects in hand, and background descriptions that are useful for dynamic, context-sensitive dialogues.} 
% Related paper for hand gestures
% https://link.springer.com/article/10.1007/s40747-023-01173-6

%Location is identified via real-time indoor location tracking using an image-to-image embedding similarity approach along with a heuristic based ensemble methods, comparing incoming frames with the spatial map representation initially created during the onboarding process.

\begin{figure*}
    \centering
    \includegraphics[scale=0.27]{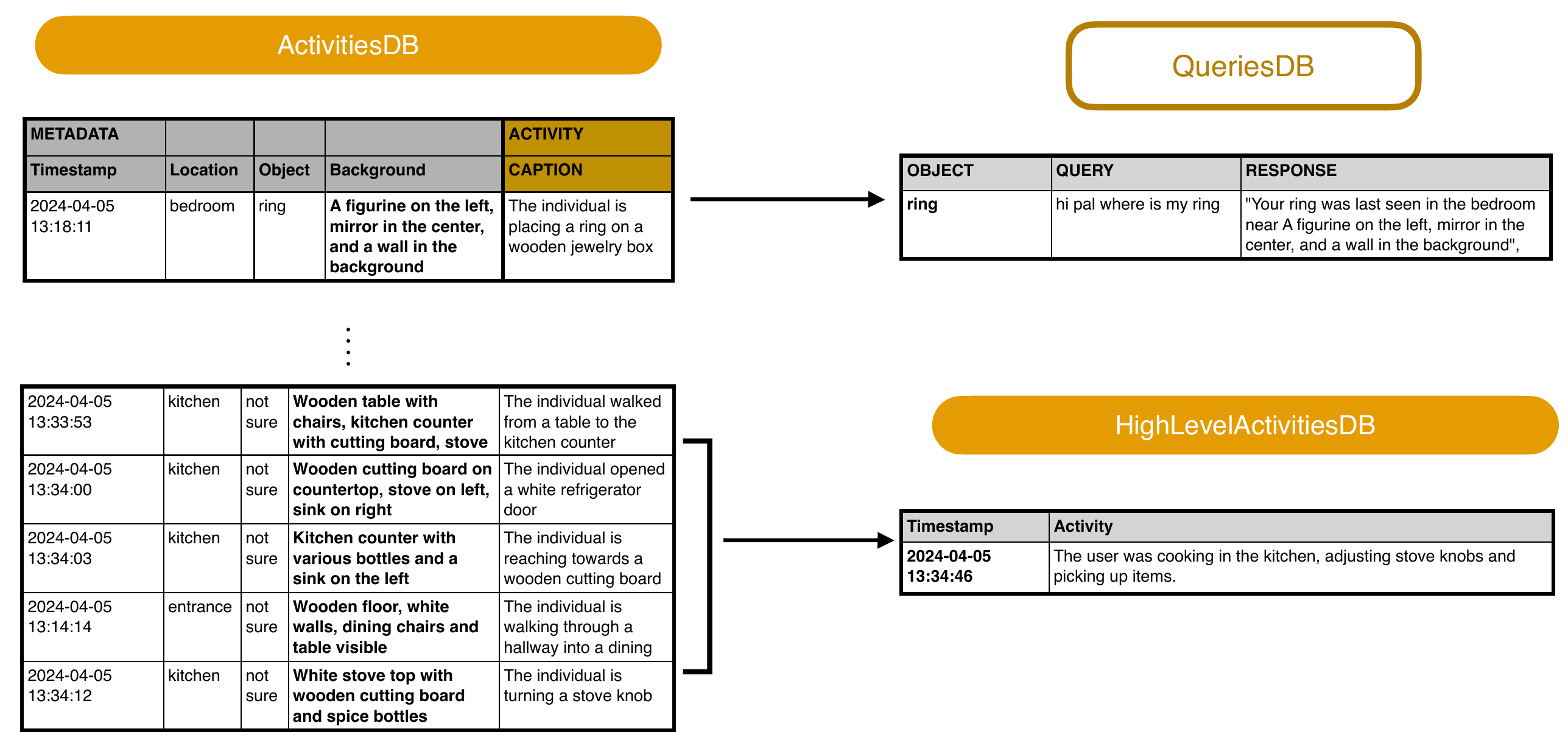}
    \caption{An example of activity log, query log, and higher level activities extracted (from Participant 5 of user study)} 
    \label{fig:activities}
\end{figure*}

The detection models produce text for each frame batch, which is embedded and stored in a vector database to create a time-sequenced activity log. The language system powered by an LLM agent described below understands user voice queries and responds accordingly using the vision context (user example in Figure \ref{fig:activities}).

The underlying technology architecture consisting of camera preprocessing, visual feature extraction, and LLM query/ processing are designed to be reused for multiple features besides object finding such as proactive safety reminders and retrospective recall of past actions. Figure 15 displays the usage of visual and language system to support context based safety reminders. Testing focused on the object finding feature.  

% add in figure of the 

\subsection{Object Retrieval}
%Once the location contexts are set up, the user can ask MemPal about any misplaced object without requiring explicit tagging or object registration and only using voice. The user would first activate MemPal through the wakeword: “Pal” before querying a specific object (Ex. "Pal, where are my keys" or "I can’t find my keys, Pal"). The voice query is processed to text, then categorized as an "object" query using an LLM, before extracting the last seen location of that object from the activity log to formulate an audio response in the following format: "Your [object] was last seen in the [detected room] near [background description of the area]". If the object was not seen, MemPal responds: "I’m not sure". 
%If users require additional information, they can pose follow-up questions without reiterating the object in search (e.g., "Pal, can you be more specific?" or "Pal, what was I doing right before I saw it?") implemented by storing chat memory. This design facilitates a naturalistic conversational interaction, mirroring human dialogue. Figure \ref{fig:object-system} demonstrates the user action and interaction sequence for object retrieval using MemPal. 
\subsubsection{User Flow}
Once the location contexts are set up during the onboarding process, MemPal allows users to voice-query about misplaced objects without needing explicit tagging or object registration. Users activate the system using the wakeword "Pal," followed by their query, such as "Pal, where are my keys?" or "I can’t find my keys, Pal." MemPal retrieves the last seen location of the extracted object from the activity log and responds in the format: "Your [object] was last seen in the [detected room] near [background description]." If the object is unlocated, MemPal replies with "I’m not sure." \textcolor{black}{LLMs can manage increasingly complex queries that go beyond simple object retrieval, such as "What did I do before I misplaced my glasses?" This functionality is essential to meeting the needs of users, especially those with memory impairments, who may need more than basic object finding.}

\begin{figure*}
    \centering
    \includegraphics[scale=0.25]{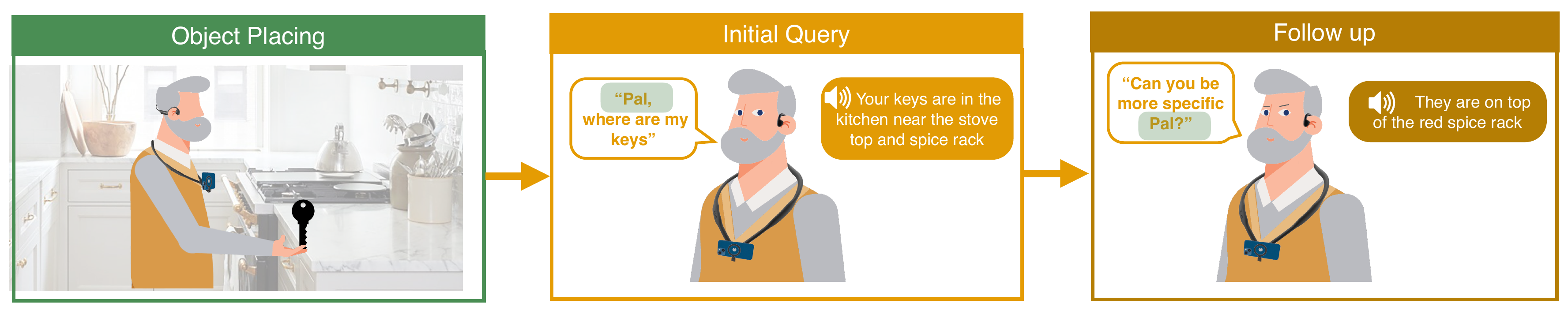}
    \caption{User flow for object retrieval. When the user places an object at a location in the house, MemPal stores this information which can be later retrieved during QA. The user can ask the location of the specified object as well as followup questions.}
    \label{fig:object-system}
\end{figure*}

Users can also ask follow-up questions without repeating the object in search, enhancing the fluidity of the conversation (e.g., "Pal, can you be more specific?" or "Pal, what was I doing right before I saw it?") \textcolor{black}{due to an LLM's ability to process long context.} This design, supported by the system's chat memory, enables interactions that mimic natural human dialogue powered by LLMs. Figure \ref{fig:object-system} illustrates the sequence of user actions and interactions for object retrieval using MemPal.

\begin{figure*}
    \centering
    \includegraphics[scale=0.27]{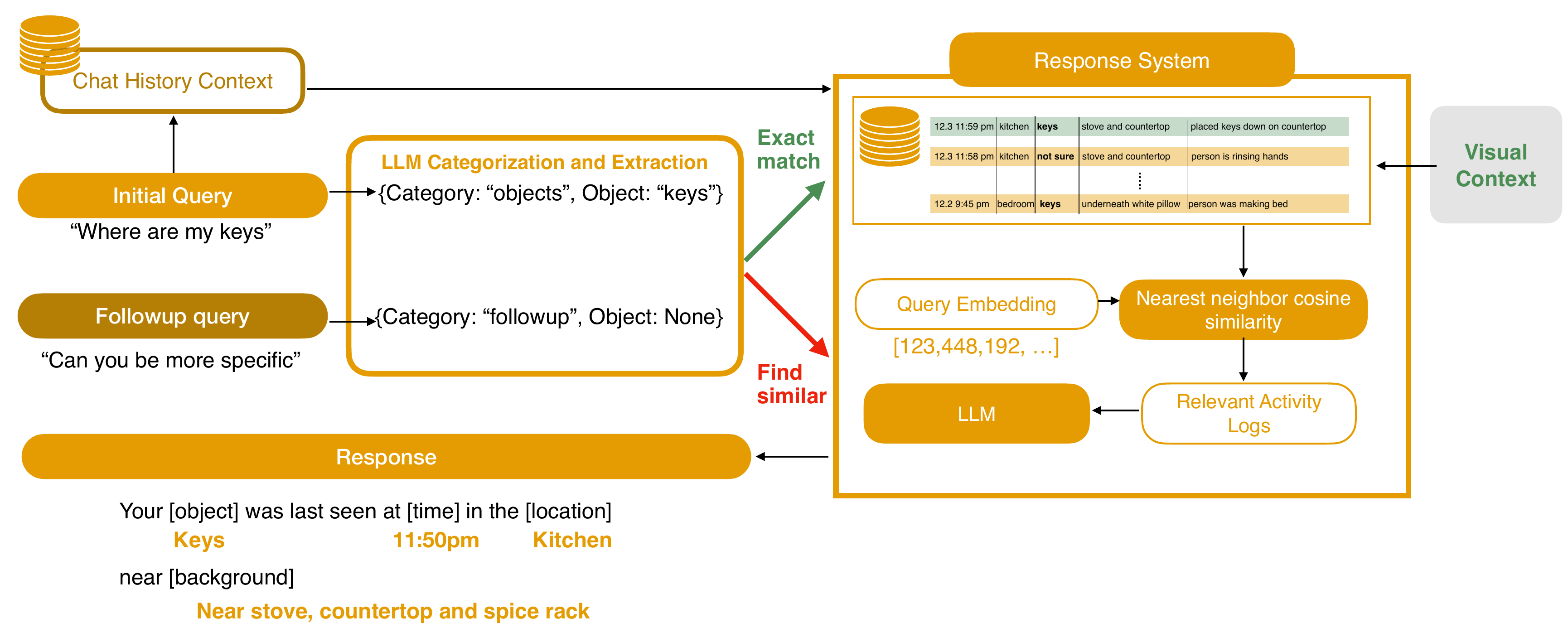}
    \caption{Object Retrieval Implementation: This workflow demonstrates the question and answer system specifically for object retrieval which uses the visual context as determined in Figure \ref{fig:visual} to respond to user queries starting with query categorization.} 
    \label{fig:object-imp}
\end{figure*}

\subsubsection{Language System Implementation}

\textcolor{black}{During object retrieval, \textit{ActivitiesDB} is filtered based on object metadata. If an exact match is found, the embeddings are chronologically sorted, and the most recent timestamp is returned. The format is: "Your [object] was last seen at [time] in the [location] near [background]." For similar objects with slightly different names, embedding similarity and retrieval-augmented generation (RAG) are used. The query is embedded using OpenAI’s Ada model (text-embedding-ada-002), and cosine similarity scores are calculated with \textit{ActivitiesDB} embeddings. The top k=10 most similar embeddings are selected, with metadata (timestamp, location, object, background) used to augment an LLM (OpenAI GPT-3) prompt for the last-seen details. If it's a follow-up question, prior chat history is included before invoking RAG. Figure \ref{fig:object-imp} illustrates the retrieval workflow.}

\subsection{Prototype Apparatus and Infrastructure}

The wearable prototype system for the study consists of an iPhone held with a magnetic clasp on a neck mount and a Bluetooth connected bone conduction headset for audio input and output shown in Figure \ref{fig:object-system}. The iPhone acts as an egocentric camera device centered between the shoulders and tilted slightly at a downward angle towards the user’s hands as it streams camera footage to a server. The bone conduction headset consists of a microphone and speaker around the ear, non-interfering with hearing aids. For the first prototype we used a neck worn camera in order to keep the camera from moving but still in a position of highest FOV for activity detection. An ideal form factor would attach on the user's clothes. We envision future iterations to the components integrated into a single magnetic clip with a mic, speaker, camera, and haptic feedback for query activation.

Vector databases for location calibration and automated diary are persisted locally on-device while location trajectory is persisted on Google Firebase Realtime database to be able to share them with authorized caregivers. User queries to MemPal are also stored locally on-device.

\section{User study}
 
We conducted a within-subjects study with 15 older adults in their own homes to evaluate the interaction, accuracy, and overall user experience of the MemPal system. %Participants walked through the location setup process and tested MemPal's object retrieval feature through a comparative study while the safety reminders and summarizer features were tested to understand user perceptions.

%\subsection{Study Design}

%To evaluate RQ1, participants engaged in an object finding task with audio assistance from MemPal and two existing methods (visual assistance and no assistance). To evaluate RQ2, participants provided subjective feedback for all features.$

\begin{figure*}
    \raggedright
    \centering
    \includegraphics[scale=0.25]{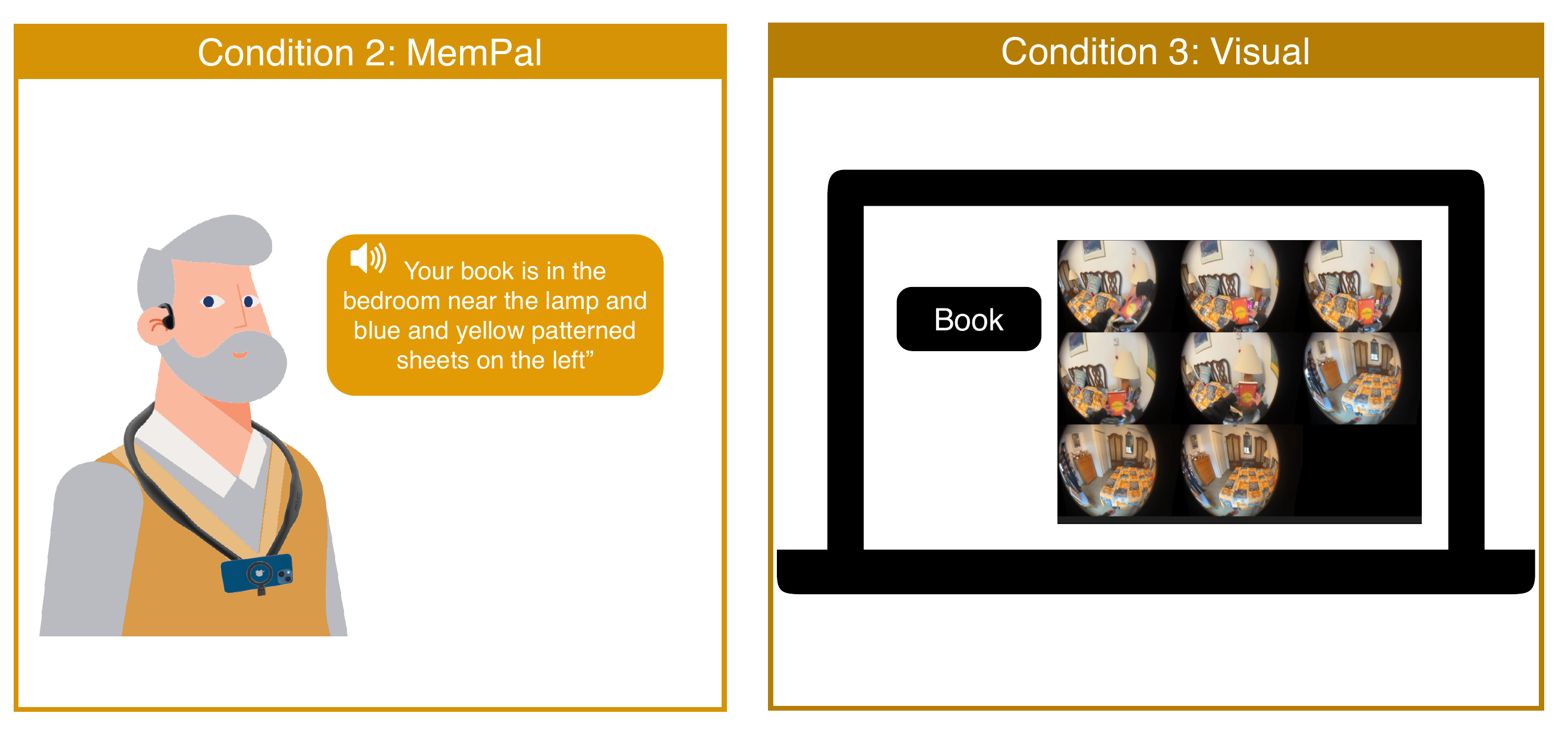}
    \caption{The user interaction for each memory assistance condition during the object search process. (Left) The MemPal condition provides audio descriptions; (Right) The Visual condition displays a tiled image of the object's last seen location.}
    \label{conditions}
\end{figure*}

%The study was divided into two parts to address the research questions. 
The study had two parts. Part 1 tested the effectiveness of a voice-enabled object retrieval feature to answer RQ1, modeled after the GoFinder study \cite{gofinder}. It had three experimental conditions and the order was counterbalanced across participants (Figure \ref{conditions}): %Three conditions were compared in Part 1—Baseline (control), MemPal (assistance), and Visual (assistance similar to GoFinder)—to assess the effects of MemPal’s system on objective performance and subjective measures, 

\begin{itemize}
    \item \textbf{Baseline}: without assistance, control condition %(current status quo). 
    \item \textbf{MemPal}: with audio assistance (verbal descriptions of object's last seen location) triggered by a voice query
    \item \textbf{Visual}: with visual assistance of last seen tiled image of object along with detected object label (a subsystem of MemPal that is used to generate audio descriptions).
\end{itemize} 

The Visual condition builds off Gofinder’s Object-based aid system (which groups objects by visual appearance) but unlike GoFinder, the objects are labelled and labels are then clustered, allowing for easier search. The images are displayed on a laptop, as older adults prefer larger screens like those of tablets over smartphones \cite{kim2022exploring} and voice descriptions were not given in this condition. We showed the participants the tiled image of the 235 degree view since a larger camera view provided more visual context background for the user to identify the location of objects. However, in future iterations we plan remove distortion during post processing to avoid confusion.

%The first condition was a \textbf{Baseline} condition where participants searched for objects without any assistance. This condition demonstrates the status quo since all participants mentioned that they don’t use any memory assistance in the form of technology to help them find objects. We also set up a third condition: Visual which is a subsystem of MemPal and builds off Gofinder’s Object-based aid system (grouping images of classified objects and ranking based on timestamp for easier search) but just with no audio output. However, GoFinder's implementation differs due to a clustering approach with no text-based object categorization whereas our visual condition categorizes by object name leading to easier search. The image during the visual condition is used by the MemPal system to generate the audio descriptions in the MemPal condition. 

Part 2 studied the experience and user perceptions of MemPal to answer RQ2. 
%Part 2 evaluated safety reminders tailored for older adults as well as viewing the Summarizer app display and receiving qualitative feedback for these three features, to answer RQ2.  

\subsection{Tasks}
% part 1 and part 2
The objective measures calculated during object retrieval were meant to answer RQ1 and the subjective user experience measures determined based on experiences with object placing and retrieval were meant to answer RQ2. 
\subsubsection{Object Finding}
\begin{figure*}
    \raggedright
    \centering
    \includegraphics[scale=0.25]{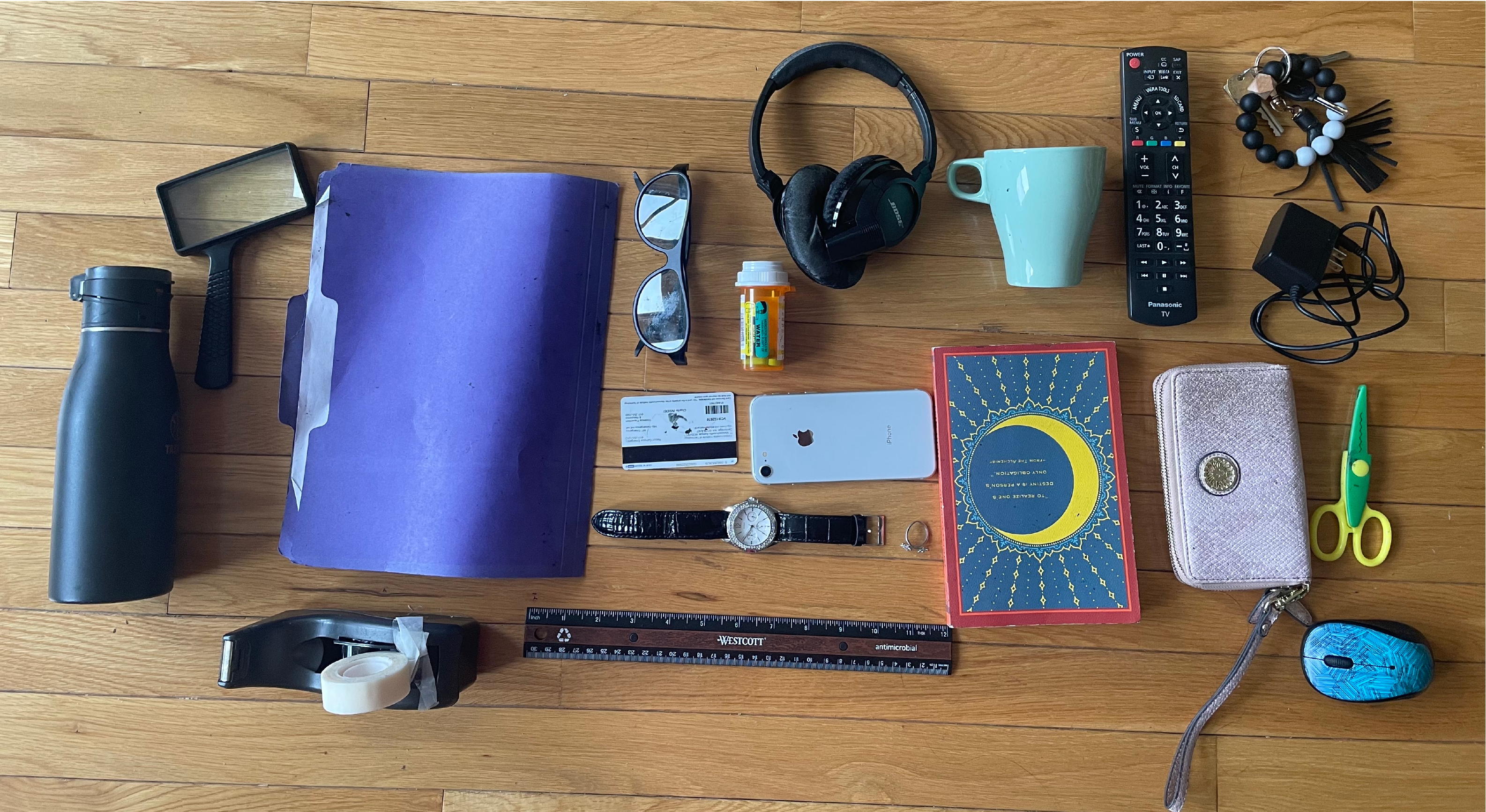}
    \caption{Object set used during the study for each participant are shown below: folder, cup, phone, bottle, medication, glasses, headphones, book, charger, remote, ID card, ring, wallet, watch, magnifying glass, tape, scissors, ruler, mouse, keys.}
    \label{fig:objects}
\end{figure*}

\textbf{Object Placing:} We collected 20 objects (Figure \ref{fig:objects}) which resembled commonly misplaced objects based on survey results of recruited participants (Appendix \ref{fig:demographics} and user research, Appendix \ref{appendixinterviews}). Occasionally, an object was replaced with an item from the user to enhance personalization. %Prior to the study we surveyed participants as well as online community forums of dementia patients for the most commonly misplaced items and among the following included: glasses, purse, phone, hearing aids, keys, coffee mug, and paperwork. 
Participants were first asked to wear the MemPal camera system and place all 20 objects in various locations around their house (for our system to auto-register the locations of objects) without any explicit labeling. Half were hidden, half were visible, and distributed evenly across rooms to avoid overcrowding (following the  GoFinder study \cite{gofinder}). No markers were used to indicate specific locations, unlike in GoFinder \cite{gofinder}, as our study was conducted in-the-wild. This setup was designed to mimic real-life scenarios of the participant themselves misplacing commonly used items in non-obvious places and to increase recall difficulty as much as possible given the study's short duration.

%\subsubsection{Object Retrieval}
\textbf{Object Retrieval:} 40 minutes after the object placing task, participants were asked to retrieve 20 objects from randomized conditions (6-7 objects each). The researcher specified the target object, and participants had up to 3 minutes to search, simulating a hurried search as in previous studies~\cite{gofinder}.

\subsection{Measures}

\subsubsection{Technical Evaluation}

%For the MemPal condition, 
We annotated whether the audio response from MemPal accurately represented each object's last seen location.
% A similar annotation was conducted for the tiled images during the Visual condition. 
Audio responses were categorized in the following (1) Correct: if the detected location and scene description of the background accurately reflected the object's location, (2) Incorrect Location: if the object was correctly categorized in the last seen timestamp so a correct description was generated but the Indoor location algorithm detected the incorrect location, (3) Object misidentified: If the object was correctly identified at some timestamp but misidentified at t-1 timestamp, (4) No object detected: if the object was not contained in the ActivitiesDB log metadata. For the Visual condition, a similar analysis was performed which evaluated the accuracy of solely the VLM system so camera image responses were categorized into the same (3) and (4) categories.   
%Camera positioning of whether the user placed the object in front of the frame before placing the object down would affect cases (2) and (4). Additionally occlusion of the object would affect the object detection subcomponent cases (3) or (4). 
Data for each participant's app screen, activity logs, and images were stored for manual review by two researchers post-study. Additionally, latency of responses were calculated from start of query to audio query response. Individual system component latency was also calculated for camera pre-processing, real-time location algorithm, VLM (GPT4-V) processing, and total time for each batch of streaming frames. The evaluation was conducted by two researchers who were independent of the data collection and blinded to the conditions.

\subsubsection{Objective Evaluation Measures}

 \begin{itemize}
     \item \textbf{Retrieval Accuracy:} We determined whether the user correctly found the object within the 3 minutes of search and then calculated a percentage of total objects found/ total objects within each condition to determine an accuracy metric.
     \item  \textbf{Path Length:} Measured as the number of rooms an individual searched (based on the calibrated locations) before finding an object within the 3 minutes. For each condition, we averaged the path length across all objects which included those that the user already remembered. 
 \end{itemize}

 % We hypothesized that path length would decrease and retrieval accuracy would increase with assistance. 

In the MemPal condition, we included only the data points where MemPal accurately delivered the correct audio response for the specified object. We excluded inaccurate responses as they caused participants to revert to the baseline condition, thereby minimizing accuracy as a confounding factor in our evaluation of the objective metrics.
Similarly, for the Visual condition, we retained only data points where the visual image produced by MemPal correctly depicted the last observed location of the object. Furthermore, we excluded instances where participants did not rely on the image for retrieval, ensuring that the analysis accurately reflects the assistance's effect on the objective metrics. We analyzed results both excluding and including these data points.

\subsubsection{Subjective Evaluation Measures}

After each condition, we measured self-perceived task load (using Raw NASA-TLX), confidence of retrieval (7-point Likert scale), and recall difficulty (7-point Likert scale) through written questionnaires as seen in Appendix \ref{appendixSelfReportedConfidence} (following the evaluation metrics in Memoro, memory augmentation device study~\cite{zulfikar2024memoro}). %Since there is significant positive correlation between memory performance and self-efficacy, we designed the questions to address RQ2.

After all conditions were completed for object finding, we measured the user experience and long-term potential use of just the MemPal condition with questions rated on a 7-point Likert scale. To evaluate the perceived helpfulness of MemPal’s audio descriptions versus visual images, we measured reliance on one type of assistance versus the other as well as overall usefulness of the camera feed both on a 7-point Likert scale. Full questionnaires are in the Appendix \ref{appendixPostObjectRetrieval}. Open-ended written feedback on overall thoughts and improvements were collected. %We asked participants two written questions of overall feedback on the specific feature regarding overall thoughts and areas of improvement.% which we then analyzed and coded as described in Section~\ref{sec_feedback}. %in order to determine how necessary camera-based images are in order to find lost objects quickly

%In addition, measures related to long-term potential use based on personal needs were evaluated with the following categories: likeliness to use and trust of the system’s responses.

%\subsubsection{Safety Reminders}
%\setcounter{secnumdepth}{4}
%\paragraph{\textbf{Technical evaluation of feature}}

%We assessed the accuracy of activity detection for safety reminder tasks based on the information displayed in the MemPal app. Accuracy for each tracked reminder was categorized into (1) Correct: MemPal app displayed the correct time and a checkmark for the completed task. (2) Incorrect activity detected: The activity was detected in ActivitiesDB, but the safety reminder LLM agent could not discern it from the text-based log. (2) No activity detected: The completion of the reminder was not found in ActivitiesDB. Data for each participant's app screen, activity logs, and images were stored for manual review by two researchers post-study.

%\setcounter{secnumdepth}{4}
%\paragraph{\textbf{Subjective Evaluation Measures}}
%We measured self-perceived user experience and usefulness of the safety reminder feature which included reminder input through the app and receiving audio-based reminders through questions rated on a 7-point Likert scale as seen in Appendix \ref{appendixPostSafety}. Open-ended written feedback on overall thoughts, improvements, and preferred format of safety reminders were collected. %also was coded and analyzed as described in Section~\ref{sec_feedback}. 

\subsubsection{Overall System Usability and Experience} After the study was complete, we asked participants to complete a system usability scale assessment (SUS) \cite{brooke1996sus} and rate the overall usefulness of the entire MemPal system in their everyday lives using a 7-point Likert scale. Then, we interviewed participants on their experience with the system, overall feedback and thoughts on privacy and data sharing. Full list of questions is in Appendix \ref{appendixOpenEndedInterview}. %Feedback was audio recorded and coded independently by two researchers following Thematic analysis method~\cite{braun2006using}. Researchers then reviewed the coded data and themes to come up with final themes and analysis. 

\subsection{Procedure}
\begin{figure*}
    \raggedright
    \includegraphics[scale=0.23]{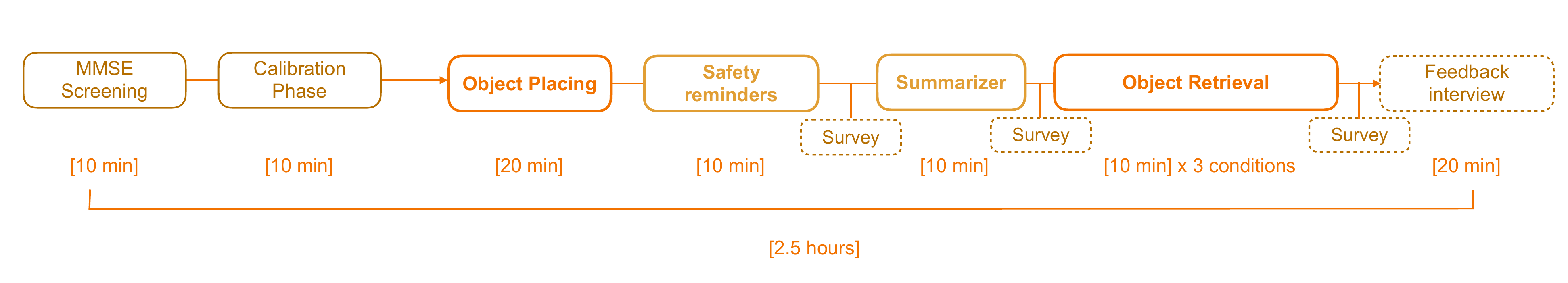}
    \caption{Procedure flow of the 2.5 hour study within each participant's home. All surveys took around 5 minutes.}
    \label{procedure}
\end{figure*}

The study was conducted in the older adults' homes within one floor and took about 2.5 hours (Figure \ref{procedure}). Demographic information was collected in the pre-screening questionnaire and follow-up emails. At the start of the study, a Mini Mental State Examination (MMSE) screening \cite{cockrell2002mini} was administered to assess baseline cognitive level. % since participants self-reported their memory condition (normal, subjective cognitive decline (SCD), MCI, Dementia) on the online screening questionnaire.
Then, participants were asked to use the MemPal app to walk through the setup phase which included reading instructions, watching an example video, and taking a video of their home while verbally labeling certain rooms as described in Figure \ref{fig:location-system}. Following that, participants performed the object placing task. Next, they took a 30 minute break to experiment with some other prototyped features like safety reminders and a summmarizer without formal testing. After another 10-minute break (which is about 40 minutes after the object placing task), participants retrieved the objects within the 3 conditions: (1) Baseline, (2) MemPal, (3) Visual. Lastly, we asked participants about their general feedback and concerns on the overall system through an interview.  
%The sequence of conditions were counterbalanced across participants and hidden/ non hidden objects were split equally between the conditions. 

\subsection{Participants}
Participants were recruited through local hospital-led memory support groups, email lists, caregiver support groups, social media, and partnering with senior communities. There were 15 participants ($9$ female, $6$ male, age range = $62$ to $96$, $M$=$74.9$, $SD$=$8.6$). Participants were fluent or native English speakers with normal or corrected-to-normal hearing and vision. All participants lived in either apartments or homes (8 in homes, 7 in apartments) with an average of $5.13$ rooms. Appendix \ref{fig:roomfreq} shows distribution of rooms across participant homes. Participants were asked about their cognitive ability and varied between normal cognitive ability (n=8), subjective cognitive decline (SCD) (n=4), and Mild Cognitive Impairment (MCI) (n=3). Additionally, participants rated the frequency of various actions, including: misplacing objects, forgetting to bring intended items, forgetting intended actions, forgetting errands, forgetting medications, and forgetting intended purchasing. The study received ethics approval from the university ethics review board, and participants gave written consent to participate with knowledge of what data was being stored and how it was processed. Participants received 45 USD worth of compensation for completing the study. Participant demographic data can be viewed in Appendix \ref{tab:demographics}. 

%Additionally, we also asked participants to list the most frequently misplaced items (Appendix). %\ref{fig:demographics}). 

\section{Results}
We analyzed the results from the user study to evaluate the overall system’s technical accuracy within different real-life settings, objective effects on object retrieval, and subjective user perceptions and experience for the overall system.

\subsection{\textbf{Technical Evaluation of System}} 
\label{sec_technical}
As part of the user study, we conducted a technical evaluation of the Object Finding system to understand the performance within different environments and in real world conditions. 

\subsubsection{Object Finding:}
%During the MemPal condition, we annotated whether the audio response from MemPal accurately represented the each object's last seen location. A similar annotation was conducted for the tiled images during the Visual condition. For MemPal, audio responses were categorized in the following (1) Correct: if the detected location and scene description of the background accurately reflected the object's location (2) Incorrect Location: if the object was correctly categorized in the last seen timestamp so a correct description was generated but the Indoor location algorithm detected the incorrect location (3) Object misidentified: If the object was correctly identified at some timestamp but misidentified at  t-1 timestamp (4) No object detected: if the object was not contained in the ActivitiesDB log metadata. Camera positioning of whether the user placed the object in front of the frame before placing the object down would affect cases (2) and (4). Additionally occlusion of the object would affect the object detection subcomponent cases (3) or (4). During the Visual condition, a similar analysis was performed which evaluates the accuracy of solely the VLM system so camera image responses were categorized into the same (3) and (4) categories.  

The results in Table~\ref{tab:objectacc} include analyses from all objects that were retrieved during the MemPal and Visual conditions. Count refers to all trials within each category and total count refers to responses within each condition. Since Visual is a subset of the MemPal condition, there were 145 trials that evaluated the accuracy of visual images and 92 trials which evaluated the accuracy of audio responses based on those images. MemPal's accuracy of audio descriptions was $72\%$  and Visual images was $53\%$. Inaccurate responses were mainly due to misidentified objects of $24\%$, which suggests more fine grained object detection. The location algorithm given the initial calibration video performed correctly $78\%$ of the time.  

\begin{table*}
    \centering
    \begin{tabular}{cccccc}
        \toprule
        \textbf{} & \textbf{Correct (MemPal)} & \textbf{Correct (Visual)} & \textbf{Incorrect location} & \textbf{No object detected} & \textbf{Object misidentified}\\
        \midrule
        Count & 66 & 28 & 20 & 18 & 22 \\
        Total Count & 92 & 53 & 92 & 145 & 92 \\
        Percent ($\%$) & 72\% & 53\% & 22\% & 12\% & 24\% \\
        \midrule
        \bottomrule
    \end{tabular}
    \caption{Object Retrieval Accuracy for MemPal and Camera Conditions}
    \label{tab:objectacc}
\end{table*}

%\subsubsection{Safety Reminders:} Table \ref{tab:safetyacc} shows the results for safety reminder accuracy. Based on the 3 reminders during the study, our system had an accuracy of $85\%$ where inaccuracies resulted mainly from the activity not being detected (usually due to camera positioning). Participants excluded from the safety reminder technical evaluation include P1-P4 due to app malfunctions during the task. 

\begin{table*}
    \centering
    \begin{tabular}{cccc}
        \toprule
        \textbf{Statistic} & \textbf{Correct} & \textbf{Incorrect activity detected} & \textbf{No activity detected} \\
        \midrule
        Mean Accuracy & 0.85 & 0.03 & 0.12 \\
        Stdev & 0.17 & 0.10 & 0.17\\
        Count Detected & 28  & 1 & 4\\ 
        Total Reminders & 33 & 33 & 33\\
        \midrule
        \bottomrule
    \end{tabular}
    \caption{Overview of the Safety Reminders Accuracy (P5-P15)}
    \label{tab:safetyacc}
\end{table*}

\subsubsection{Latency:} We evaluated the latency for the language system and vision system separately. Latency for the language system was calculated by the query response time (time of audio output) - query start time (time of audio input). From 204 interactions across all user studies the mean of query-response time was $2.17s$ and SD of $0.68s$. The vision system latency was calculated by first evaluating latency for each individual component and total processing time of each query to then be inserted in the database (Table \ref{tab:processtime}). 

\begin{table*}
    \centering
    \begin{tabular}{cccc}
        \toprule
        \textbf{} & \textbf{Locations} & \textbf{VLM} & \textbf{Total Time}\\
        \midrule
        Mean Process Time (s) & 0.429 & 11 & 5.689\\
        Stdev Process Time (s)  & 0.328 & 4.105 & 1.975\\
        Total Process Calls (avg/user) & 724.333 & 85.266 & 97.267\\
        \midrule
        \textbf{Total Pre-processing Time (s)} & 9.82\\
        \textbf{Total Full Time (s)} & 26.16\\
        \bottomrule
    \end{tabular}
    \caption{Device Processing Time}
    \label{tab:processtime}
\end{table*}

\subsection{Object Retrieval}
% it didn't work just for two people 

%% we didn't see differences in measures 
% limitations and discussions
% we think that it's because 

% accuracy of audio descriptions 
% we want to look at the ones that are accurate 
% in cases that it worked well, path length was....
%% including all the data and how much of the data were excluded 

% when the system did not give and 
% The accuracy of the system depends on the user’s home configuration and could drastically affect users' opinions on the feature, so we excluded participants $P3$, $P13$ for which the accuracy of MemPal was below $40\% $ when determining the SUS score.  Accuracy is calculated as percent of correct responses for the object finding task during the MemPal condition.

%\subsubsection{Objective}

Based on Section~\ref{sec_technical}, $28\%$ of data from the MemPal condition and $47\%$ from the Visual condition were inaccurate and thus, excluded from analysis to accurately assess the effects of assistance modes on the objective measures: retrieval accuracy and path length. We report the results for this set of data and any differences in results with all data (regardless of accuracy). 

\subsubsection{Retrieval Accuracy higher with MemPal than Baseline} According to the Shapiro-Wilk test, the retrieval accuracy was not normally distributed ($p$<$.05$). A Friedman test showed a significant difference in retrieval accuracy between conditions ($X^2$=$14.8$, $p$<$.001$). %p=.000624$
A post-hoc analysis using Wilcoxon signed-rank tests after Bonferroni correction showed that users had significantly higher retrieval accuracy with MemPal than Baseline ($p$=$.015$) but no significant differences for the other condition pairs: Baseline-Visual ($p$=$.087$), MemPal-Visual ($p$=$1.03$). Baseline ($M$=$.81$, $SD$=$.18$), MemPal ($M$=$.97$, $SD$=$.07$), Visual ($M$=$.95$, $SD$=$.11$). When including all data regardless of accuracy, differences between Baseline and MemPal loses significance ($p$=$.054$), and difference between Baseline-Visual is significant ($p$=$.02$). 

\subsubsection{Path Length lower with MemPal and Visual than Baseline:} 
\begin{figure*}
    \centering
    \includegraphics[scale=0.2]{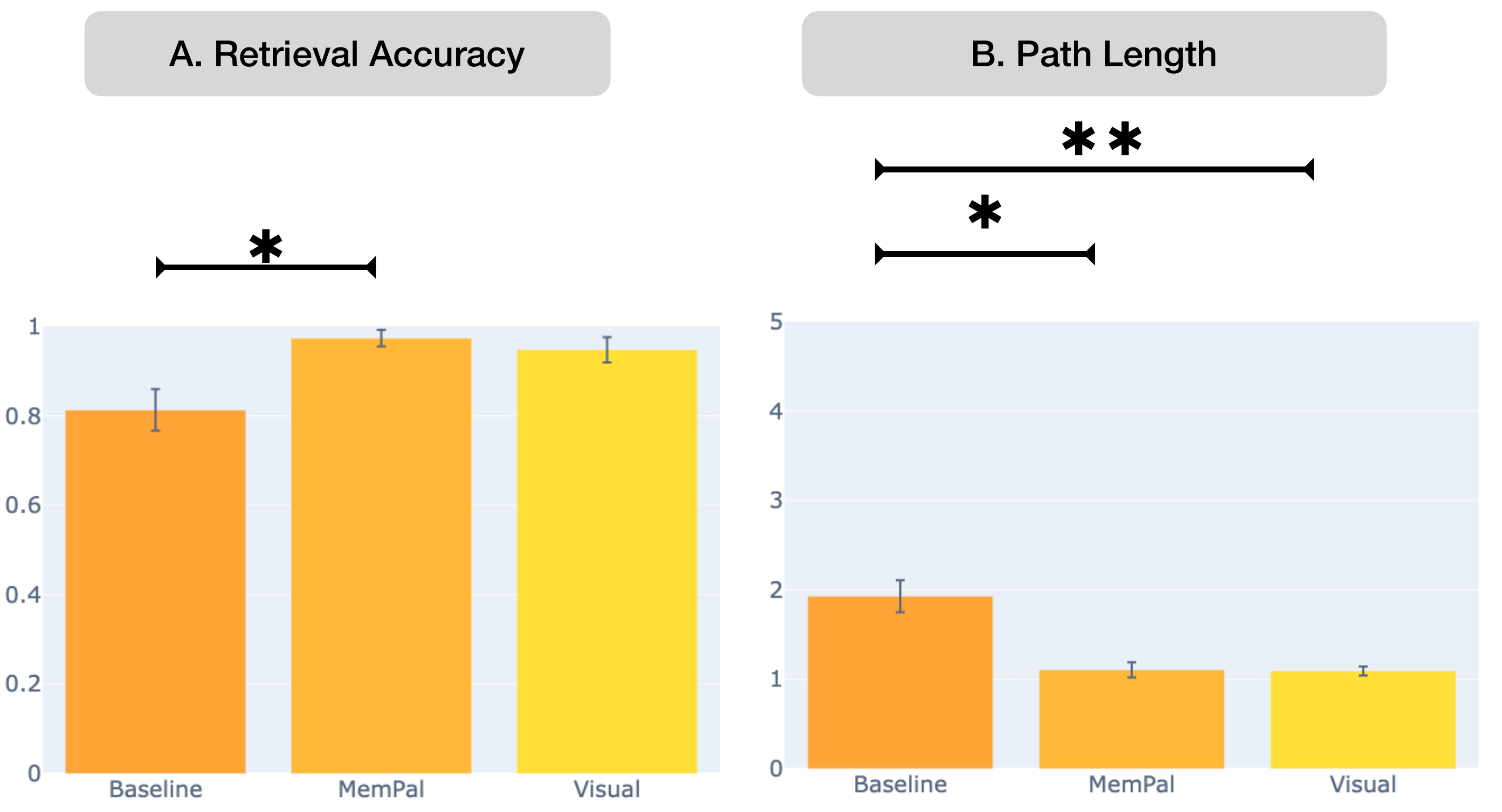}
    \caption{Retrieval Accuracy and Path length. $*$: p<.05, $**$: p<.001}
    \label{fig:objective}
   
\end{figure*}

Path length did not meet the normality assumption ($p$<$.05$). A Friedman test showed statistical differences between conditions for this measure ($X^2$=$17.7$, $p$<$.001$). 
%p=0.000143$
A post-hoc analysis using a Wilcoxon signed-rank test after Bonferroni correction showed that the path length for MemPal was significantly lower than Baseline ($p$=$.014$), and path length for Visual was significantly lower than Baseline ($p$=$.0066$). There was no significant difference between MemPal-Visual ($p$=$2.053$). Figure \ref{fig:objective} consists of the bar charts and mean of these objective metrics for each condition. Baseline ($M$=$1.93$, $SD$=$.69$), MemPal ($M$=$1.10$, $SD$=$.33$), Visual ($M$=$1.09$, $SD$=$.19$). When including all data regardless of accuracy, the same condition pairs remain significant ($p$<$.05$) with slight adjustments to the level of significance: Baseline-MemPal ($p$=$.007$) and Baseline-Visual ($p$=$.01$). 

%The results above were calculated looking only at trials where MemPal gave accurate information (so 26 trials out of 99 were excluded for the MemPal condition and 25 out of 53 trials were excluded for the Visual condition). When we included all trials, the differences in retrieval accuracy between Baseline and MemPal was no longer significant and instead difference between Baseline and Visual gained significance ($p$=$.02$). Path length significance remained the same.   

\subsubsection{No difference in Task Load between conditions} A repeated measures ANOVA ($p$=$.031$) showed significant differences in task load (RTLX) scores between conditions. A post-hoc analysis using pairwise t-tests with Bonferroni correction indicated that no condition pairs were significant: Baseline-MemPal ($t$=$1.47$, $p$=$0.486$) Baseline-Visual ($t$=$2.44$, $p$=$.087$)  MemPal-Visual ($t$=$1.60$, $p$=$0.398$). Baseline ($M$=$56.1$, $SD$=$18.3$), MemPal ($M$=$50.5$, $SD$=$17.8$), Visual ($M$=$44.4$, $SD$=$13.4$).

\subsubsection{No difference in rated Confidence between conditions} There was a significant main effect in the confidence of finding objects between conditions from a Friedman test ($X^2$=$8.84$, $p$=$0.012$). However, a post hoc Wilcoxon signed rank analysis after Bonferroni correction determined no condition pairs were significant: Baseline-MemPal ($X^2$=$26.0$, $p$=$0.91$) Baseline-Visual  ($X^2$=$13.0$, $p$=$0.06$)  MemPal-Visual $(X^2$=$2.0$, $p$=$0.212$). Baseline ($M$=$4.87$, $SD$=$1.19$), MemPal ($M$=$5.27$, $SD$=$1.53$), Visual ($M$=$5.87$, $SD$=$0.92$).

\subsubsection{Recall Difficulty lower with MemPal and Visual than Baseline} 
\begin{figure*}
    \centering
    \includegraphics[scale=0.23]{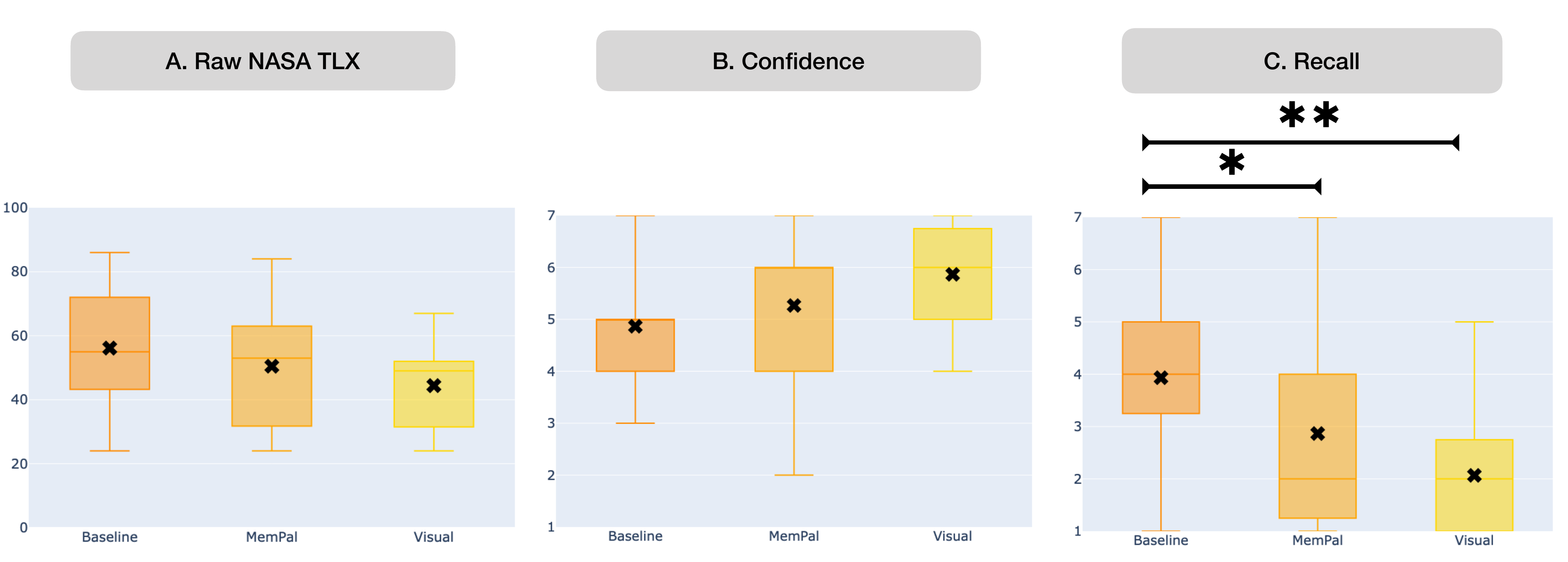}
    \caption{Raw NASA TLX for task load [0-100], confidence [1-7], difficulty to recall [1-7].  $*$: p<.05, $**$: p<.001}
    \label{fig:subjective}
   
\end{figure*}

A Friedman test showed significant differences in the recall difficulty of finding objects between conditions ($X^2$=$15.3$, $p$=$.0005$, Figure~\ref{fig:subjective}. A post hoc Wilcoxon signed rank analysis after Bonferroni correction showed that the recall difficulty for both MemPal ($p$=$.022$) and Visual ($p$=$.0064$) was significantly lower than Baseline. %, Baseline-MemPal ($p$=$0.022$), Baseline-Visual ($p$=$0.0064$). 
However, recall difficulty for MemPal was not significantly different than Visual ($p$=$0.119$). Baseline ($M$=$3.93$, $SD$=$1.67$), MemPal ($M$=$2.87$, $SD$=$1.81$), Visual ($M$=$2.07$, $SD$=$1.10$).

\subsection{User Experience for MemPal}
\begin{figure*}
    \centering
    \includegraphics[scale=0.25]{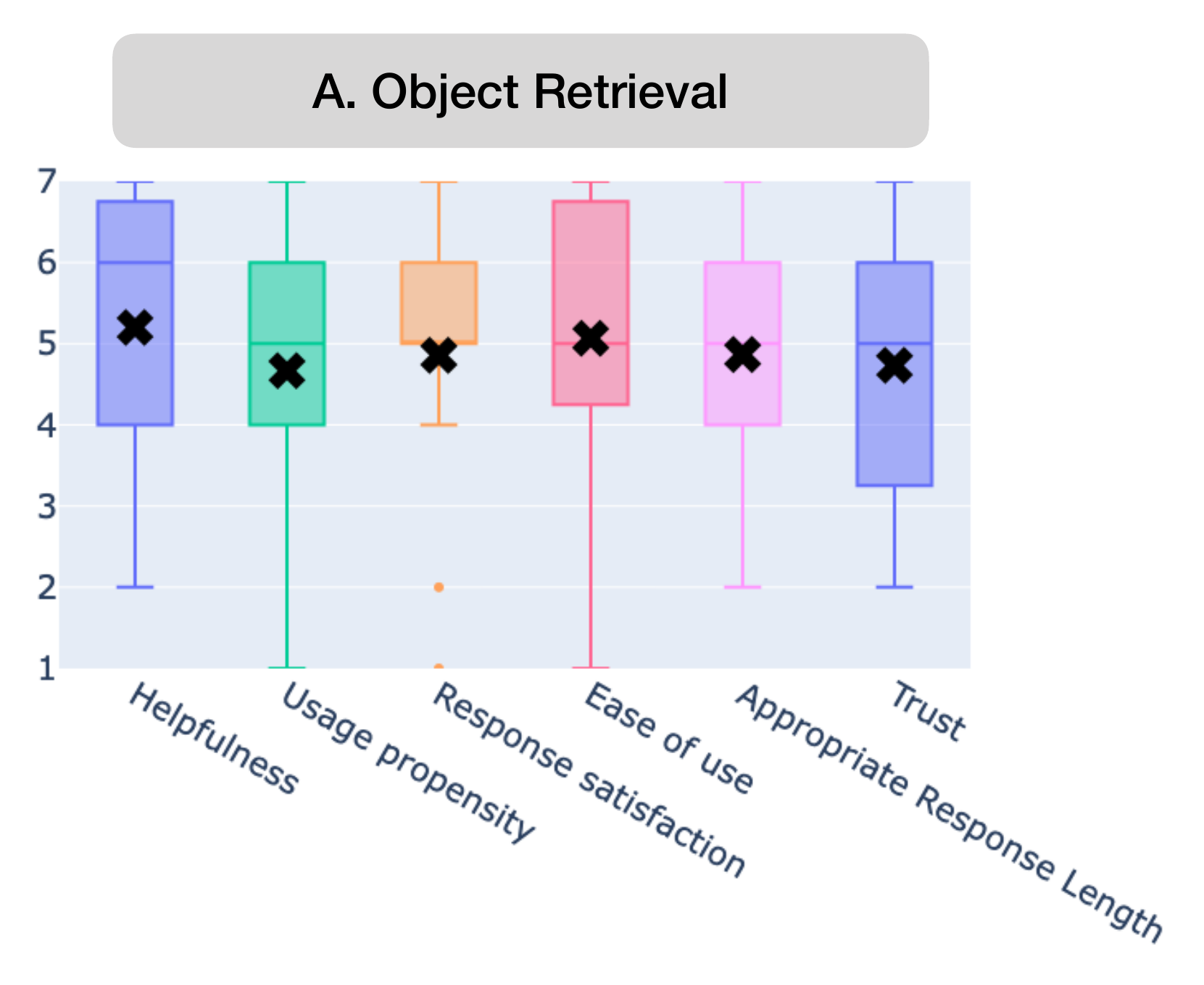}
    \caption{Categorized user perception and experience levels using [1-7] Likert scale.}
    \label{fig:features}
   
\end{figure*}

We analyzed overall user experience as shown in Figure \ref{fig:features}. 

For the Object Retrieval feature, just using MemPal’s audio descriptions, we found the mean scores across the categories ranked highest to lowest: Helpfulness ($M$=$5.20$, $SD$=$1.66$), Ease of use ($M$=$5.07$, $SD$=$1.87$), Usage propensity ($M$=$4.67$, $SD$=$2.02$), Response satisfaction ($M$=$4.86$, $SD$=$1.61$), Appropriate Response Length ($M$=$4.87$, $SD$=$1.36$), and Trust ($M$=$4.73$, $SD$=$1.83$). Additionally, we found the mean scores for usefulness of the Visual condition ($M$=$5.57$, $SD$=$0.97$) and reliance on Visual vs MemPal ($M$=$4.93$, $SD$=$1.49$). We found weak correlations between users’ MMSE results and their other perceptions of the system (Helpfulness r = -0.171, Ease of use  r = -0.356, Usage propensity r = -0.0442, Response satisfaction r = 0.0802, Appropriate Response Length r = -0.136, Trust r = 0.0235, usefulness r = 0.252) using Pearson's correlation tests. 

%\subsubsection{Safety Reminder:} For the Safety Reminder feature, we similarly evaluated the mean scores across the categories highest to lowest: Timing Satisfaction ($M$=$6.27$, $SD$=$1.03$), Helpfulness ($M$=$6.13$, $SD$=$1.25$), Usage propensity ($M$=$5.60$, $SD$=$2.20$), and Ease of use ($M$=$5.33$, $SD$=$2.13$).

%\subsubsection{Feature differences for helpfulness and usage propensity:} Lastly, we evaluated the difference of helpfulness between all features to determine if one or more features were more helpful than others (Figure \ref{fig:features-compare}). We conducted a Friedman test and found no significant differences ($X^2$=$4.68$, $p$=$.0962$): Object: ($M$=$5.20$, $SD$=$1.66$), Safety: ($M$=$6.13$, $SD$=$1.25$), Summarizer: ($M$=$5.53$, $SD$=$1.25$). Similarly for the usage propensity across all features, we found a significant difference across categories ($p$=$0.0064$). A post hoc analysis using Wilcoxon signed-rank test corrected by Bonferroni determined that there was a significant difference for Object-Safety ($p$=$.0177$) and Object-Summarizer ($p$=$0.0447$), but not for Safety-Summarizer ($p$=$2.18$). Object: ($M$=$4.67$, $SD$=$2.02$), Safety: ($M$=$5.60$, $SD$=$2.20$), Summarizer: ($M$=$5.73$, $SD$=$1.44$).

%\begin{figure}[h!]
   % \centering
   % \includegraphics[scale=0.25]%%%{charts/features-compare.pdf}
  %  \caption{Helpfulness and Usage propensity across all features. (Left-blue): Object Retrieval (Middle-Green): Safety, (Right-yellow): Summarizer. $*$: p<.05}
 %   \label{fig:features-compare}
%%\end{figure}

\subsection{System Usability}
MemPal’s overall system had a mean system usability score (SUS) of $69.375$ ($SD=17.3$) and overall usefulness in everyday life had a mean value of $6.0$ ($SD=1.51$). Following Go-Finder's study report on usability~\cite{gofinder}, we summarize each participant's SUS scores, age and average accuracy of MemPal in Table~\ref{tab:sus}. We did not have SUS scores for P3, P9 and P13 as they thought that the system was too early in development and opted to not complete the scale. A lower accuracy of MemPal as well as higher MMSE scores might have affected the overall SUS (usability) score. We found a moderate negative Pearson’s correlation between MMSE results and SUS scores (r = -0.6064, p = 0.04795). This indicates that participants who scored higher in MMSE (better cognitive function) tended to give the system lower SUS scores.

% \begin{table}[h!]
%   \centering
%   \input{tables/tables_sus}
%   \caption{SUS}
% \end{table}

%\documentclass{article}
%\usepackage{booktabs} % For professional looking tables

%\begin{document}

\begin{table*}
    \centering
    \begin{tabular}{cccc}
        \toprule
        \textbf{Participants} & \textbf{Overall SUS} & \textbf{Age} & \textbf{Accuracy} \\
        \midrule
        Participant 1 & 67.5 & 77 & 0.67 \\
        Participant 2 & 42.5 & 62 & 0.71 \\
        Participant 3 & N/A & 73 & 0.29 \\
        Participant 4 & 82.5 & 67 & 0.67 \\
        Participant 5 & 62.5 & 78 & 0.86 \\
        Participant 6 & 55 & 76 & 0.71 \\
        Participant 7 & 70 & 73 & 0.71 \\
        Participant 8 & 85 & 80 & 0.67 \\
        Participant 9 & N/A & 70 & 0.83 \\
        Participant 10 & 82.5 & 77 & 1.00 \\
        Participant 11 & 40 & 96 & 0.67 \\
        Participant 12 & 90 & 66 & 1.00 \\
        Participant 13 & N/A & 87 & 0.33 \\
        Participant 14 & 90 & 67 & 1.00 \\
        Participant 15 & 65 & 74 & 0.43 \\
        \midrule
        \textbf{Mean} & 69.375 & 74.27 & 0.71 \\
        \textbf{Standard Deviation} & 17.32460809 & 7.45 & 0.21 \\
        \bottomrule
    \end{tabular}
    \caption{Summary of Users' SUS Scores, Age, and Accuracy of MemPal. Participants with N/A for SUS opted not to complete the SUS scale. Lower accuracy of MemPal might have affected SUS score.}
    \label{tab:sus}
\end{table*}

%\end{document}

\textcolor{black}{\subsection{Qualitative Feedback}}
\label{sec_feedback}

Open feedback and interview transcripts were coded independently by two researchers following thematic analysis method~\cite{braun2006using} to
generate initial themes. The researchers then reviewed the coded data and themes to form our final analysis and themes.

% instead of paragraph form should we do bullet point and describe limitations and themes in bullets? 
\subsubsection{\textbf{Positive System Feedback}}
Several participants reacted positively to the system (P11, P14, P7, P13): ``I think this is so phenomenal and easy to use'' (P14) and ``at some point everyone is going to need a MemPal'' (P11). Participants thought the system onboarding phase was ``ok'' (P1, P7, P9, P10) and P10 liked that he could create ``personal labels'' of locations in his home. Participants felt that the object retrieval system was `helpful'' (P7, P10, P11, P14, P8) and P4 remarked that it was a``great feature''. P15 commented that ``It's really amazing that it can do that, that the system can know where my objects are''. A few also mentioned that just wearing MemPal would increase awareness of placing objects (P6, P2, P13): ``A new system makes you more conscious about hiding things'' (P6). Additionally, participants preempted potential other positive impacts of the system like ``learning your behavior and help you have better habits''(P2) and that it could ``give suggestion to reduce forgetting objects'' (P14). 

\subsubsection{\textbf{Improve wearable device comfort}} Although MemPal was not designed for optimal comfort, participants mentioned that the camera was bulky and not as lightweight (P5, P7, P10, P14, P12) but still thought the current``pendant idea is nice'' (P5) and bone conduction headset '' did not interfere with hearing aid '' (P5), and glasses can be confusing'' (P14).
\subsubsection{\textbf{Improve accuracy}} (1) Speech Recognition: Participants mentioned that speech recognition could be improved especially for the 'Pal' wake-word detection (P7, P9, P13) sometimes queries "taking multiple tries" (P8) especially for those with "poor pronunciation" (P11). (2) Insufficient audio descriptions: Many preferred the descriptions to be more specific but still succinct, such as ``knowing which drawer'' (P14), being able to ``identify colors'' (P1, P4), differentiate between similar objects, hidden versus not hidden objects, and ``height from the ground'' (P3) which the current system couldn't provide. However, P14 said ``There's a fine balance between how detailed it should be (P14)``. (3) Inaccurate Responses: A few participants felt that MemPal was still an ``early stage development'' (P11) and ``not accurate enough yet'' (P13). This also made it ``too hard to tell'' how useful the system could be (P3).
\subsubsection{\textbf{Improve system learning curve}}: Participants (P6, P7, P4) also felt like they needed ``to get used to the device'' since ``everything is so new'' (P6), specifically for object retrieval where P7 felt the ``need to be instructed for how to place things'' so that the object is in sight of the camera and the system onboarding phase where they would need ``more instruction and demonstration'' (P11, P14) or potentially ``would need a caregiver to do it'' (P8).
\subsubsection{\textbf{Optionality in visual vs audio assistance}}:Participants found that having the option of using MemPal or Visual or both would to be most useful (P2, P3, P9, P11, P14, P15) since both modalities offered their own benefits and drawbacks. %even ``push a button for a picture (would like a picture on tablet), push button for voice/ description.'' 
For the visual mode, users noted that "fisheye lenses are difficult to understand" (P3), suggesting that the lens distortion may make it harder for them to match the view to their real-world perception of the space in their home. Additionally, visuals provide more information than a short audio description could provide like ''[capturing] exactly which drawer'' the object was at (P14). For the audio mode, users commented that the assistance is ``less helpful when relying only on verbal description. Pictures are also helpful'' (P2) even though verbal descriptions provided the better speed needed for object finding: ``speed is as important as accuracy” (P14) and users might ``sometimes forget the name of the object you are looking for'' (P7).

\subsubsection{\textbf{Privacy and Data Sharing}} 
 
Most participants were comfortable with their data about their 'activity log' being shared with their caregiver(s) and physician(s) (P2, P4-P10, P12-P15). Participants thought that it was ``necessary'' (P5, P15), P10 remarked that it was about weighing ``privacy versus safety issues'' and P14 felt that ``privacy is a thing of the past''. P15 mentioned ``when you go to your doctor, you sort of like aren’t quite sure about it so you want the data and they really know'' and P14 said ``definitely want the caregiver to know just [when] like you have a child''. However, a few said it depends on various factors (P1, P3, P11): ``I wouldn’t want to have constant monitoring. It just depends on level of cognition and degree of supervision. If my husband was living alone then constant supervision is needed'' (P1), and ``maybe in the future with doctors or caretakers but now [I] only [want it] personal'' (P3). Additionally participants felt that body worn cameras were less intrusive than smart home cameras (P1,P14).

\section{Discussion} 

\subsection{MemPal Helpfulness for Object Finding} 
    
Regarding RQ1: ``What are the effects of using the MemPal voice-based system for object retrieval on retrieval accuracy, path length, cognitive task load, retrieval confidence and recall difficulty for older adults, compared to using visual cues or no system?'', showed that MemPal's audio descriptions improve object retrieval performance compared to using no system by increasing the rate of correct object identification (retrieval accuracy) and reducing the average number of rooms searched (path length) as well reduced perceived difficulty in remembering object locations (recall difficulty). Both assistance modes supported users in similar ways, suggesting that participants could retrieve objects effectively using either visual or audio cues. %Moreover, the addition of MemPal's last-seen object images also decreased path length. 
The visual condition decreased path length and recall difficulty compared to using no system. This is consistent with the results from the GoFinder study which also showed retrieval performance improvements with their visual condition compared to no system~\cite{gofinder}. %Participants rated it less difficult to recall where objects were with assistance (both with Mempal Audio and Visual).
We note that the results discussed are based on our analysis that included trials where participants used the correct feedback from each of the assistance modes. Reduced path length, a proxy for time spent searching, suggests that specifying the room helps significantly narrow the search, a benefit derived from the room localization method but as noted in Qualitative Feedback, descriptive context is most valuable either with better descriptions or high quality wide-angle images.   In addition, there were no significant differences in path length nor recall difficulty between the two assistance modes, indicating that participants performed similarly for visual and audio.

Task load and retrieval confidence did not have significant differences across conditions. The lack of difference in task load may stem from speech recognition issues in the MemPal condition and difficulties in object localization within the visual images in the Visual condition, which could be blurry or distorted by fish-eye lenses as noted in participants' Qualitative Feedback. Despite these issues, participants valued the optional visual aid for its clarity in providing room-specific details and some found it useful ($M$=$5.57$ of 7). Similar to findings from previous works~\cite{liu2023older}, solely using visual aids may not be ideal for older adult users who prefer voice-driven systems. Moreover, visual aids raise concerns about privacy and increased storage needs which MemPal aimed to avoid. Therefore, users who have vocal deterioration or have trouble with language like P9 and P11 may prefer visual images and users who have poor eyesight or do not like interacting with tablet devices may prefer audio output. Future work can include diving into preferences and perceptions of older adult users regarding using an optimal combination of both audio and visual aids in object retrieval tasks to tailor more user-friendly assistance technologies. 

\subsection{MemPal Feature experience} 
Regarding RQ2: ``What are older adults' perceptions and experiences of using a voice-enabled wearable assistant for object retrieval?'', older adults rated MemPal with an average usability (SUS) score of 69.375. This score is slightly above the average SUS benchmark of 68~\cite{bangor2008empirical,brooke1996sus} and approaches the threshold of 70, which is considered indicative of good usability \cite{lewis2018system}. \textcolor{black}{However low usability scores are attributed to potentially more critical feedback from those with high MMSE scores which positively suggests those with lower MMSE (MCI and SCD) find the system more useful.} This is an encouraging result, particularly given that MemPal is in its early stages of development. Comparatively, MemPal's usability score is modestly lower than that of GoFinder \cite{gofinder}, which recorded a SUS score of 75.4. However, it is important to note that GoFinder was not evaluated amongst older adults or those with limited technology experience. The lower usability scores observed with MemPal may be partly attributed to its initial form factor, which some participants found uncomfortable, as well as varying levels of technological familiarity among the participants. As participants mentioned, due to hearing aid constraints, vision impairments, and different types of hair within this pilot population, a body-worn camera with either a clip or magnetic attachment seemed to be most feasible.

MemPal is designed to enhance independence by addressing critical pain points identified through our initial interviews. Given the diverse preferences regarding format of descriptions and modality of assistance, the system should offer options and customization. This approach will allow us to tailor the system more closely to user needs. 

In designing a wearable camera system that uses visual context, we y stored textual information \textcolor{black}{(anoymized and securely stored in a protected database)} and creating a voice interface that could easily interpret this textual data~\cite{davies2015security}. \textcolor{black}{The system also complies with ethical guidelines for LLMs in healthcare \cite{harrer2023attention}.} We found that participants were very comfortable sharing this text-based data with their caregivers and providers as well as storing only text-based data on the system, whereas they mentioned that having alternative smart home camera systems that are not egocentric or systems that store continuous camera data may feel uncomfortable (P1). Additionally, ease of use amongst old adults is imperative which requires voice-activation and we found that MemPal had a high ease of use score which proves that voice-based systems for object retrieval are easy to use M=5.38. Participants were also satisfied by the responses (M=5.17) due to its low latency with an average response time of 2.2s as well as appropriate length (M=5.17). Lastly, we wanted to ensure a smooth setup process that is user-friendly especially since systems like FMT did not test the initial installation stage that requires users to set up markers around their home to track activity \cite{fmt} so by introducing participants to the system onboarding stage we were able to develop and pilot a quick setup process. Most participants were satisfied with the location setup without needing much direction. 

% \cite{davies2015security}. Therefore, design decisions revolving around privacy were prioritized. These include the following: (1) No storage of videos and photos, instead only low dimensional embeddings of text captions and initial home calibration on device. (2) Transparency of activity data to patients through summarized analytics. (3) Local database storage on the user’s device instead of an external server and only synced with authorized caregivers after each day.
    %\item \textbf{Limited onboarding} Recognizing older adults users' preference for simplicity, our system requires minimal setup, featuring a quick five-minute home tour video for initial configuration. 
 %   \item \textbf{Socially acceptable form factor} Neck-worn devices are more likely worn by older adults compared to other form factors (e.g., smart glasses) \cite{schwind2020anticipated}.

\section{Future Work and Limitations}

The aim of this work was to create a first, working prototype that can be tested by the target user group, older adults,in their homes for feedback and suggest future design guidelines rather than create the most accurate system.  

\subsection{Limitations}

\subsubsection{Technical Accuracy and System Limitations} 
There were technical limitations in the accuracy of the system particularly room localization, object detection, and speech recognition, which affected user experiences as well as system limitations such as low wearable device comfort, system learning curve, and less descriptive responses.  While the location algorithm worked for a variety of layouts and lighting, it was not generally adaptable to every layout we encountered in our user study. The positioning of the camera also presented challenges, as it influenced the detection of objects and locations, depending on the FOV \textcolor{black}{especially for hidden locations}. The participant feedback revealed that users must be consciously aware of placing objects in front of the wearable camera-based system for it to effectively assist them. As a voice-enabled system, it struggled to adapt to various accents and slurred speech, affecting its ability to understand queries from certain users. 
\subsubsection{Study Design}
One limitation of the object retrieval study design was that participants occasionally remembered where objects were placed since they placed the objects themselves and retrieved within 30 minutes. Although an ideal study design would have increased the time length to days between placing and retrieval, this led to other confounding factors. 

\subsection{Future Work}
%\subsubsection{Future System Additions}

\subsubsection{Future System Improvements}
Future iterations of a wearable memory aid system could incorporate a queryless interface for older adults who may have trouble with speech. A queryless system for object retrieval could use biomarkers or eye tracking to detect confusion, thereby initiating assistance automatically. Real-time task help tailored to the activities of each participant and verbal feedback loops could further improve user interaction. The calibration experience could be refined by including specific furniture items and sub-room areas that are unique to each household. To improve privacy and battery life, future enhancements could include moving machine learning processing to on-device smaller models, which would eliminate the need for processing video through closed-source models as was done in this early-stage study, which participants consented to. Lastly, to adapt to the large diversity in older adults' preferences and needs, such systems should be able to adapt to suit different stages of memory conditions (ex. providing both Visual and Audio modes as options for object retrieval and multiple context options for safety reminders). 

\subsubsection{Future Study Design}
A long term study to extend the feature set of testing including proactive safety reminders and retrospective task recall and summarization will prove the system's effectiveness of general task assistance beyond object retrieval. A longitudinal study exploring long term impacts on memory due to potential over-reliance on wearable memory assistants and unintended uses potentially as social companions will help in more robust, user-centric designs of the system. Testing on specific sub-populations such as people with dementia and Alzheimer's patients who have advanced-stage memory conditions can further provide useful insight. Long term use could also facilitate the learning of behavior in order to suggest better habits and placements of objects to reduce the frequency of memory errors. 

\subsection{Future Directions for MemPal}
As both a Remote Patient Monitoring (RPM) tool and a wearable agent, MemPal has significant potential for expansion. The safety reminder workflow designed in Figure 15 could integrate with voice-inputted daily schedules or calendars for context-based proactive event reminders or task sequence boards for proactive task assistance. These could also be implemented using synthesized voices of people whom older adults are familiar with, such as friends or family members~\cite{10.1145/3479590}. MemPal could facilitate everyday tasks for individuals with memory impairments by leveraging MemPal's task logging and LLM query-response infrastructure to understand actions, activities, and predict the next sequence of steps. Additionally, more fine-grained activity features (such as time taken per activity step) could be used to objectively assess Instrumental Activities of Daily Living (IADL) performance, a common measure for memory-impaired individuals \cite{guo2020instrumental}. MemPal's query log of voice interactions could also be utilized to diagnose conditions such as dementia or depression \cite{kumar2022dementia} by analyzing speech-to-text patterns. Considering the challenges older adults users may face with voice queries, alternative input modalities like haptic touch or silent speech should be explored. Lastly, physicians and speech therapists have highlighted the ability of using a system like MemPal in aiding cognitive rehabilitation through a quiz format, improving memory rather than just supporting it. 

\subsection{Implications of Future Assistive Technologies for Older Adults}

With the rise of LLMs, MemPal demonstrates how multimodal, context-aware technology using LLMs and VLMs can enhance assistive systems for older adults. Future research should further explore the use of multimodal large models to support multifaceted daily living needs. Advancements should prioritize minimalist interfaces and even queryless functionality, addressing unique challenges identified in our study, such as language variations and speech issues among older adults. Additionally, systems should aim to facilitate behavior change and memory training~\cite{10.1145/3351235,chan2022augmenting}, considering older adults' reluctance to seek help and their tendency to overestimate their memory. Finally, ensuring privacy is crucial; future models should focus on protecting user data by limiting information sharing to text only and avoiding image-based data to maintain user confidentiality.

\section{Conclusion}

In this work, we designed and evaluated a wearable camera-based and voice-enabled memory assistant, MemPal, with older adults within their own homes to address their most common pain points regarding ensuring independence like assisting with object finding based on analysis from pre-study interviews, forums and literature. Through investigating the effect of voice-enabled queries with MemPal for object retrieval and comparing it to no assistance and visual assistance (RQ1), findings show that MemPal improved retrieval performance and reduced recall difficulty compared to no aid  but similar to visual aid. Although a prototype, the results also show that older adults found MemPal usable and highly useful (RQ2) but customization is important. Through this work, we contribute towards understanding user design specifications of older adults in their own homes and designing a LLM-based multimodal wearable system that facilitates object retrieval assistance. We provide the opportunity to extend upon MemPal's technology and user interface, enabling general memory assistance. This work represents a crucial first step toward a general personal memory agent for older adults, supporting independent and safe living at home.

\begin{acks}
The authors would like to thank Aditya Suri, Rachel Park, Cayden Pierce, Nigel Norman, Arnav Kapur, Roy Shilkrot,  Nathan Whitmore, and Leyla Buljina for their valuable support during the course of this project. 
\end{acks}
\bibliographystyle{ACM-Reference-Format}
\bibliography{sample-base}


\begin{thebibliography}{79}

%%% the Web address.

\ifx \showCODEN    \undefined \def \showCODEN     #1{\unskip}     \fi
\ifx \showDOI      \undefined \def \showDOI       #1{#1}\fi
\ifx \showISBNx    \undefined \def \showISBNx     #1{\unskip}     \fi
\ifx \showISBNxiii \undefined \def \showISBNxiii  #1{\unskip}     \fi
\ifx \showISSN     \undefined \def \showISSN      #1{\unskip}     \fi
\ifx \showLCCN     \undefined \def \showLCCN      #1{\unskip}     \fi
\ifx \shownote     \undefined \def \shownote      #1{#1}          \fi
\ifx \showarticletitle \undefined \def \showarticletitle #1{#1}   \fi
\ifx \showURL      \undefined \def \showURL       {\relax}        \fi
% invisible to TeX
\providecommand\bibfield[2]{#2}
\providecommand\bibinfo[2]{#2}
\providecommand\natexlab[1]{#1}
\providecommand\showeprint[2][]{arXiv:#2}

\bibitem[sta(2022)]%
        {statista2022agepopulation}
 \bibinfo{year}{2022}\natexlab{}.
\newblock \bibinfo{title}{U.S. - seniors as a percentage of the population 1950-2050}.
\newblock
\newblock
\urldef\tempurl%
\url{https://www.statista.com/statistics/457822/share-of-old-age-population-in-the-total-us-population/}
\showURL{%
\tempurl}
\newblock
\shownote{Accessed: 2023-12-13}.


\bibitem[Abowd et~al\mbox{.}(2002)]%
        {abowd2002aware}
\bibfield{author}{\bibinfo{person}{Gregory~D Abowd}, \bibinfo{person}{Aaron~F Bobick}, \bibinfo{person}{Irfan~A Essa}, \bibinfo{person}{Elizabeth~D Mynatt}, {and} \bibinfo{person}{Wendy~A Rogers}.} \bibinfo{year}{2002}\natexlab{}.
\newblock \showarticletitle{The aware home: A living laboratory for technologies for successful aging}. In \bibinfo{booktitle}{\emph{Proceedings of the AAAI-02 Workshop “Automation as Caregiver}}. \bibinfo{pages}{1--7}.
\newblock


\bibitem[Ali et~al\mbox{.}(2018)]%
        {ali2018aging}
\bibfield{author}{\bibinfo{person}{Mohammad~Rafayet Ali}, \bibinfo{person}{Kimberly Van~Orden}, \bibinfo{person}{Kimberly Parkhurst}, \bibinfo{person}{Shuyang Liu}, \bibinfo{person}{Viet-Duy Nguyen}, \bibinfo{person}{Paul Duberstein}, {and} \bibinfo{person}{M~Ehsan Hoque}.} \bibinfo{year}{2018}\natexlab{}.
\newblock \showarticletitle{Aging and engaging: A social conversational skills training program for older adults}. In \bibinfo{booktitle}{\emph{23rd International Conference on Intelligent User Interfaces}}. \bibinfo{pages}{55--66}.
\newblock


\bibitem[{Alzheimer's Association}(2023)]%
        {AlzheimersFactsFigures2023}
\bibfield{author}{\bibinfo{person}{{Alzheimer's Association}}.} \bibinfo{year}{2023}\natexlab{}.
\newblock \showarticletitle{2023 Alzheimer's disease facts and figures}.
\newblock \bibinfo{journal}{\emph{Alzheimer's \& Dementia}}  \bibinfo{volume}{19} (\bibinfo{year}{2023}), \bibinfo{pages}{1598--1695}.
\newblock
\urldef\tempurl%
\url{https://doi.org/10.1002/alz.13016}
\showDOI{\tempurl}


\bibitem[Arno et~al\mbox{.}(1999)]%
        {arno1999economic}
\bibfield{author}{\bibinfo{person}{Peter~S Arno}, \bibinfo{person}{Carol Levine}, {and} \bibinfo{person}{Margaret~M Memmott}.} \bibinfo{year}{1999}\natexlab{}.
\newblock \showarticletitle{The Economic Value Of Informal Caregiving: President Clinton's proposal to provide relief to family caregivers opens a long-overdue discussion of this “invisible” health care sector.}
\newblock \bibinfo{journal}{\emph{Health Affairs}} \bibinfo{volume}{18}, \bibinfo{number}{2} (\bibinfo{year}{1999}), \bibinfo{pages}{182--188}.
\newblock


\bibitem[Balota et~al\mbox{.}(2000)]%
        {balota2000memory}
\bibfield{author}{\bibinfo{person}{David~A Balota}, \bibinfo{person}{Patrick~O Dolan}, {and} \bibinfo{person}{Janet~M Duchek}.} \bibinfo{year}{2000}\natexlab{}.
\newblock \showarticletitle{Memory changes in healthy young and older adults}.
\newblock \bibinfo{journal}{\emph{The Oxford handbook of memory}} (\bibinfo{year}{2000}), \bibinfo{pages}{395--410}.
\newblock


\bibitem[Bangor et~al\mbox{.}(2008)]%
        {bangor2008empirical}
\bibfield{author}{\bibinfo{person}{Aaron Bangor}, \bibinfo{person}{Philip~T Kortum}, {and} \bibinfo{person}{James~T Miller}.} \bibinfo{year}{2008}\natexlab{}.
\newblock \showarticletitle{An empirical evaluation of the system usability scale}.
\newblock \bibinfo{journal}{\emph{Intl. Journal of Human--Computer Interaction}} \bibinfo{volume}{24}, \bibinfo{number}{6} (\bibinfo{year}{2008}), \bibinfo{pages}{574--594}.
\newblock


\bibitem[Bickmore et~al\mbox{.}(2005)]%
        {bickmore2005s}
\bibfield{author}{\bibinfo{person}{Timothy~W Bickmore}, \bibinfo{person}{Lisa Caruso}, \bibinfo{person}{Kerri Clough-Gorr}, {and} \bibinfo{person}{Tim Heeren}.} \bibinfo{year}{2005}\natexlab{}.
\newblock \showarticletitle{‘It’s just like you talk to a friend’relational agents for older adults}.
\newblock \bibinfo{journal}{\emph{Interacting with Computers}} \bibinfo{volume}{17}, \bibinfo{number}{6} (\bibinfo{year}{2005}), \bibinfo{pages}{711--735}.
\newblock


\bibitem[Braun and Clarke(2006)]%
        {braun2006using}
\bibfield{author}{\bibinfo{person}{Virginia Braun} {and} \bibinfo{person}{Victoria Clarke}.} \bibinfo{year}{2006}\natexlab{}.
\newblock \showarticletitle{Using thematic analysis in psychology}.
\newblock \bibinfo{journal}{\emph{Qualitative research in psychology}} \bibinfo{volume}{3}, \bibinfo{number}{2} (\bibinfo{year}{2006}), \bibinfo{pages}{77--101}.
\newblock


\bibitem[Brewer and Piper(2017)]%
        {brewer2017xpress}
\bibfield{author}{\bibinfo{person}{Robin~N Brewer} {and} \bibinfo{person}{Anne~Marie Piper}.} \bibinfo{year}{2017}\natexlab{}.
\newblock \showarticletitle{xPress: Rethinking design for aging and accessibility through an IVR blogging system}.
\newblock \bibinfo{journal}{\emph{Proceedings of the ACM on Human-Computer Interaction}} \bibinfo{volume}{1}, \bibinfo{number}{CSCW} (\bibinfo{year}{2017}), \bibinfo{pages}{1--17}.
\newblock


\bibitem[Brooke et~al\mbox{.}(1996)]%
        {brooke1996sus}
\bibfield{author}{\bibinfo{person}{John Brooke} {et~al\mbox{.}}} \bibinfo{year}{1996}\natexlab{}.
\newblock \showarticletitle{SUS-A quick and dirty usability scale}.
\newblock \bibinfo{journal}{\emph{Usability evaluation in industry}} \bibinfo{volume}{189}, \bibinfo{number}{194} (\bibinfo{year}{1996}), \bibinfo{pages}{4--7}.
\newblock


\bibitem[Brown et~al\mbox{.}(2020)]%
        {brown2020language}
\bibfield{author}{\bibinfo{person}{Tom Brown}, \bibinfo{person}{Benjamin Mann}, \bibinfo{person}{Nick Ryder}, \bibinfo{person}{Melanie Subbiah}, \bibinfo{person}{Jared~D Kaplan}, \bibinfo{person}{Prafulla Dhariwal}, \bibinfo{person}{Arvind Neelakantan}, \bibinfo{person}{Pranav Shyam}, \bibinfo{person}{Girish Sastry}, \bibinfo{person}{Amanda Askell}, {et~al\mbox{.}}} \bibinfo{year}{2020}\natexlab{}.
\newblock \showarticletitle{Language models are few-shot learners}.
\newblock \bibinfo{journal}{\emph{Advances in neural information processing systems}}  \bibinfo{volume}{33} (\bibinfo{year}{2020}), \bibinfo{pages}{1877--1901}.
\newblock


\bibitem[Buzzelli et~al\mbox{.}(2020)]%
        {buzzelli2020vision}
\bibfield{author}{\bibinfo{person}{Marco Buzzelli}, \bibinfo{person}{Alessio Alb{\'e}}, {and} \bibinfo{person}{Gianluigi Ciocca}.} \bibinfo{year}{2020}\natexlab{}.
\newblock \showarticletitle{A vision-based system for monitoring elderly people at home}.
\newblock \bibinfo{journal}{\emph{Applied Sciences}} \bibinfo{volume}{10}, \bibinfo{number}{1} (\bibinfo{year}{2020}), \bibinfo{pages}{374}.
\newblock


\bibitem[Chan et~al\mbox{.}(2020)]%
        {chan2020prompto}
\bibfield{author}{\bibinfo{person}{Samantha~WT Chan}, \bibinfo{person}{Shardul Sapkota}, \bibinfo{person}{Rebecca Mathews}, \bibinfo{person}{Haimo Zhang}, {and} \bibinfo{person}{Suranga Nanayakkara}.} \bibinfo{year}{2020}\natexlab{}.
\newblock \showarticletitle{Prompto: Investigating receptivity to prompts based on cognitive load from memory training conversational agent}.
\newblock \bibinfo{journal}{\emph{Proceedings of the ACM on interactive, mobile, wearable and ubiquitous technologies}} \bibinfo{volume}{4}, \bibinfo{number}{4} (\bibinfo{year}{2020}), \bibinfo{pages}{1--23}.
\newblock


\bibitem[Chan et~al\mbox{.}(2019b)]%
        {chan2019prospero}
\bibfield{author}{\bibinfo{person}{Samantha~WT Chan}, \bibinfo{person}{Haimo Zhang}, {and} \bibinfo{person}{Suranga Nanayakkara}.} \bibinfo{year}{2019}\natexlab{b}.
\newblock \showarticletitle{Prospero: A personal wearable memory coach}. In \bibinfo{booktitle}{\emph{Proceedings of the 10th Augmented Human International Conference 2019}}. \bibinfo{pages}{1--5}.
\newblock


\bibitem[Chan(2020)]%
        {10.1145/3334480.3375031}
\bibfield{author}{\bibinfo{person}{Samantha W.~T. Chan}.} \bibinfo{year}{2020}\natexlab{}.
\newblock \showarticletitle{Biosignal-Sensitive Memory Improvement and Support Systems}. In \bibinfo{booktitle}{\emph{Extended Abstracts of the 2020 CHI Conference on Human Factors in Computing Systems}} (Honolulu, HI, USA) \emph{(\bibinfo{series}{CHI EA '20})}. \bibinfo{publisher}{Association for Computing Machinery}, \bibinfo{address}{New York, NY, USA}, \bibinfo{pages}{1–7}.
\newblock
\showISBNx{9781450368193}
\urldef\tempurl%
\url{https://doi.org/10.1145/3334480.3375031}
\showDOI{\tempurl}


\bibitem[Chan et~al\mbox{.}(2019a)]%
        {10.1145/3351235}
\bibfield{author}{\bibinfo{person}{Samantha W.~T. Chan}, \bibinfo{person}{Thisum Buddhika}, \bibinfo{person}{Haimo Zhang}, {and} \bibinfo{person}{Suranga Nanayakkara}.} \bibinfo{year}{2019}\natexlab{a}.
\newblock \showarticletitle{ProspecFit: In Situ Evaluation of Digital Prospective Memory Training for Older Adults}.
\newblock \bibinfo{journal}{\emph{Proc. ACM Interact. Mob. Wearable Ubiquitous Technol.}} \bibinfo{volume}{3}, \bibinfo{number}{3}, Article \bibinfo{articleno}{77} (\bibinfo{date}{Sept.} \bibinfo{year}{2019}), \bibinfo{numpages}{20}~pages.
\newblock
\urldef\tempurl%
\url{https://doi.org/10.1145/3351235}
\showDOI{\tempurl}


\bibitem[Chan et~al\mbox{.}(2021)]%
        {10.1145/3479590}
\bibfield{author}{\bibinfo{person}{Sam W.~T. Chan}, \bibinfo{person}{Tamil~Selvan Gunasekaran}, \bibinfo{person}{Yun~Suen Pai}, \bibinfo{person}{Haimo Zhang}, {and} \bibinfo{person}{Suranga Nanayakkara}.} \bibinfo{year}{2021}\natexlab{}.
\newblock \showarticletitle{KinVoices: Using Voices of Friends and Family in Voice Interfaces}.
\newblock \bibinfo{journal}{\emph{Proc. ACM Hum.-Comput. Interact.}} \bibinfo{volume}{5}, \bibinfo{number}{CSCW2}, Article \bibinfo{articleno}{446} (\bibinfo{date}{Oct.} \bibinfo{year}{2021}), \bibinfo{numpages}{25}~pages.
\newblock
\urldef\tempurl%
\url{https://doi.org/10.1145/3479590}
\showDOI{\tempurl}


\bibitem[Chan(2022)]%
        {chan2022augmenting}
\bibfield{author}{\bibinfo{person}{Wei Ting~Samantha Chan}.} \bibinfo{year}{2022}\natexlab{}.
\newblock \emph{\bibinfo{title}{Augmenting Human Prospective Memory through Cognition-Aware Technologies}}.
\newblock \bibinfo{thesistype}{Ph.\,D. Dissertation}. \bibinfo{school}{ResearchSpace@ Auckland}.
\newblock


\bibitem[Cheng et~al\mbox{.}(2023)]%
        {vidcompare}
\bibfield{author}{\bibinfo{person}{Sijie Cheng}, \bibinfo{person}{Zhicheng Guo}, \bibinfo{person}{Jingwen Wu}, \bibinfo{person}{Kechen Fang}, \bibinfo{person}{Peng Li}, \bibinfo{person}{Huaping Liu}, {and} \bibinfo{person}{Yang Liu}.} \bibinfo{year}{2023}\natexlab{}.
\newblock \showarticletitle{Can Vision-Language Models Think from a First-Person Perspective?}
\newblock \bibinfo{journal}{\emph{arXiv preprint arXiv:2311.15596}} (\bibinfo{year}{2023}).
\newblock


\bibitem[Cockrell and Folstein(2002)]%
        {cockrell2002mini}
\bibfield{author}{\bibinfo{person}{Joseph~R Cockrell} {and} \bibinfo{person}{Marshal~F Folstein}.} \bibinfo{year}{2002}\natexlab{}.
\newblock \showarticletitle{Mini-mental state examination}.
\newblock \bibinfo{journal}{\emph{Principles and practice of geriatric psychiatry}} (\bibinfo{year}{2002}), \bibinfo{pages}{140--141}.
\newblock


\bibitem[Collerton et~al\mbox{.}(2014)]%
        {collerton2014exploratory}
\bibfield{author}{\bibinfo{person}{Daniel Collerton}, \bibinfo{person}{Emily Forster}, {and} \bibinfo{person}{Derek Packham}.} \bibinfo{year}{2014}\natexlab{}.
\newblock \showarticletitle{An exploratory study of the effectiveness of memory aids for older people living in supported accommodation}.
\newblock \bibinfo{journal}{\emph{Journal of Applied Gerontology}} \bibinfo{volume}{33}, \bibinfo{number}{8} (\bibinfo{year}{2014}), \bibinfo{pages}{963--981}.
\newblock


\bibitem[Corbett et~al\mbox{.}(2021)]%
        {corbett2021voice}
\bibfield{author}{\bibinfo{person}{Cynthia~F Corbett}, \bibinfo{person}{Pamela~J Wright}, \bibinfo{person}{Kate Jones}, {and} \bibinfo{person}{Michael Parmer}.} \bibinfo{year}{2021}\natexlab{}.
\newblock \showarticletitle{Voice-activated virtual home assistant use and social isolation and loneliness among older adults: mini review}.
\newblock \bibinfo{journal}{\emph{Frontiers in Public Health}}  \bibinfo{volume}{9} (\bibinfo{year}{2021}), \bibinfo{pages}{742012}.
\newblock


\bibitem[Dahmen et~al\mbox{.}(2018)]%
        {dahmen2018design}
\bibfield{author}{\bibinfo{person}{Jessamyn Dahmen}, \bibinfo{person}{Bryan Minor}, \bibinfo{person}{DJ Cook}, \bibinfo{person}{Thao Vo}, {and} \bibinfo{person}{Maureen Schmitter-Edgecombe}.} \bibinfo{year}{2018}\natexlab{}.
\newblock \showarticletitle{Design of a smart home-driven digital memory notebook to support self-management of activities for older adults}.
\newblock \bibinfo{journal}{\emph{Gerontechnology}} (\bibinfo{year}{2018}).
\newblock


\bibitem[Davies et~al\mbox{.}(2015)]%
        {davies2015security}
\bibfield{author}{\bibinfo{person}{Nigel Davies}, \bibinfo{person}{Adrian Friday}, \bibinfo{person}{Sarah Clinch}, \bibinfo{person}{Corina Sas}, \bibinfo{person}{Marc Langheinrich}, \bibinfo{person}{Geoff Ward}, {and} \bibinfo{person}{Albrecht Schmidt}.} \bibinfo{year}{2015}\natexlab{}.
\newblock \showarticletitle{Security and privacy implications of pervasive memory augmentation}.
\newblock \bibinfo{journal}{\emph{IEEE Pervasive Computing}} \bibinfo{volume}{14}, \bibinfo{number}{1} (\bibinfo{year}{2015}), \bibinfo{pages}{44--53}.
\newblock


\bibitem[De~Miguel et~al\mbox{.}(2017)]%
        {de2017home}
\bibfield{author}{\bibinfo{person}{Koldo De~Miguel}, \bibinfo{person}{Alberto Brunete}, \bibinfo{person}{Miguel Hernando}, {and} \bibinfo{person}{Ernesto Gambao}.} \bibinfo{year}{2017}\natexlab{}.
\newblock \showarticletitle{Home camera-based fall detection system for the elderly}.
\newblock \bibinfo{journal}{\emph{Sensors}} \bibinfo{volume}{17}, \bibinfo{number}{12} (\bibinfo{year}{2017}), \bibinfo{pages}{2864}.
\newblock


\bibitem[{Deepgram}(2024)]%
        {deepgramspeech}
\bibfield{author}{\bibinfo{person}{{Deepgram}}.} \bibinfo{year}{2024}\natexlab{}.
\newblock \bibinfo{title}{Deepgram Speech-to-Text API}.
\newblock
\newblock
\urldef\tempurl%
\url{https://deepgram.com/product/speech-to-text}
\showURL{%
\tempurl}
\newblock
\shownote{Accessed: 2024-04-30}.


\bibitem[Dubourg et~al\mbox{.}(2016)]%
        {dubourg2016sensecam}
\bibfield{author}{\bibinfo{person}{Lydia Dubourg}, \bibinfo{person}{Ana~Rita Silva}, \bibinfo{person}{Christophe Fitamen}, \bibinfo{person}{Chris~JA Moulin}, {and} \bibinfo{person}{C{\'e}line Souchay}.} \bibinfo{year}{2016}\natexlab{}.
\newblock \showarticletitle{SenseCam: A new tool for memory rehabilitation?}
\newblock \bibinfo{journal}{\emph{Revue Neurologique}} \bibinfo{volume}{172}, \bibinfo{number}{12} (\bibinfo{year}{2016}), \bibinfo{pages}{735--747}.
\newblock


\bibitem[Farivar et~al\mbox{.}(2020)]%
        {farivar2020wearable}
\bibfield{author}{\bibinfo{person}{Samira Farivar}, \bibinfo{person}{Mohamed Abouzahra}, {and} \bibinfo{person}{Maryam Ghasemaghaei}.} \bibinfo{year}{2020}\natexlab{}.
\newblock \showarticletitle{Wearable device adoption among older adults: A mixed-methods study}.
\newblock \bibinfo{journal}{\emph{International Journal of Information Management}}  \bibinfo{volume}{55} (\bibinfo{year}{2020}), \bibinfo{pages}{102209}.
\newblock


\bibitem[Fontana and Gunderson(2002)]%
        {fontana2002ultra}
\bibfield{author}{\bibinfo{person}{Robert~J Fontana} {and} \bibinfo{person}{Steven~J Gunderson}.} \bibinfo{year}{2002}\natexlab{}.
\newblock \showarticletitle{Ultra-wideband precision asset location system}. In \bibinfo{booktitle}{\emph{2002 IEEE Conference on Ultra Wideband Systems and Technologies (IEEE Cat. No. 02EX580)}}. IEEE, \bibinfo{pages}{147--150}.
\newblock


\bibitem[Gasteiger et~al\mbox{.}(2021)]%
        {gasteiger2021friends}
\bibfield{author}{\bibinfo{person}{Norina Gasteiger}, \bibinfo{person}{Kate Loveys}, \bibinfo{person}{Mikaela Law}, {and} \bibinfo{person}{Elizabeth Broadbent}.} \bibinfo{year}{2021}\natexlab{}.
\newblock \showarticletitle{Friends from the future: a scoping review of research into robots and computer agents to combat loneliness in older people}.
\newblock \bibinfo{journal}{\emph{Clinical interventions in aging}} (\bibinfo{year}{2021}), \bibinfo{pages}{941--971}.
\newblock


\bibitem[Gelonch et~al\mbox{.}(2019)]%
        {gelonch2019acceptability}
\bibfield{author}{\bibinfo{person}{Olga Gelonch}, \bibinfo{person}{Mireia Ribera}, \bibinfo{person}{N{\'u}ria Codern-Bov{\'e}}, \bibinfo{person}{S{\'\i}lvia Ramos}, \bibinfo{person}{Maria Quintana}, \bibinfo{person}{Gloria Chico}, \bibinfo{person}{Noem{\'\i} Cerulla}, \bibinfo{person}{Paula Lafarga}, \bibinfo{person}{Petia Radeva}, {and} \bibinfo{person}{Maite Garolera}.} \bibinfo{year}{2019}\natexlab{}.
\newblock \showarticletitle{Acceptability of a lifelogging wearable camera in older adults with mild cognitive impairment: a mixed-method study}.
\newblock \bibinfo{journal}{\emph{BMC geriatrics}} \bibinfo{volume}{19}, \bibinfo{number}{1} (\bibinfo{year}{2019}), \bibinfo{pages}{1--10}.
\newblock


\bibitem[Guo and Sapra(2020)]%
        {guo2020instrumental}
\bibfield{author}{\bibinfo{person}{Hui~Jun Guo} {and} \bibinfo{person}{Amit Sapra}.} \bibinfo{year}{2020}\natexlab{}.
\newblock \showarticletitle{Instrumental activity of daily living}.
\newblock  (\bibinfo{year}{2020}).
\newblock


\bibitem[Hamilton et~al\mbox{.}(2021)]%
        {10.1145/3463914.3463918}
\bibfield{author}{\bibinfo{person}{Matthew~Allan Hamilton}, \bibinfo{person}{Anthony~Paul Beug}, \bibinfo{person}{Howard~John Hamilton}, {and} \bibinfo{person}{Wil~James Norton}.} \bibinfo{year}{2021}\natexlab{}.
\newblock \showarticletitle{Augmented Reality Technology for People Living with Dementia and their Care Partners}. In \bibinfo{booktitle}{\emph{Proceedings of the 2021 5th International Conference on Virtual and Augmented Reality Simulations}} (Melbourne, VIC, Australia) \emph{(\bibinfo{series}{ICVARS '21})}. \bibinfo{publisher}{Association for Computing Machinery}, \bibinfo{address}{New York, NY, USA}, \bibinfo{pages}{21–30}.
\newblock
\showISBNx{9781450389327}
\urldef\tempurl%
\url{https://doi.org/10.1145/3463914.3463918}
\showDOI{\tempurl}


\bibitem[Harrer(2023)]%
        {harrer2023attention}
\bibfield{author}{\bibinfo{person}{Stefan Harrer}.} \bibinfo{year}{2023}\natexlab{}.
\newblock \showarticletitle{Attention is not all you need: the complicated case of ethically using large language models in healthcare and medicine}.
\newblock \bibinfo{journal}{\emph{EBioMedicine}}  \bibinfo{volume}{90} (\bibinfo{year}{2023}).
\newblock


\bibitem[Harris(2006)]%
        {harris2006experience}
\bibfield{author}{\bibinfo{person}{Phyllis~Braudy Harris}.} \bibinfo{year}{2006}\natexlab{}.
\newblock \showarticletitle{The experience of living alone with early stage Alzheimer's disease: What are the person's concerns?}
\newblock \bibinfo{journal}{\emph{Alzheimer's Care Today}} \bibinfo{volume}{7}, \bibinfo{number}{2} (\bibinfo{year}{2006}), \bibinfo{pages}{84--94}.
\newblock


\bibitem[Hasan et~al\mbox{.}(2019)]%
        {hasan2019real}
\bibfield{author}{\bibinfo{person}{Moh~Khalid Hasan}, \bibinfo{person}{Md Shahjalal}, \bibinfo{person}{Mostafa~Zaman Chowdhury}, {and} \bibinfo{person}{Yeong~Min Jang}.} \bibinfo{year}{2019}\natexlab{}.
\newblock \showarticletitle{Real-time healthcare data transmission for remote patient monitoring in patch-based hybrid OCC/BLE networks}.
\newblock \bibinfo{journal}{\emph{Sensors}} \bibinfo{volume}{19}, \bibinfo{number}{5} (\bibinfo{year}{2019}), \bibinfo{pages}{1208}.
\newblock


\bibitem[He and Chan(2015)]%
        {he2015wi}
\bibfield{author}{\bibinfo{person}{Suining He} {and} \bibinfo{person}{S-H~Gary Chan}.} \bibinfo{year}{2015}\natexlab{}.
\newblock \showarticletitle{Wi-Fi fingerprint-based indoor positioning: Recent advances and comparisons}.
\newblock \bibinfo{journal}{\emph{IEEE Communications Surveys \& Tutorials}} \bibinfo{volume}{18}, \bibinfo{number}{1} (\bibinfo{year}{2015}), \bibinfo{pages}{466--490}.
\newblock


\bibitem[Hodges et~al\mbox{.}(2006)]%
        {hodges2006sensecam}
\bibfield{author}{\bibinfo{person}{Steve Hodges}, \bibinfo{person}{Lyndsay Williams}, \bibinfo{person}{Emma Berry}, \bibinfo{person}{Shahram Izadi}, \bibinfo{person}{James Srinivasan}, \bibinfo{person}{Alex Butler}, \bibinfo{person}{Gavin Smyth}, \bibinfo{person}{Narinder Kapur}, {and} \bibinfo{person}{Ken Wood}.} \bibinfo{year}{2006}\natexlab{}.
\newblock \showarticletitle{SenseCam: A retrospective memory aid}. In \bibinfo{booktitle}{\emph{UbiComp 2006: Ubiquitous Computing: 8th International Conference, UbiComp 2006 Orange County, CA, USA, September 17-21, 2006 Proceedings 8}}. Springer, \bibinfo{pages}{177--193}.
\newblock


\bibitem[Hong et~al\mbox{.}(2024)]%
        {hong2024aconect}
\bibfield{author}{\bibinfo{person}{Junyuan Hong}, \bibinfo{person}{Wenqing Zheng}, \bibinfo{person}{Han Meng}, \bibinfo{person}{Siqi Liang}, \bibinfo{person}{Anqing Chen}, \bibinfo{person}{Hiroko~H. Dodge}, \bibinfo{person}{Jiayu Zhou}, {and} \bibinfo{person}{Zhangyang Wang}.} \bibinfo{year}{2024}\natexlab{}.
\newblock \showarticletitle{A-{CONECT}: Designing {AI}-based Conversational Chatbot for Early Dementia Intervention}. In \bibinfo{booktitle}{\emph{ICLR 2024 Workshop on Large Language Model (LLM) Agents}}.
\newblock
\urldef\tempurl%
\url{https://openreview.net/forum?id=rACfuoNKBU}
\showURL{%
\tempurl}


\bibitem[Jones et~al\mbox{.}(2021)]%
        {jones2021reducing}
\bibfield{author}{\bibinfo{person}{Valerie~K Jones}, \bibinfo{person}{Michael Hanus}, \bibinfo{person}{Changmin Yan}, \bibinfo{person}{Marcia~Y Shade}, \bibinfo{person}{Julie Blaskewicz~Boron}, {and} \bibinfo{person}{Rafael Maschieri~Bicudo}.} \bibinfo{year}{2021}\natexlab{}.
\newblock \showarticletitle{Reducing loneliness among aging adults: the roles of personal voice assistants and anthropomorphic interactions}.
\newblock \bibinfo{journal}{\emph{Frontiers in public health}}  \bibinfo{volume}{9} (\bibinfo{year}{2021}), \bibinfo{pages}{750736}.
\newblock


\bibitem[Kasper et~al\mbox{.}(2015)]%
        {kasper2015disproportionate}
\bibfield{author}{\bibinfo{person}{Judith~D Kasper}, \bibinfo{person}{Vicki~A Freedman}, \bibinfo{person}{Brenda~C Spillman}, {and} \bibinfo{person}{Jennifer~L Wolff}.} \bibinfo{year}{2015}\natexlab{}.
\newblock \showarticletitle{The disproportionate impact of dementia on family and unpaid caregiving to older adults}.
\newblock \bibinfo{journal}{\emph{Health Affairs}} \bibinfo{volume}{34}, \bibinfo{number}{10} (\bibinfo{year}{2015}), \bibinfo{pages}{1642--1649}.
\newblock


\bibitem[Kim et~al\mbox{.}(2022)]%
        {kim2022exploring}
\bibfield{author}{\bibinfo{person}{Sunyoung Kim}, \bibinfo{person}{Willow Yao}, {and} \bibinfo{person}{Xiaotong Du}.} \bibinfo{year}{2022}\natexlab{}.
\newblock \showarticletitle{Exploring older adults’ adoption and use of a tablet computer during COVID-19: longitudinal qualitative study}.
\newblock \bibinfo{journal}{\emph{JMIR aging}} \bibinfo{volume}{5}, \bibinfo{number}{1} (\bibinfo{year}{2022}), \bibinfo{pages}{e32957}.
\newblock


\bibitem[Kumar et~al\mbox{.}(2022)]%
        {kumar2022dementia}
\bibfield{author}{\bibinfo{person}{M~Rupesh Kumar}, \bibinfo{person}{Susmitha Vekkot}, \bibinfo{person}{S Lalitha}, \bibinfo{person}{Deepa Gupta}, \bibinfo{person}{Varasiddhi~Jayasuryaa Govindraj}, \bibinfo{person}{Kamran Shaukat}, \bibinfo{person}{Yousef~Ajami Alotaibi}, {and} \bibinfo{person}{Mohammed Zakariah}.} \bibinfo{year}{2022}\natexlab{}.
\newblock \showarticletitle{Dementia detection from speech using machine learning and deep learning architectures}.
\newblock \bibinfo{journal}{\emph{Sensors}} \bibinfo{volume}{22}, \bibinfo{number}{23} (\bibinfo{year}{2022}), \bibinfo{pages}{9311}.
\newblock


\bibitem[Lewis(2018)]%
        {lewis2018system}
\bibfield{author}{\bibinfo{person}{James~R Lewis}.} \bibinfo{year}{2018}\natexlab{}.
\newblock \showarticletitle{The system usability scale: past, present, and future}.
\newblock \bibinfo{journal}{\emph{International Journal of Human--Computer Interaction}} \bibinfo{volume}{34}, \bibinfo{number}{7} (\bibinfo{year}{2018}), \bibinfo{pages}{577--590}.
\newblock


\bibitem[Li et~al\mbox{.}(2019)]%
        {fmt}
\bibfield{author}{\bibinfo{person}{Franklin~Mingzhe Li}, \bibinfo{person}{Di~Laura Chen}, \bibinfo{person}{Mingming Fan}, {and} \bibinfo{person}{Khai~N Truong}.} \bibinfo{year}{2019}\natexlab{}.
\newblock \showarticletitle{FMT: A wearable camera-based object tracking memory aid for older adults}.
\newblock \bibinfo{journal}{\emph{Proceedings of the ACM on Interactive, Mobile, Wearable and Ubiquitous Technologies}} \bibinfo{volume}{3}, \bibinfo{number}{3} (\bibinfo{year}{2019}), \bibinfo{pages}{1--25}.
\newblock


\bibitem[Lin et~al\mbox{.}(2005)]%
        {lin2005object}
\bibfield{author}{\bibinfo{person}{Chi-yau Lin}, \bibinfo{person}{Chia-nan Ke}, \bibinfo{person}{Shao-you Cheng}, \bibinfo{person}{Jane Yung-jen Hsu}, {and} \bibinfo{person}{Hao-hua Chu}.} \bibinfo{year}{2005}\natexlab{}.
\newblock \showarticletitle{Object reminder and safety alarm}. In \bibinfo{booktitle}{\emph{Embedded and Ubiquitous Computing--EUC 2005 Workshops: EUC 2005 Workshops: UISW, NCUS, SecUbiq, USN, and TAUES, Nagasaki, Japan, December 6-9, 2005. Proceedings}}. Springer, \bibinfo{pages}{499--508}.
\newblock


\bibitem[Liu et~al\mbox{.}(2016)]%
        {liu2016smart}
\bibfield{author}{\bibinfo{person}{Lili Liu}, \bibinfo{person}{Eleni Stroulia}, \bibinfo{person}{Ioanis Nikolaidis}, \bibinfo{person}{Antonio Miguel-Cruz}, {and} \bibinfo{person}{Adriana~Rios Rincon}.} \bibinfo{year}{2016}\natexlab{}.
\newblock \showarticletitle{Smart homes and home health monitoring technologies for older adults: A systematic review}.
\newblock \bibinfo{journal}{\emph{International journal of medical informatics}}  \bibinfo{volume}{91} (\bibinfo{year}{2016}), \bibinfo{pages}{44--59}.
\newblock


\bibitem[Liu et~al\mbox{.}(2023)]%
        {liu2023older}
\bibfield{author}{\bibinfo{person}{Mingzhou Liu}, \bibinfo{person}{Caixia Wang}, {and} \bibinfo{person}{Jing Hu}.} \bibinfo{year}{2023}\natexlab{}.
\newblock \showarticletitle{Older adults’ intention to use voice assistants: Usability and emotional needs}.
\newblock \bibinfo{journal}{\emph{Heliyon}} \bibinfo{volume}{9}, \bibinfo{number}{11} (\bibinfo{year}{2023}).
\newblock


\bibitem[Majumdar et~al\mbox{.}(2023)]%
        {majumdar2023findthis}
\bibfield{author}{\bibinfo{person}{Arjun Majumdar}, \bibinfo{person}{Fei Xia}, \bibinfo{person}{Dhruv Batra}, \bibinfo{person}{Leonidas Guibas}, {et~al\mbox{.}}} \bibinfo{year}{2023}\natexlab{}.
\newblock \showarticletitle{Findthis: Language-driven object disambiguation in indoor environments}. In \bibinfo{booktitle}{\emph{7th Annual Conference on Robot Learning}}.
\newblock


\bibitem[Marchant and Williams(2011)]%
        {marchant2011memory}
\bibfield{author}{\bibinfo{person}{Jenna Asha~Malini Marchant} {and} \bibinfo{person}{Kristine~N Williams}.} \bibinfo{year}{2011}\natexlab{}.
\newblock \showarticletitle{Memory matters in assisted living}.
\newblock \bibinfo{journal}{\emph{Rehabilitation Nursing}} \bibinfo{volume}{36}, \bibinfo{number}{2} (\bibinfo{year}{2011}), \bibinfo{pages}{83--88}.
\newblock


\bibitem[Mehrotra et~al\mbox{.}(2016)]%
        {mehrotra2016embodied}
\bibfield{author}{\bibinfo{person}{Siddharth Mehrotra}, \bibinfo{person}{Vivian~Genaro Motti}, \bibinfo{person}{Helena Frijns}, \bibinfo{person}{Tugce Akkoc}, \bibinfo{person}{Sena~B{\"u}{\c{s}}ra Yenge{\c{c}}}, \bibinfo{person}{Oguz Calik}, \bibinfo{person}{Marieke~MM Peeters}, {and} \bibinfo{person}{Mark~A Neerincx}.} \bibinfo{year}{2016}\natexlab{}.
\newblock \showarticletitle{Embodied conversational interfaces for the elderly user}. In \bibinfo{booktitle}{\emph{Proceedings of the 8th Indian Conference on Human-Computer Interaction}}. \bibinfo{pages}{90--95}.
\newblock


\bibitem[Mogle et~al\mbox{.}(2023)]%
        {mogle2023individual}
\bibfield{author}{\bibinfo{person}{Jacqueline Mogle}, \bibinfo{person}{Jennifer~R Turner}, \bibinfo{person}{Sakshi Bhargava}, \bibinfo{person}{Robert~S Stawski}, \bibinfo{person}{David~M Almeida}, {and} \bibinfo{person}{Nikki~L Hill}.} \bibinfo{year}{2023}\natexlab{}.
\newblock \showarticletitle{Individual differences in frequency and impact of daily memory lapses: results from a national lifespan sample}.
\newblock \bibinfo{journal}{\emph{BMC geriatrics}} \bibinfo{volume}{23}, \bibinfo{number}{1} (\bibinfo{year}{2023}), \bibinfo{pages}{670}.
\newblock


\bibitem[Molesworth and Sharrock(2016)]%
        {molesworth2016evaluation}
\bibfield{author}{\bibinfo{person}{Sue Molesworth} {and} \bibinfo{person}{Lisa Sharrock}.} \bibinfo{year}{2016}\natexlab{}.
\newblock \showarticletitle{An Evaluation of" Autographer plus Flo}.
\newblock  (\bibinfo{year}{2016}).
\newblock


\bibitem[Nabati and Ghorashi(2023)]%
        {nabati2023real}
\bibfield{author}{\bibinfo{person}{Mohammad Nabati} {and} \bibinfo{person}{Seyed~Ali Ghorashi}.} \bibinfo{year}{2023}\natexlab{}.
\newblock \showarticletitle{A real-time fingerprint-based indoor positioning using deep learning and preceding states}.
\newblock \bibinfo{journal}{\emph{Expert Systems with Applications}}  \bibinfo{volume}{213} (\bibinfo{year}{2023}), \bibinfo{pages}{118889}.
\newblock


\bibitem[Nagarajan et~al\mbox{.}(2024)]%
        {nagarajan2024egoenv}
\bibfield{author}{\bibinfo{person}{Tushar Nagarajan}, \bibinfo{person}{Santhosh~Kumar Ramakrishnan}, \bibinfo{person}{Ruta Desai}, \bibinfo{person}{James Hillis}, {and} \bibinfo{person}{Kristen Grauman}.} \bibinfo{year}{2024}\natexlab{}.
\newblock \showarticletitle{EgoEnv: Human-centric environment representations from egocentric video}.
\newblock \bibinfo{journal}{\emph{Advances in Neural Information Processing Systems}}  \bibinfo{volume}{36} (\bibinfo{year}{2024}).
\newblock


\bibitem[Nilsson and Akenine-M{\"o}ller(2020)]%
        {ssim}
\bibfield{author}{\bibinfo{person}{Jim Nilsson} {and} \bibinfo{person}{Tomas Akenine-M{\"o}ller}.} \bibinfo{year}{2020}\natexlab{}.
\newblock \showarticletitle{Understanding ssim}.
\newblock \bibinfo{journal}{\emph{arXiv preprint arXiv:2006.13846}} (\bibinfo{year}{2020}).
\newblock


\bibitem[O'Brien et~al\mbox{.}(2022)]%
        {o2022optimizing}
\bibfield{author}{\bibinfo{person}{Katherine O'Brien}, \bibinfo{person}{Sophia~W Light}, \bibinfo{person}{Sara Bradley}, {and} \bibinfo{person}{Lee Lindquist}.} \bibinfo{year}{2022}\natexlab{}.
\newblock \showarticletitle{Optimizing voice-controlled intelligent personal assistants for use by home-bound older adults}.
\newblock \bibinfo{journal}{\emph{Journal of the American Geriatrics Society}} \bibinfo{volume}{70}, \bibinfo{number}{5} (\bibinfo{year}{2022}), \bibinfo{pages}{1504--1509}.
\newblock


\bibitem[Olson et~al\mbox{.}(2011)]%
        {olson2011diffusion}
\bibfield{author}{\bibinfo{person}{Katherine~E Olson}, \bibinfo{person}{Marita~A O’Brien}, \bibinfo{person}{Wendy~A Rogers}, {and} \bibinfo{person}{Neil Charness}.} \bibinfo{year}{2011}\natexlab{}.
\newblock \showarticletitle{Diffusion of technology: frequency of use for younger and older adults}.
\newblock \bibinfo{journal}{\emph{Ageing international}} \bibinfo{volume}{36}, \bibinfo{number}{1} (\bibinfo{year}{2011}), \bibinfo{pages}{123--145}.
\newblock


\bibitem[{OpenCV}(2023)]%
        {opencv_calibration}
\bibfield{author}{\bibinfo{person}{{OpenCV}}.} \bibinfo{year}{2023}\natexlab{}.
\newblock \bibinfo{title}{Camera Calibration}.
\newblock \bibinfo{howpublished}{\url{https://docs.opencv.org/4.x/dc/dbb/tutorial_py_calibration.html}}.
\newblock
\newblock
\shownote{Accessed: 2024-04-29}.


\bibitem[Opfermann et~al\mbox{.}(2017)]%
        {opfermann2017communicative}
\bibfield{author}{\bibinfo{person}{Christiane Opfermann}, \bibinfo{person}{Karola Pitsch}, \bibinfo{person}{Ramin Yaghoubzadeh}, {and} \bibinfo{person}{Stefan Kopp}.} \bibinfo{year}{2017}\natexlab{}.
\newblock \showarticletitle{The Communicative Activity of" Making Suggestions" as an Interactional Process: Towards a Dialog Model for HAI}. In \bibinfo{booktitle}{\emph{Proceedings of the 5th International Conference on Human Agent Interaction}}. \bibinfo{pages}{161--170}.
\newblock


\bibitem[Oshimi et~al\mbox{.}(2023)]%
        {oshimi2023locatar}
\bibfield{author}{\bibinfo{person}{Hiroto Oshimi}, \bibinfo{person}{Monica Perusqu{\'\i}a-Hern{\'a}ndez}, \bibinfo{person}{Naoya Isoyama}, \bibinfo{person}{Hideaki Uchiyama}, {and} \bibinfo{person}{Kiyoshi Kiyokawa}.} \bibinfo{year}{2023}\natexlab{}.
\newblock \showarticletitle{LocatAR: An AR Object Search Assistance System for a Shared Space}. In \bibinfo{booktitle}{\emph{Proceedings of the Augmented Humans International Conference 2023}}. \bibinfo{pages}{66--76}.
\newblock


\bibitem[Pradhan et~al\mbox{.}(2019)]%
        {pradhan2019phantom}
\bibfield{author}{\bibinfo{person}{Alisha Pradhan}, \bibinfo{person}{Leah Findlater}, {and} \bibinfo{person}{Amanda Lazar}.} \bibinfo{year}{2019}\natexlab{}.
\newblock \showarticletitle{" Phantom Friend" or" Just a Box with Information" Personification and Ontological Categorization of Smart Speaker-based Voice Assistants by Older Adults}.
\newblock \bibinfo{journal}{\emph{Proceedings of the ACM on human-computer interaction}} \bibinfo{volume}{3}, \bibinfo{number}{CSCW} (\bibinfo{year}{2019}), \bibinfo{pages}{1--21}.
\newblock


\bibitem[Ramos et~al\mbox{.}(2016)]%
        {ramos2016designing}
\bibfield{author}{\bibinfo{person}{Laura Ramos}, \bibinfo{person}{Elise Van Den~Hoven}, {and} \bibinfo{person}{Laurie Miller}.} \bibinfo{year}{2016}\natexlab{}.
\newblock \showarticletitle{Designing for the Other'Hereafter' When Older Adults Remember about Forgetting}. In \bibinfo{booktitle}{\emph{Proceedings of the 2016 CHI Conference on Human Factors in Computing Systems}}. \bibinfo{pages}{721--732}.
\newblock


\bibitem[Rom{\'a}n et~al\mbox{.}({[n.\,d.]})]%
        {romanvoice}
\bibfield{author}{\bibinfo{person}{Adri{\'a}n~Valera Rom{\'a}n}, \bibinfo{person}{DP Mart{\i}nez}, \bibinfo{person}{AL Murciego}, \bibinfo{person}{Diego~M Jim{\'e}nez-Bravo}, {and} \bibinfo{person}{Juan~F de Paz}.} \bibinfo{year}{[n.\,d.]}\natexlab{}.
\newblock \bibinfo{title}{Voice assistant application for avoiding sedentarism in elderly people based on iot technologies, Electron. 10 (8)(2021)}.
\newblock
\newblock


\bibitem[Schryer and Ross(2013)]%
        {schryer2013use}
\bibfield{author}{\bibinfo{person}{Emily Schryer} {and} \bibinfo{person}{Michael Ross}.} \bibinfo{year}{2013}\natexlab{}.
\newblock \showarticletitle{The use and benefits of external memory aids in older and younger adults}.
\newblock \bibinfo{journal}{\emph{Applied Cognitive Psychology}} \bibinfo{volume}{27}, \bibinfo{number}{5} (\bibinfo{year}{2013}), \bibinfo{pages}{663--671}.
\newblock


\bibitem[Schwind and Henze(2020)]%
        {schwind2020anticipated}
\bibfield{author}{\bibinfo{person}{Valentin Schwind} {and} \bibinfo{person}{Niels Henze}.} \bibinfo{year}{2020}\natexlab{}.
\newblock \showarticletitle{Anticipated User Stereotypes Systematically Affect the Social Acceptability of Mobile Devices}. In \bibinfo{booktitle}{\emph{Proceedings of the 11th Nordic Conference on Human-Computer Interaction: Shaping Experiences, Shaping Society}}. \bibinfo{pages}{1--12}.
\newblock


\bibitem[Seo et~al\mbox{.}(2021)]%
        {seo2021overthere}
\bibfield{author}{\bibinfo{person}{Hyunggoog Seo}, \bibinfo{person}{Jaedong Kim}, \bibinfo{person}{Kwanggyoon Seo}, \bibinfo{person}{Bumki Kim}, {and} \bibinfo{person}{Junyong Noh}.} \bibinfo{year}{2021}\natexlab{}.
\newblock \showarticletitle{Overthere: A simple and intuitive object registration method for an absolute mid-air pointing interface}.
\newblock \bibinfo{journal}{\emph{Proceedings of the ACM on Interactive, Mobile, Wearable and Ubiquitous Technologies}} \bibinfo{volume}{5}, \bibinfo{number}{3} (\bibinfo{year}{2021}), \bibinfo{pages}{1--24}.
\newblock


\bibitem[Sidner et~al\mbox{.}(2018)]%
        {sidner2018creating}
\bibfield{author}{\bibinfo{person}{Candace~L Sidner}, \bibinfo{person}{Timothy Bickmore}, \bibinfo{person}{Bahador Nooraie}, \bibinfo{person}{Charles Rich}, \bibinfo{person}{Lazlo Ring}, \bibinfo{person}{Mahni Shayganfar}, {and} \bibinfo{person}{Laura Vardoulakis}.} \bibinfo{year}{2018}\natexlab{}.
\newblock \showarticletitle{Creating new technologies for companionable agents to support isolated older adults}.
\newblock \bibinfo{journal}{\emph{ACM Transactions on Interactive Intelligent Systems (TiiS)}} \bibinfo{volume}{8}, \bibinfo{number}{3} (\bibinfo{year}{2018}), \bibinfo{pages}{1--27}.
\newblock


\bibitem[Stigall et~al\mbox{.}(2019)]%
        {stigall2019older}
\bibfield{author}{\bibinfo{person}{Brodrick Stigall}, \bibinfo{person}{Jenny Waycott}, \bibinfo{person}{Steven Baker}, {and} \bibinfo{person}{Kelly Caine}.} \bibinfo{year}{2019}\natexlab{}.
\newblock \showarticletitle{Older adults' perception and use of voice user interfaces: a preliminary review of the computing literature}. In \bibinfo{booktitle}{\emph{Proceedings of the 31st Australian Conference on Human-Computer-Interaction}}. \bibinfo{pages}{423--427}.
\newblock


\bibitem[Tsiourti et~al\mbox{.}(2014)]%
        {tsiourti2014virtual}
\bibfield{author}{\bibinfo{person}{Christiana Tsiourti}, \bibinfo{person}{Emilie Joly}, \bibinfo{person}{Cindy Wings}, \bibinfo{person}{Maher~Ben Moussa}, {and} \bibinfo{person}{Katarzyna Wac}.} \bibinfo{year}{2014}\natexlab{}.
\newblock \showarticletitle{Virtual assistive companions for older adults: qualitative field study and design implications}. In \bibinfo{booktitle}{\emph{Proceedings of the 8th International Conference on Pervasive Computing Technologies for Healthcare}}. \bibinfo{pages}{57--64}.
\newblock


\bibitem[Ventura et~al\mbox{.}(2014)]%
        {ventura2014global}
\bibfield{author}{\bibinfo{person}{Jonathan Ventura}, \bibinfo{person}{Clemens Arth}, \bibinfo{person}{Gerhard Reitmayr}, {and} \bibinfo{person}{Dieter Schmalstieg}.} \bibinfo{year}{2014}\natexlab{}.
\newblock \showarticletitle{Global localization from monocular slam on a mobile phone}.
\newblock \bibinfo{journal}{\emph{IEEE transactions on visualization and computer graphics}} \bibinfo{volume}{20}, \bibinfo{number}{4} (\bibinfo{year}{2014}), \bibinfo{pages}{531--539}.
\newblock


\bibitem[Victor(2017)]%
        {Dibia2017}
\bibfield{author}{\bibinfo{person}{Dibia Victor}.} \bibinfo{year}{2017}\natexlab{}.
\newblock \showarticletitle{HandTrack: A Library For Prototyping Real-time Hand TrackingInterfaces using Convolutional Neural Networks}.
\newblock \bibinfo{journal}{\emph{GitHub repository}} (\bibinfo{year}{2017}).
\newblock
\urldef\tempurl%
\url{https://github.com/victordibia/handtracking/tree/master/docs/handtrack.pdf}
\showURL{%
\tempurl}


\bibitem[Wang et~al\mbox{.}(2023)]%
        {wang2023enabling}
\bibfield{author}{\bibinfo{person}{Bryan Wang}, \bibinfo{person}{Gang Li}, {and} \bibinfo{person}{Yang Li}.} \bibinfo{year}{2023}\natexlab{}.
\newblock \showarticletitle{Enabling conversational interaction with mobile ui using large language models}. In \bibinfo{booktitle}{\emph{Proceedings of the 2023 CHI Conference on Human Factors in Computing Systems}}. \bibinfo{pages}{1--17}.
\newblock


\bibitem[{World Health Organization}(2024)]%
        {WHO2024}
\bibfield{author}{\bibinfo{person}{{World Health Organization}}.} \bibinfo{year}{2024}\natexlab{}.
\newblock \bibinfo{title}{Ageing and Health}.
\newblock \bibinfo{howpublished}{\url{https://www.who.int/news-room/fact-sheets/detail/ageing-and-health}}.
\newblock
\newblock
\shownote{Accessed: 2024-05-01}.


\bibitem[Xygkou et~al\mbox{.}(2024)]%
        {xygkou2024mindtalker}
\bibfield{author}{\bibinfo{person}{Anna Xygkou}, \bibinfo{person}{Chee~Siang Ang}, \bibinfo{person}{Panote Siriaraya}, \bibinfo{person}{Jonasz~Piotr Kopecki}, \bibinfo{person}{Alexandra Covaci}, \bibinfo{person}{Eiman Kanjo}, {and} \bibinfo{person}{Wan-Jou She}.} \bibinfo{year}{2024}\natexlab{}.
\newblock \showarticletitle{MindTalker: Navigating the Complexities of AI-Enhanced Social Engagement for People with Early-Stage Dementia}. In \bibinfo{booktitle}{\emph{Proceedings of the CHI Conference on Human Factors in Computing Systems}}. \bibinfo{pages}{1--15}.
\newblock


\bibitem[Yagi et~al\mbox{.}(2021)]%
        {gofinder}
\bibfield{author}{\bibinfo{person}{Takuma Yagi}, \bibinfo{person}{Takumi Nishiyasu}, \bibinfo{person}{Kunimasa Kawasaki}, \bibinfo{person}{Moe Matsuki}, {and} \bibinfo{person}{Yoichi Sato}.} \bibinfo{year}{2021}\natexlab{}.
\newblock \showarticletitle{GO-finder: a registration-free wearable system for assisting users in finding lost objects via hand-held object discovery}. In \bibinfo{booktitle}{\emph{26th International Conference on Intelligent User Interfaces}}. \bibinfo{pages}{139--149}.
\newblock


\bibitem[Yan et~al\mbox{.}(2022)]%
        {yan2022camfi}
\bibfield{author}{\bibinfo{person}{Ge Yan}, \bibinfo{person}{Chao Zhang}, \bibinfo{person}{Jiadi Wang}, \bibinfo{person}{Zheng Xu}, \bibinfo{person}{Jianhui Liu}, \bibinfo{person}{Jintao Nie}, \bibinfo{person}{Fangtian Ying}, {and} \bibinfo{person}{Cheng Yao}.} \bibinfo{year}{2022}\natexlab{}.
\newblock \showarticletitle{CamFi: An AI-driven and Camera-based System for Assisting Users in Finding Lost Objects in Multi-Person Scenarios}. In \bibinfo{booktitle}{\emph{CHI Conference on Human Factors in Computing Systems Extended Abstracts}}. \bibinfo{pages}{1--7}.
\newblock


\bibitem[Zulfikar et~al\mbox{.}(2024)]%
        {zulfikar2024memoro}
\bibfield{author}{\bibinfo{person}{Wazeer Zulfikar}, \bibinfo{person}{Samantha Chan}, {and} \bibinfo{person}{Pattie Maes}.} \bibinfo{year}{2024}\natexlab{}.
\newblock \showarticletitle{Memoro: Using Large Language Models to Realize a Concise Interface for Real-Time Memory Augmentation}.
\newblock \bibinfo{journal}{\emph{arXiv preprint arXiv:2403.02135}} (\bibinfo{year}{2024}).
\newblock


\end{thebibliography}

\appendix
\section{Appendix}  

\subsection{Interviews}\label{appendixinterviews}

For the live interviews, our feedback group included four neurologists (D) and one speech pathologist (SP) who have extensive experience with over 100 patients facing neurogenic disorders. Additionally, we gathered insights from caregiver support groups comprising 12 caregivers (C) where 8 were healthy older adults themselves and memory support groups including 12 individuals with memory decline (E), facilitated through a hospital network. We also engaged with individuals with memory loss (OE) and caregivers (OC) through anonymous online forums. 

Guiding questions for each demographic were used to extract insights: 

For caregivers, speech pathologists, and physicians: 
\begin{itemize}
    \item What are the current memory support tools used? 
    \item What is the most difficult part of caregiving and provide scenarios? 
    \item What data do you wish you had to make an accurate assessment of the patient's memory? 
    \item What type of form factor is the best for constant monitoring?
\end{itemize}

For older adults: 
\begin{itemize}
    \item What objects are frequently lost? 
    \item  Describe some scenarios that caused frustration when you did not remember something.
\end{itemize}

Insights are coded and thematically analyzed in the following categories:

\begin{itemize}

\item Issue 1: Object-finding

Description: Finding lost objects is frustrating and time consuming
Examples: Scenarios of losing purse in home, concerns about not recognizing own possessions, examples of frequently lost objects  
Quotes: 
\begin{itemize}
    \item OC8: "My MIL was always missing her purse and was sure someone had stolen it. She would ransack her room all night long. It would be in the closet. This happened often."
    \item OC9: “Hearing aids and purse. Sometimes hearing aids are in the purse. Drives my poor dad nuts.” 
    \item OC10:  “The phone chargers! Always misplaces her iPhone cord, looks for another one, takes someone else's type C, then replaces that with some random micro USB or whatever... ugh... The issue here is that she doesn't know what is hers and what isn't, and ends up taking and keeping or throwing away other people's belongings.” 
    \item OC11: “Glasses and dentures”  
\end{itemize}

\item Issue 2: Safety reminders

2A: Anecdotal Scenarios
Description: Anecdotal scenarios of daily tasks that would be often forgotten and could cause safety concerns
Examples: scenarios of short term memory loss like forgetting to take important items when leaving the home, quotes from speech pathologists or doctors referencing certain scenarios where safety reminders would be most useful 
Quotes: 
\begin{itemize}
    \item D1: “You have the most leverage when developing systems to ensure safety, especially during activities like cooking.” 
    \item E6: “Didn’t have a wallet or purse when I left the house” 
    \item E7: “I forgot to put milk back in the fridge and my wife got mad”
    \item E8: “I have issues with short term memory- when going to the doctor if you’re asked what did you do today being like I have no idea, it’s hard to remember the day to day” 
\end{itemize}

2B: Current safety monitoring and reminders
Description: Current methods of how caregivers give reminders to older adults including verbal or written reminders, how older adults would respond to the current system and problems associated 
Examples: type of reminder provided and reaction to that reminder from older adults, opinions from caregivers about how current systems lead to caregiver burden
Quotes: 
\begin{itemize}
    \item SP1: “Speech pathologists use sequence boards to help with safety trained tasks so a log of safety related tasks can reduce our burden as well as caregiver burden” 
    \item C10: “extra burden of decision making is being deferred to caregivers” 
    \item C12: “use scraps of paper to write a schedule, then I got a portable whiteboard and big letters that I print out first thing in the morning” 
    \item OC1: “I put up 'reminders' for my husband suffering from dementia and they didn't do a damn bit of good”  
    \item OC2: “My written reminders got ignored as their reality says it’s not needed. My verbal reminders often are met with combative passive aggressive responses as their reality says it’s not needed. The reality is the reminders are needed for them to function properly." 
    \item OC3: “I found it necessary to remove things, place things out of reach, and supervise with cameras and movement alarms to keep my person safe and healthy.” 
    \item OC4: “The medication was also a problem for my person who forgets what was taken or not even with a labeled pill dispenser. For safety I choose to dispense all medications”  
    \item OC5: “Too many signs or signs up too long become white noise. Be selective. She still needs some verbal prompts though. You can certainly try your post to note reminders about safety, but I'd expect limited returns and possible irritation.”  
    \item OC6: “We have notes everywhere, but he doesn't pay attention to any of them. Whenever he is performing any action, I just watch over him and make sure he has what he needs. It can be exhausting. If he's going to the shower, I ask, "Do you have your towel, clothes, deodorant?" He'll answer yes every time, then go back and forth getting what he forgot” 
    \item OC7: “had a whiteboard directly in front of the chair where my mom slept. It listed the basic facts of her life (her name and location included) and what was about to happen next. If someone came over, I put the name and schedule on the board. After a while, that board was a lifeline; she knew to look at it when she was confused.” 
    \item OC8: "Some other things might be to lock the door, make sure faucets are turned off, charge your phone (if cell phone), don’t eat expired food in fridge, never give out information to anyone over phone / email (however to convey she should avoid scammers)."
\end{itemize}

\item Issue 3: Confabulation of past events 
Description: inability to remember longer sequences of actions or tasks in the past and approaches to currently solve  
Examples: physicians providing suggestions that passive remote tracking is useful, caregivers and speech pathologists providing anecdotes of ways to combat this 
Quotes: 
\begin{itemize}
    \item D1:  “remote patient monitoring is very helpful and tracking is helpful since we use a lot of cognitive assessment. Patients thinks they’re fine but their spouse comes in and you can’t target anyone in denial”
    \item D2: “Passive sensors are better since caregivers are usually filling out subjective questionnaires of activities of daily living form and functional assessment”
    \item E6: My biggest issue with his memory is that he'll tell us he has something when he doesn't. He lies daily.
    \item E7: While it is technically lying, the term is confabulation. He likely doesn't remember if he grabbed an item or finished a task or if he's done the ask repeatedly, his days are mixed up and he thinks he did it. Confabulation is used to fill in a memory the person doesn't have or a blank in a series of memories. It's very frustrating. My mom deals with the same thing. Not as intensely but it's especially evident when having conversations about certain things.
    \item E8: When I left for short trips, I would give her a printed page of instructions: where I was, how to ring the alarm button, when I would call, what she should do. Reviewing the whiteboard with her became part of our morning routine. Her daily schedule at that point was reduced to TV and aides coming into the house. I would write the aide's name, the hours, where I was, and when I would be back. TV, especially sports, became a highlight of the day at the scheduled time. When she moved to a nursing home at the end, I got a second white board and propped both up on the desk, because her questions had doubled. I also laminated a copy of the information for the aides to read to her at night. 
    \item SP1: A memory book is a diary of daily activities that can help patients remember events. The memory book can be used as a therapy tool and a functional guide.
\end{itemize}
    
\end{itemize}
%To enable the features described above, MemPal consists of a vision system and language system (Figure \ref{fig:overall}). The vision system creates an automated diary of activity (example seen in \ref{fig:activities} and the language system powered by an LLM agent understands user voice queries and responds accordingly using the vision context. The underlying technology architecture consisting of camera preprocessing, visual feature extraction, and LLM query/ processing are reused for all features with slight modifications to fit to the feature. Vector databases pertaining to location calibration (CalibrationDB), automated diary (ActivitiesDB), and summarized diary (HigherLevelDB) are persisted locally on-device while location trajectory and safety reminder history are persisted on Google Firebase Realtime database to be able to share them with authorized caregivers. User queries to MemPal (QueriesDB) are also stored locally on-device. 

\section{System Implementation}\label{systemimp}
\subsection{Vision System}

\begin{figure*}
    \centering
    \includegraphics[scale=0.27]{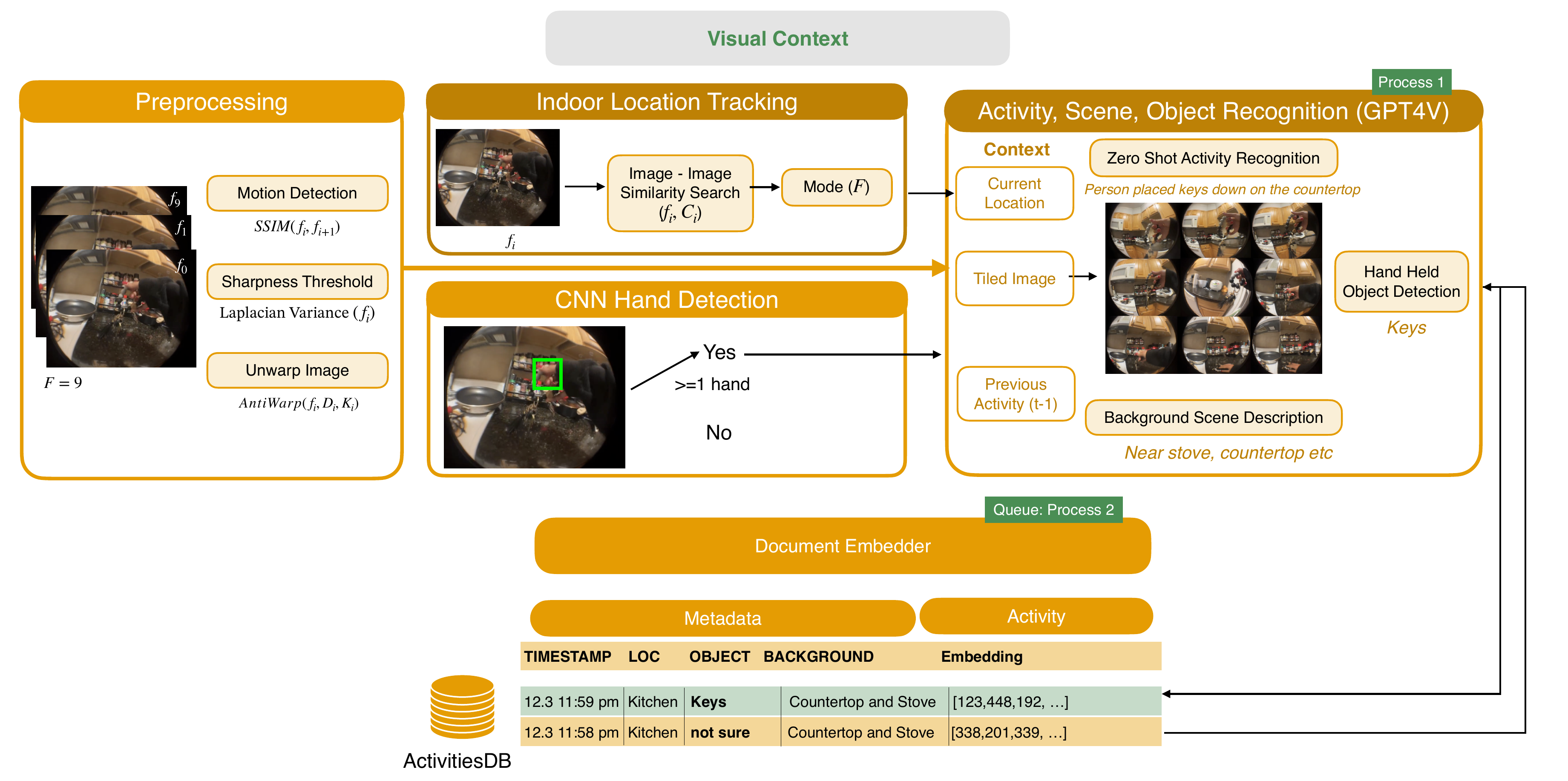}
    \caption{Vision system workflow: Every frame in a frame batch of nine is first pre-processed to create a tiled image (Left) before indoor room localization. Then, if the user's hand(s) is detected in the frame of the wearable camera (Center Bottom), a Vision Language Model (VLM) specifically GPT4-V (Right) that uses the detected location context and previous activity, describes the hand-held object, activity, and background.}
    \label{fig:visual}
\end{figure*}

\subsubsection{Camera Pre-processing}
 To minimize noise and reduce unnecessary computation, frames with low sharpness (Laplacian variance < 25) and adjacent frames with high similarity (structural similarity index score (SSIM) <0.95) are discarded \cite{ssim}. A 235$^{\circ}$ fish-eye lens is used to increase FOV and standard camera calibration was applied to correct distortions from the lens \cite{opencv_calibration}.

\subsubsection{Indoor Location Tracking}

\begin{figure*}
    \centering
    \includegraphics[scale=0.27]{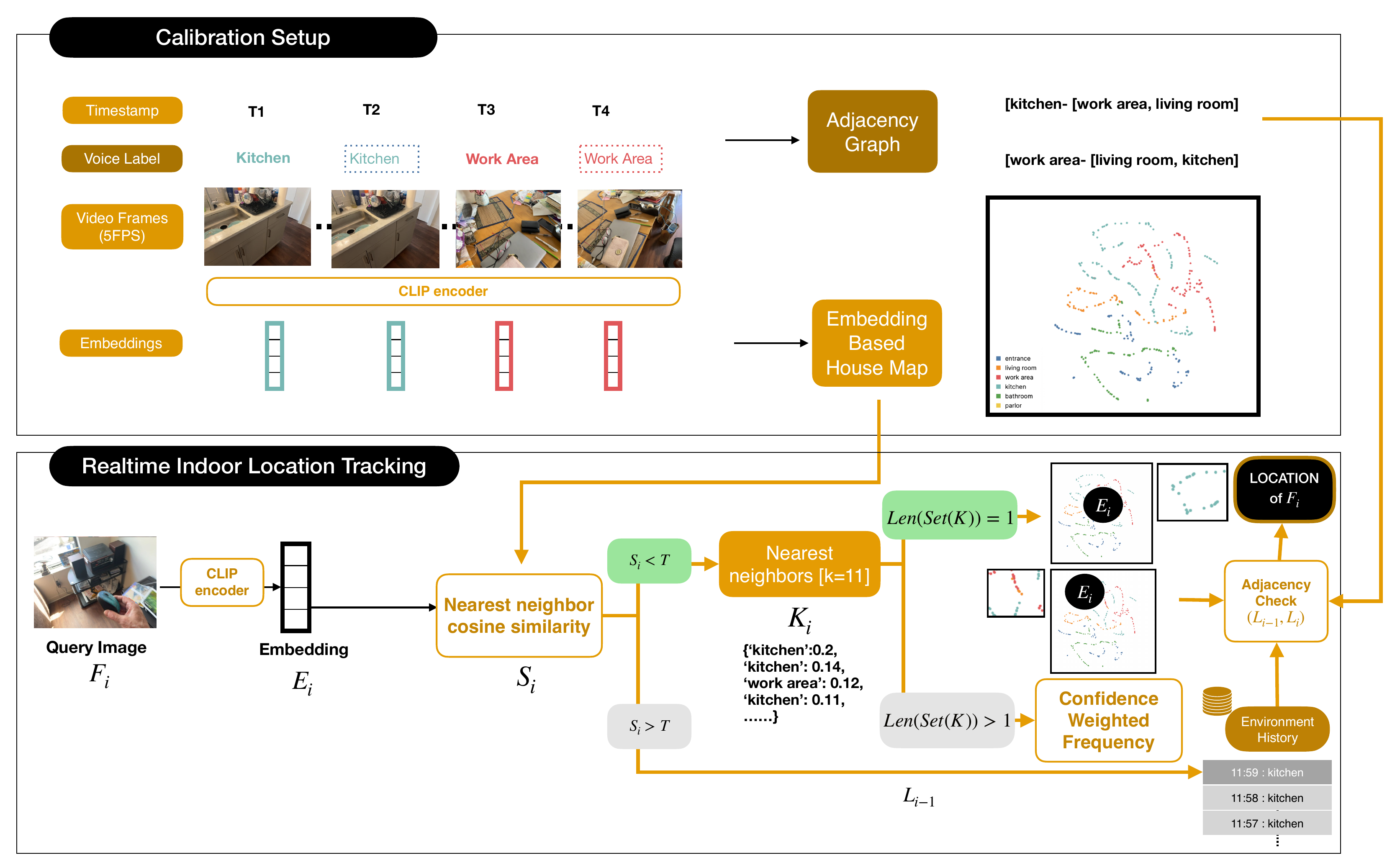}
    \caption{Details of the indoor location tracking algorithm including calibration phase and localization.}
    \label{fig:location-im}
\end{figure*}

MemPal uses the monocular egocentric camera to perform embedding based real-time indoor location tracking without requiring other sensors like GPS, unlike high compute algorithms like SLAM \cite{ventura2014global}. Embeddings capture a semantic representation within the context of the user’s home allowing for a more personalized search. A calibration video is first processed into (1) an embedding map and (2) a room adjacency list to represent an initial spatial map of the user's home. This home representation is used as context for real time localization within the map. The algorithm is described in Figure \ref{fig:location-im}. 

\begin{itemize}
    \item \textbf{Calibration Setup}: The calibration video frames are uniformly sampled and embedded using a pretrained CLIP-ViT-B-32 transformer model. These embeddings are stored in vector database named \textit{CalibrationDB} for fast retrieval along with the voice-labeled room name as metadata for each embedding. The temporal sequence of the room labels are used to generate an adjacency list that maps each room to a set of connected or adjacent rooms, ensuring that new locations detected every second is adjacent to the previous one. With additional usage, accumulating data enriches the location embedding map allowing it to adapt to the home's configuration environment and improve its robustness.

    %Current methodologies for camera based real-time mapping and localization that achieve high accuracy, such as SLAM (Simultaneous Localization and Mapping) (ex. ORB-SLAM\cite{mur2015orb}), Structure from Motion (SfM) (ex. COLMAP\cite{fisher2021colmap}), or 3D point cloud generation \cite{li2021image}, require substantial computational resources, are sensitive to motion and changes in illumination, or require extensive and precise initial calibration. This poses challenges for implementation on wearable devices and deployment for older adults, particularly when aiming to preserve privacy and perform processing on-devices. %Bluetooth beacons coupled with ML triangulation methods require complex installation processes are not cost-efficient for large-scale deployment.

    %A recent technique, EgoEnv\cite{nagarajan2024egoenv}, leverages egocentric video walkthrough to create human-centric environment representations for initial calibration but doesn't allow for customized labels based on individual home layouts. A more accurate method, EgoVLP\cite{qinghong2022egocentric}, integrates egocentric video with language pretraining models to enhance the semantic understanding of the environment, providing contextual descriptions for objects, but requires significantly more training data. 
    \item \textbf{Real-time Location Tracking}: During localization, a heuristic-based ensemble approach identifies the correct room the user is in in relation to the calibration map and adjacency room list. Algorithm is illustrated in \ref{fig:location-im} and listed below.

    \begin{itemize}
            \item \textbf{Step 1}: Calculate image similarities: Cosine similarity metric is calculated between the image embedding of $F_{i}$ and the calibration map embeddings (range:[0,2] where 0 denotes the same image and 2 denotes orthogonal images), to represent the distance between embeddings. 
            \item \textbf{Step 2}: When performing similarity search, if the cosine similarity between the current location’s embedding and the top k=1 nearest neighbor is higher than the confidence threshold (T) = 0.22 (which means less confident), then the previous location is set $L_(i-1)$. 
            \item \textbf{Step 3}: If the confidence threshold of the top k=1 image is within an appropriate level, the unique locations from the top k=11 location candidates are determined. 
                \subitem If the unique locations within the set of top candidates is equal to 1, this signifies high detection confidence which means that the query location seems to be within a cluster of location points for that area. 
                \subitem If the unique locations within the set of top candidates is greater than 1, this signifies lower confidence which means that potentially the image is in the boundary between two clusters of images or location clusters are overlapping which often happens if rooms within a household have very similar visual resemblance. The most frequent element in the list of location choices is then returned, weighted by confidence level.
        
            Environment features are encoded through the history of previously detected locations. To reduce the embedding process time, we decided to use the compressed history of locations rather than concatenated embeddings like EgoEnv \cite{nagarajan2024egoenv}. If the resulting new location determined is not adjacent to the previous location, then the current location detected is set to the previous confident location.
            \item \textbf{Step 4}: The mode among the list of tiled locations (len<=9) detected within each frame is then set as the final location.   
            
        \end{itemize}

\end{itemize}

\subsubsection{Hand Detection}

 A deep neural network trained on the EgoHands Dataset \cite{Dibia2017} is implemented in Tensorflow 2.0 which tracks the presence and relative location of both hands in the video frame. Once 1 or 2 hands are detected within an image frame, then a tiled image is created for object and activity detection described in the next section. A tiled image compresses a batch of 9 sequential processed frames into a 3x3 matrix, arranging them from the top-left to the bottom-right, reducing the input size by a factor of 9 for efficient VLM processing.  

\subsubsection{Object and Activity Detection}

A vision-language model then analyzes the tiled image to determine the (1) hand-held object, (2) activity, and (3) background scene description including furniture positions and colors in a zero-shot manner. Objects and activities do not need to be tagged or trained prior to usage, to ensure generalizability and personalization. A textual description of the activity is generated by discerning the temporal image sequence in the tiled image along with the background objects in the frame of view. We use GPT-4 vision to extract this information, due to its better performance in comparison to other models for egocentric video understanding \cite{vidcompare}. Context about the previous activity and the current location is provided to improve the accuracy of the detection of the current activity. Once the real-time description is generated, the textual data is embedded using the all-MiniLM-L6-v2 Sentence Transformer model and stored within a 384 dimension dense vector space along with metadata including timestamp, location, object, background description (\textit{ActivitiesDB}).

\subsection{Language System}

To respond to explicit user queries about objects or past events or implicit queries during safety task monitoring, a question and answer system is created which includes query categorization and feature-specific response formation utilizing the text-based ActivitiesDB. \newline

\subsubsection{Query Processing}
First, the user query is transcribed through Deepgram’s streaming speech-to-text API \cite{deepgramspeech}. Then after the wakeword: “Pal” is detected, queries are categorized into one of the following categories using an LLM model (OpenAI GPT-3.5-turbo): (i) objects, (ii) past, (iii) future, (iv) followup, or (v) none. If the query is categorized as “objects,” the object in question is extracted within the sentence. Query categorization allows users to converse with MemPal in a naturalistic way while ensuring to respond to only memory-based relevant questions.

\subsubsection{Safety Tasks}
%\newline
\begin{figure*}
    \centering
    \includegraphics[scale=0.27]{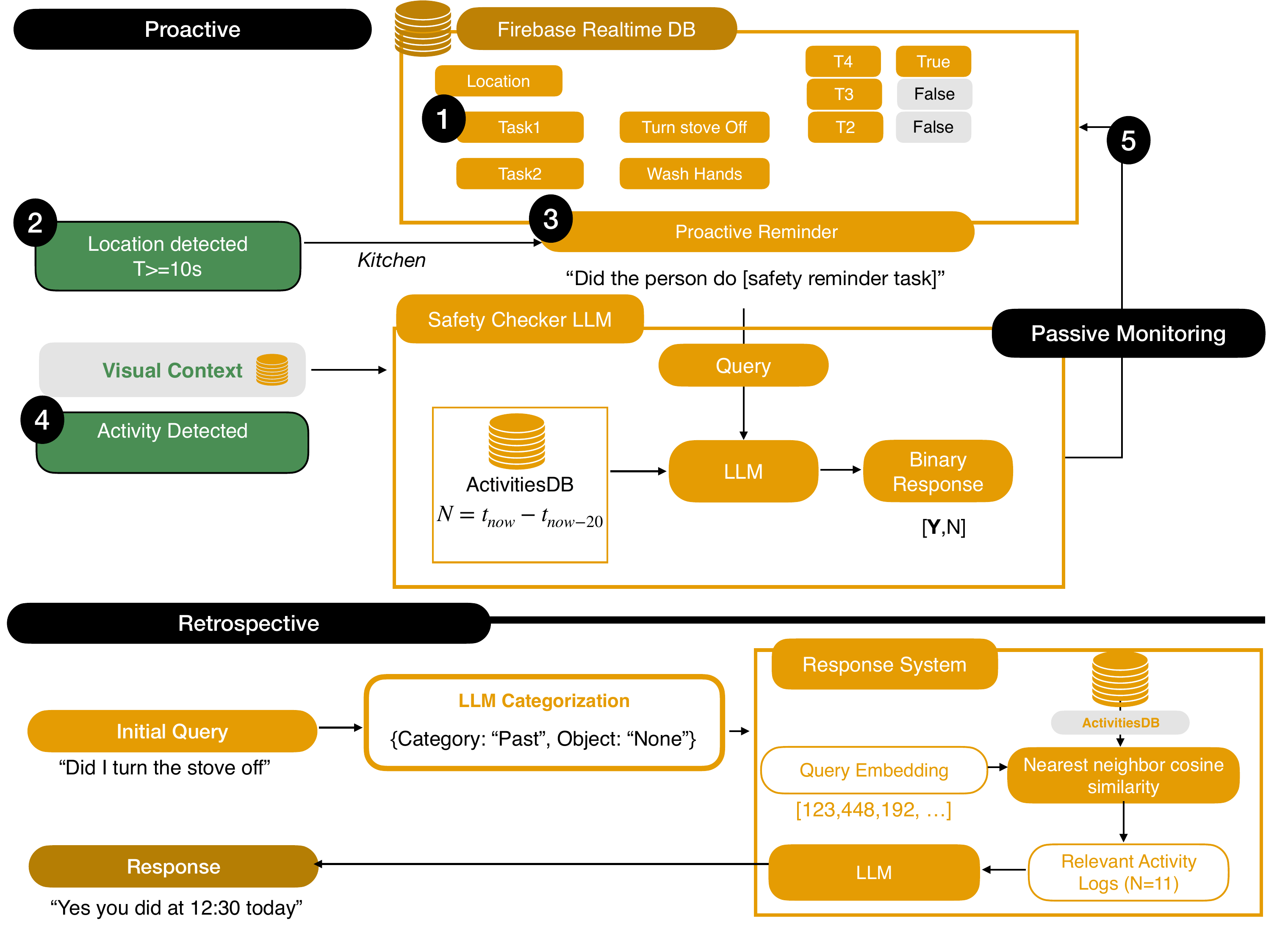}
    \caption{Safety Reminders Implementation demonstrating proactive reminders, retrospective task remembering, and passive monitoring of these safety reminders. This workflow first demonstrates how (1) reminders that are inputted within the MemPal app are stored (2) If location is detected (3) a context-based proactive reminder is triggered and (4) status of the reminder updated on the app upon activity is completed (logged in ActivitiesDB) and  (5) how the system uses a similar query response backend as object retrieval to answer user questions about tasks.}   
    \label{fig:enter-label}
\end{figure*}

\textbf{Proactive}: Text-based safety reminders that are inputted in the app and grouped by location are pushed to the Firebase Database which can be accessed by the API. These reminders are formatted in a question form (ex. "Did you remember to [safety reminder]") and provided as audio output using OpenAI Whisper’s text-to-speech model depending on the user's context. A queue of reminders for each location is set and popped upon completion of the task. 

\textbf{Retrospective}: If individuals query the MemPal system about the completion of these safety tasks or any past action, the query is first categorized into the "past" category before another RAG chain is invoked (similar to Section \ref{objectretrieval}).

\textbf{Passive task completion monitoring}: A safety checker LLM agent using a RAG chain approach similar to Section \ref{objectretrieval} assesses the completion of safety tasks in real-time using the last 2 minutes of activity data from ActivitiesDB. A question in the form of “Did the person complete the safety task: [reminder]?” is set as the query for the agent. The output is a binary 'y' or 'n' response, transmitted to the Firebase app, which updates both the safety task status and completion time.

\label{fig:demographics}
\begin{table*}
    \centering
    \begin{tabular}{p{0.05\linewidth}p{0.1\linewidth}p{0.09\linewidth}p{0.09\linewidth}p{0.08\linewidth}p{0.08\linewidth}p{0.13\linewidth}p{0.25\linewidth}}
        \toprule
        \textbf{ID} & \textbf{Misplace objects} & \textbf{Leave behind objects} & \textbf{Forget intended actions} & \textbf{Forget errands} & \textbf{Forget medication} & \textbf{Forget to buy something} & \textbf{Objects frequently misplaced} \\
        \midrule
        P1 & 2 & 1 & 2 & 1 & 1 & 1 & Reading glasses, keys (house or car)\\
        P2 & 2 & 2 & 2 & 2 & 1 & 3 & cooking objects, desks \\
        P3 & 3 & 3 & 3 & 3 & 1 & 2 & Glasses\\
        P4 & 1 & 2 & 2 & 2 & 1 & 2 & glasses, keys\\
        P5 & 2 & 2 & 2 & 2 & 2 & 2 & -\\
        P6 & 3 & 3 & 3 & 1 & 1 & 1 & Nothing\\
        P7 & 2 & 3 & 3 & 2 & 2 & 2 & Phone\\
        P8 & 3 & 3 & 1 & 1 & 1 & 1 & -\\
        P9 & 2 & 2 & 2 & 1 & 1 & 1 & phone, tea cup, glasses, headband\\
        P10 & 2 & 2 & 2 & 1 & 1 & 2 & reading glasses, cell phone\\
        P11 & 4 & 2 & 4 & 2 & 2 & 2 & shoes/slippers, ipad, phone, calendar, coffee mug, tools\\
        P12 & 2 & 2 & 2 & 2 & 1 & 2 & -\\
        P13 & 2 & 2 & 3 & 2 & 1 & 2 & phone\\
        P14 & 2 & 1 & 1 & 2 & 1 & 2 & Reading glasses, phone, keys, documents, cash\\
        P15 & - & - & - & - & - & - & papers, post-up notes, glasses\\

        \midrule
        \bottomrule
    \end{tabular}
    \caption{Frequency of forgetting events: 1: Never, 2: Few times a month, 3: 2-6 times a week, 4: Multiple Times a day}
    \label{tab:demographics}
\end{table*}

\begin{figure*}
  \centering
  \includegraphics[scale=0.40]{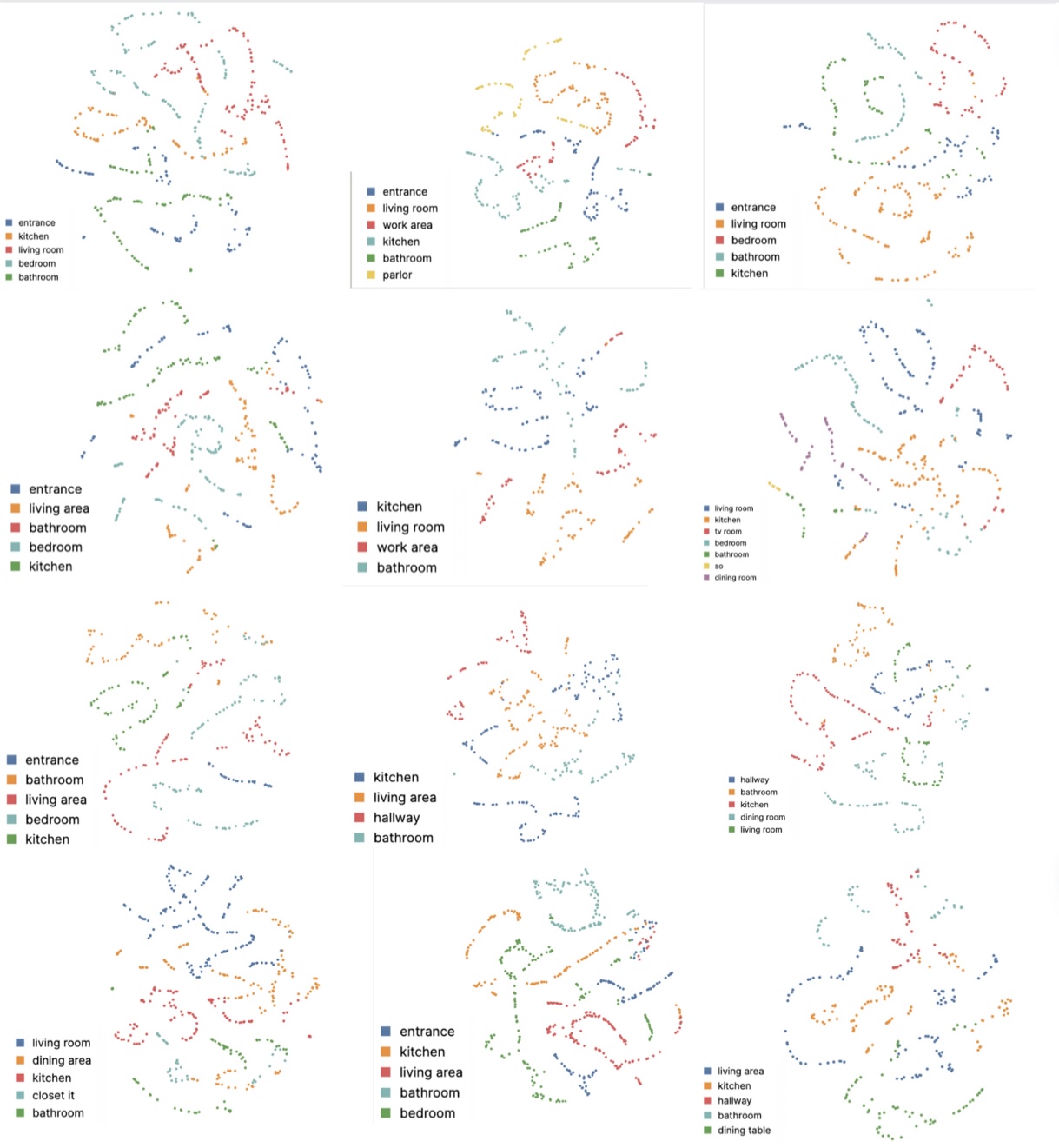}
  \caption{The participant house embedding maps (computed by Fig. \ref{fig:location-im}) show home variability with no clear clusters, indicating that standard clustering or CNN room detection models are unsuitable. The map displays 2D image embeddings, color-coded by room after dimensionality reduction.}
  
  \label{roomdistribution}
\end{figure*}

\begin{figure*}
    \centering
    \includegraphics[scale=0.35]{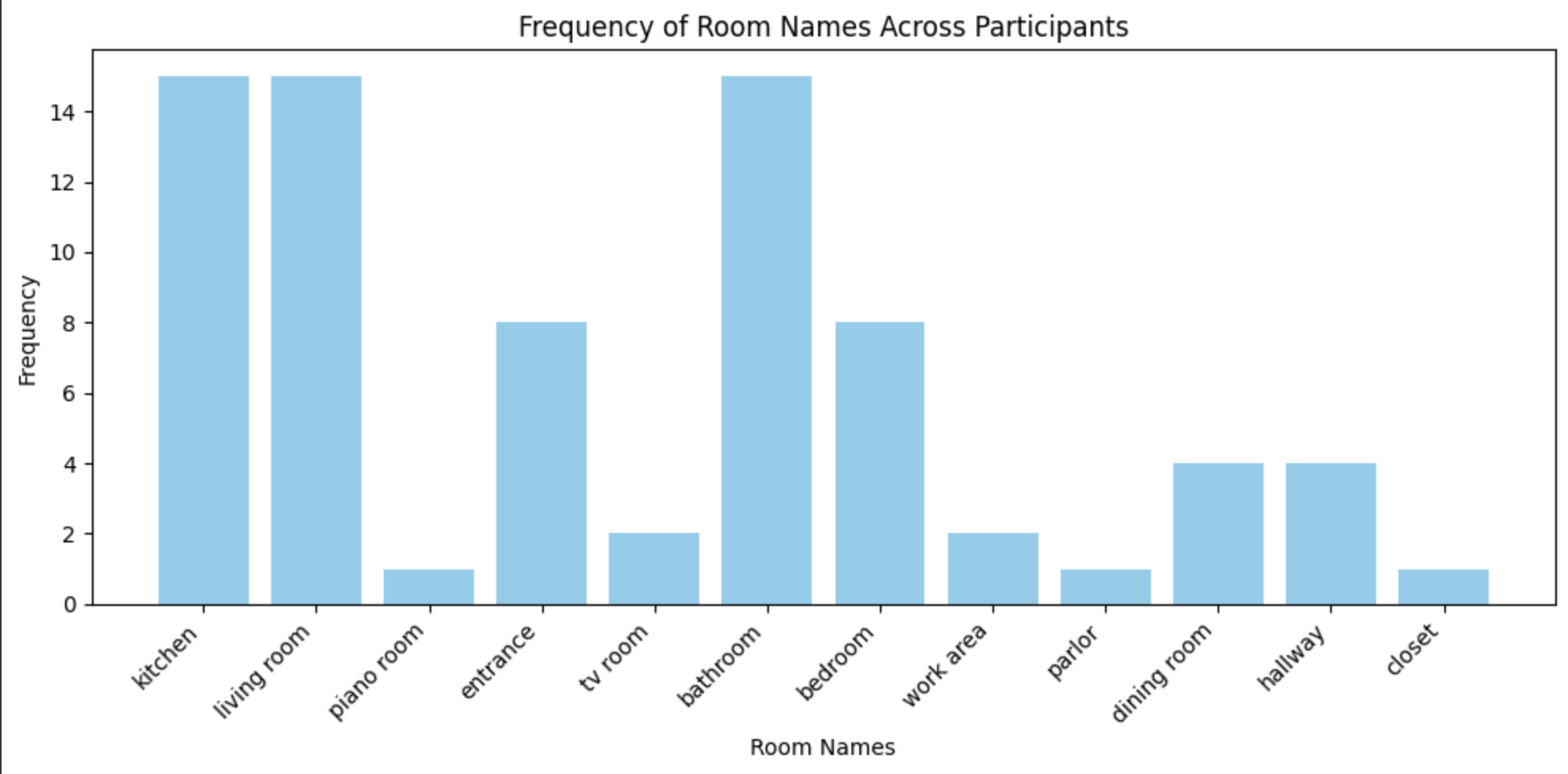}
    \caption{This shows the frequency distribution of different rooms across the various participant homes to illustrate high variation among names and the need for personal room labeling.}
    \label{fig:roomfreq}
\end{figure*}

\section{Questionnaires}
\subsection{Post Object Retrieval Task Questionnaire}
\label{appendixPostObjectRetrieval}

We measured eight aspects using a 7-point Likert scale (1=strongly disagree, 7=strongly agree).

\begin{enumerate}
    \item \textbf{Helpfulness of Object Retrieval Feature}: ``How helpful was this object retrieval feature?''
    \item \textbf{Likelihood of Future Use}: ``How likely are you to use MemPal for finding objects in the future?''
    \item \textbf{Satisfaction with Responses}: ``How satisfied were you with the responses from MemPal?''
    \item \textbf{Ease of Use}: ``How easy was the device to use?''
    \item \textbf{Appropriateness of Answer Length}: ``I felt that the length of the answers was appropriate.''
    \item \textbf{Trust in System's Responses}: ``How much do you trust the systems’ responses?''
    \item \textbf{Usefulness of Camera Feed}: ``Was the camera feed useful?''
    \item \textbf{Reliance on Camera Feed vs Background Descriptions}: ``How much did you rely on the camera feed vs the background descriptions from MemPal to find the object?''
\end{enumerate}

We also included open-ended questions to gather additional insights.

\begin{enumerate}
    \item ``What do you think about this object retrieval feature?''
    \item ``How do you think it could be improved or what could be added?''
\end{enumerate}

\subsection{Open-Ended Interview}
\label{appendixOpenEndedInterview}
We included open-ended questions to gather additional insights from Interview.\label{secinterview}

\begin{enumerate}
    \item \textbf{Overall Thoughts}: ``What are your overall thoughts on the system?''
    \item \textbf{Troubleshooting}: ``Did you have any trouble using the system? If so, what kind of trouble did you have?''
    \item \textbf{Comfort in Sharing Data}: ``How comfortable are you with having data about safety reminders and summary of activities shared with your caregiver, family, or doctor? (Comfort in sharing data)''
    \item \textbf{Confusion During Calibration}: ``For the calibration phase, is there any part of the flow where you were confused?''
    \item \textbf{Instructions for Calibration}: ``Do you feel like you have enough instruction on how to perform the calibration, and if not, what instructions would be most helpful?''
    \item \textbf{System Changes}: ``What would you change in the system? What do you want the system to do and not do?''
    \item \textbf{Current Memory Aids}: ``What memory aids do you currently use?''
\end{enumerate}

\subsection{Post NASA-TLX Questions}
\label{appendixSelfReportedConfidence}
We measured two aspects using a 7-point Likert scale (1=strongly disagree, 7=strongly agree).\label{secinterview}

\begin{enumerate}
    \item  ''I was confident about finding the objects with [condition].''
    \item ''I found it difficult to recall where the objects were with [condition]''
\end{enumerate}

\end{document}